# A statistical mechanical study of mechanical dual molecular braiding

**Dominic Lee**[1,a.)]

[1]Department of Chemistry, Imperial College London, SW7 2AZ, London

## Abstract

This is an archival document; the work contains theoretical development to be used in new publications and to stimulate further development. Here, we develop a statistical mechanical treatment of a braid formed of two molecules subject to an external pulling force and an applied moment at one of the braid ends. The braids, considered here, are those whose axis is straight when thermal fluctuations are absent. Firstly, the geometry and topology of the fluctuating braid are characterized and an energy functional is built, including a generalized interaction potential between the two molecules. To obtain expressions for the free energy, we use variational approximations for two cases. In one case, the interaction potential is considered weak and steric interactions between the two molecules determine the mean squared amplitude of fluctuations in the braid radius. In the second case, where the interactions are strong, the mean squared amplitude is self consistently calculated and the effects of the intermolecular interaction on the twisting of the two molecules in the braid are considered. In the last case, two specific examples are considered. The first of these is two uniformly charged rods with a repulsive mean field electrostatic interaction, as well as a shorter ranged attractive component. This second component may model correlation forces. The second example includes possible helix specific forces in a braid formed of two DNA molecules.

## 0. Outline

The purpose of this work is to develop a theoretical treatment of the braiding of two molecules. We will consider braids in which the braid axis is straight in the ground state (i.e. when there are no thermal fluctuations). In our treatment, we will want to include the effects of the thermal undulations of the two molecules, the elastic energy of the two molecules described by the elastic rod model, steric interactions due to the hard core radius of the molecules, interactions between two molecules, and external forces that constrain both the average linking number (braid turns) and the end to end distance of the fluctuating braid.

The work is divided into the following sections:

**Section 1:** We characterize the fluctuating braid geometry, developing expressions, in terms of local geometric parameters, for the end to end distance and the tangent vectors of the

[a.)] Electronic Mail: domolee@hotmail.com

molecular centre lines. These results will be used later analysis. The local geometric parameters used are the angle between the two tangent vectors of the molecular centre lines (tilt angle), the distance between the two molecular centre lines, the lateral displacement of the braid axis away from the straight line, ground state configuration; all considered at each point along the braid. We assume that the fluctuations of the braid axis, braid radius and tilt angle are all slowly varying.

**Section 2:** We define and discuss the topological quantities of braid twist, molecular twist (as well as the twist densities), braid writhe, molecular writhe, braid linking number and molecular linking number. We give expressions for these.

**Section 3:** We consider the elastic energy of the molecules in the braid using the elastic rod model. Expressions for both the bending and twisting energies are derived in terms of the local geometric parameters of the braid and molecular twist density.

**Section 4:** The coupling of the braid to an external pulling force is included to produce a required end to end distance of the braid, as well as an external moment to fix the average linking number of the braid. Also, we write down an expression for a generalized interaction potential between the two molecules in the braid. Lastly, we expand out our expressions for small fluctuations in the tilt angle.

**Section 5:** We discuss how one may approximate steric interactions due to each molecule in the braid having a hard core radius $a$, using an approach originally developed by [1] and adapted to DNA assemblies [2] and braiding [3].

**Section 6:** This focuses on a variational approximation for the case when the interaction between molecules can be considered weak. This case relies on assumption that the mean squared amplitude of fluctuations in the braid radius is primarily determined by steric interactions. From this expression, we are able to write down an expression for the free energy; though an evaluation of the correction due to coupling of the external moment to braid writhe is treated in the next section.

**Section 7:** Here, we deal with the calculation of the correction to the free energy due to coupling of the braid writhe to the external moment acting on the braid.

**Section 8:** We present the equations that determine the local geometric parameters of the braid, minimizing the free energy in the case where interactions are weak. We also show expressions for end to end distance and number of braid turns. Lastly, a simplifying approximation to these expressions is presented. This should work well when the mean squared fluctuation amplitude in the braid radius is small enough.

**Section 9:** We show how the equations of the previous section simplify when there is no interaction between molecules, except for steric ones.

**Section 10:** Now, we build a variational approximation where we self-consistently determine the mean squared fluctuation amplitude in the braid radius, and the degree of fluctuations in the twist of the molecules. This is needed when the interactions between the molecules are strong, as the interactions should determine the extent of these fluctuations.

**Section 11:** We present the equations that minimize the free energy of the previous section.

**Section 12:** We look at the particular example of the molecules behaving as uniform rods, with mean-field electrostatic interactions and a shorter ranged attractive interaction, attributable to correlation forces.

**Section 13:** The case of possible helix specific forces between DNA molecules is examined here. Firstly, we focus on general considerations about the non-ideal structure of DNA, relevant for such forces. Secondly, we describe the explicit form of the equations for the mean field electrostatic interaction potential between charged helices, the Kornyshev-Leikin theory.

## 1. Describing the geometry of a molecular braid

We label the two molecules forming the braid with index $\mu = 1, 2$. We suppose that length of the braided sections of both molecules is of average length $L$. However, through bending of the molecules in the braid, we must allow for the lengths of two molecules to fluctuate. Therefore, we define the lengths of molecule 1 and 2 to be $L_1 = L + \delta L$ and $L_2 = L + \delta L$, where we have for the thermal average that $\langle \delta L \rangle = 0$. We may neglect $\delta L$ in the thermodynamic limit when the molecules are infinitely long, and for our present case, these fluctuations will not be important. We will write $L = (L_1 + L_2)/2$, and in all other instances one may neglect $\delta L$, as we shall consider $L$ to be sufficiently long.

To describe the braid we start by constructing chords of length $R$ that run between two sets of successive points that run along the molecular centrelines. We will suppose that, for the braid in its ground (non-fluctuating) state, all of these chords have the same length and have the shortest distance between the two molecular centre lines. However, for a fluctuating braid, they are indeed allowed to vary in length. In the direction of the chords we may assign unit vectors $\hat{\mathbf{d}}_A$. Bisecting and perpendicular to each chord, we define a curve with position vector $\mathbf{r}_A$ that defines a point on what we define as the braid axis. We define the unit tangent vector of this curve as $\hat{\mathbf{t}}_A$ with the requirement that

$$\hat{\mathbf{d}}_A . \hat{\mathbf{t}}_A = 0. \tag{1.1}$$

We parameterize the distance along the braid axis by unit arc length $\tau$, which in turn corresponds to unit arc-length $s_\mu$ of molecule $\mu$ in the braid. We define the position vector of the trajectory of the centre lines of molecule $\mu$ as $\mathbf{r}_\mu(s_\mu)$, and their unit tangent vectors are $\hat{\mathbf{t}}_\mu(s_\mu)$. By definition of unit arc length we require that

$$\hat{\mathbf{t}}_A(\tau) = \frac{d\mathbf{r}_A(\tau)}{d\tau}, \qquad \hat{\mathbf{t}}_\mu(s_\mu) = \frac{d\mathbf{r}_\mu(s_\mu)}{ds_\mu}. \tag{1.2}$$

For the braid, $\tau$ runs from $-L_A/2$ to $L_A/2$, whereas $s_\mu$ from $-L_\mu/2$ to $L_\mu/2$. We may define $\tilde{\eta}(\tau)$, an angle between the two tangent vectors of the molecular centre lines as a function of $\tau$, so that

$$\hat{\mathbf{t}}_1.\hat{\mathbf{t}}_2 = \cos\tilde{\eta}(\tau). \tag{1.3}$$

We choose $\tilde{\eta}(\tau)$ to be positive for a left handed braid and negative for a right handed braid. Next, we may write down a relationship between $\mathbf{r}_A(\tau)$ and $\mathbf{r}_\mu(s_\mu)$

$$\mathbf{r}_\mu(s_\mu) = \tilde{\mathbf{r}}_\mu(\tau) = \mathbf{r}_A(\tau) + (-1)^\mu \frac{\tilde{R}(\tau)\hat{\mathbf{d}}_A(\tau)}{2}, \tag{1.4}$$

where $R = \tilde{R}(\tau)$. In the ground state both $\tilde{R}(\tau)$ and $\tilde{\eta}(\tau)$ are constant with respect to $\tau$. We also introduce the following vectors

$$\hat{\mathbf{n}}_\mu(s_\mu) = \frac{\hat{\mathbf{t}}_\mu(s_\mu)\times\hat{\mathbf{d}}_A(\tau(s_\mu))}{\left|\hat{\mathbf{t}}_1(s_\mu)\times\hat{\mathbf{d}}(\tau(s_\mu))\right|}, \qquad \hat{\mathbf{d}}_\mu(s_\mu) = \hat{\mathbf{n}}_\mu(s_\mu)\times\hat{\mathbf{t}}_\mu(s_\mu). \tag{1.5}$$

Using these vectors we may construct two local orthogonal frames called the *braid frames* [4], spanned by the basis sets $\{\hat{\mathbf{d}}_1,\hat{\mathbf{n}}_1,\hat{\mathbf{t}}_1\}$ and $\{\hat{\mathbf{d}}_2,\hat{\mathbf{n}}_2,\hat{\mathbf{t}}_2\}$.

We can parameterize $\mathbf{r}_A(\tau)$ in the following way

$$\mathbf{r}_A(\tau) = \tilde{x}_A(\tau)\hat{\mathbf{i}} + \tilde{y}_A(\tau)\hat{\mathbf{j}} + \tilde{z}_A(\tau)\hat{\mathbf{k}}. \tag{1.6}$$

Indeed, through the unitary requirement (Eq. (1.2)), $\tilde{x}_A(\tau)$, $\tilde{y}_A(\tau)$ and $\tilde{z}_A(\tau)$ are not independent of each other. We have that

$$\hat{\mathbf{t}}_A(\tau) = \frac{d\tilde{x}_A(\tau)}{d\tau}\hat{\mathbf{i}} + \frac{d\tilde{y}_A(\tau)}{d\tau}\hat{\mathbf{j}} + \sqrt{1-\left(\frac{d\tilde{x}_A(\tau)}{d\tau}\right)^2-\left(\frac{d\tilde{y}_A(\tau)}{d\tau}\right)^2}\,\hat{\mathbf{k}}, \tag{1.7}$$

as well as

$$\frac{d\tilde{z}_A(\tau)}{d\tau} = \sqrt{1-\left(\frac{d\tilde{x}_A(\tau)}{d\tau}\right)^2-\left(\frac{d\tilde{y}_A(\tau)}{d\tau}\right)^2}. \tag{1.8}$$

In the absence of fluctuations, the ground state, we choose that $\hat{\mathbf{t}}_A(\tau)=\hat{\mathbf{k}}$. Next, we may parameterize $\hat{\mathbf{d}}_A(\tau)$ in the following way

$$\hat{\mathbf{d}}_A(\tau)=h(\tau)\left(\cos\tilde{\theta}(\tau)\hat{\mathbf{i}}+\sin\tilde{\theta}(\tau)\hat{\mathbf{j}}+g(\tau)\hat{\mathbf{k}}\right), \tag{1.9}$$

where, when $\tilde{x}'(\tau)=\tilde{y}'(\tau)=0$, $g(\tau)=0$ and $h(\tau)=1$ (here, and throughout, the prime refers to the derivative of a function with respect to its argument). Through Eq. (1.1) and requirement that $\hat{\mathbf{d}}_A(\tau)$ must be unitary we find that

$$g(\tau)=-\left(\frac{d\tilde{x}_A(\tau)}{d\tau}\cos\theta(\tau)+\frac{d\tilde{y}_A(\tau)}{d\tau}\sin\theta(\tau)\right)\left(1-\left(\frac{d\tilde{x}_A(\tau)}{d\tau}\right)^2-\left(\frac{d\tilde{y}_A(\tau)}{d\tau}\right)^2\right)^{-1/2}, \tag{1.10}$$

$$h(\tau)=\frac{\left(1-\left(\frac{d\tilde{x}_A(\tau)}{d\tau}\right)^2-\left(\frac{d\tilde{y}_A(\tau)}{d\tau}\right)^2\right)}{\left[\left(\frac{d\tilde{x}_A(\tau)}{d\tau}\cos\theta(\tau)+\frac{d\tilde{y}_A(\tau)}{d\tau}\sin\theta(\tau)\right)^2+1-\left(\frac{d\tilde{x}_A(\tau)}{d\tau}\right)^2-\left(\frac{d\tilde{y}_A(\tau)}{d\tau}\right)^2\right]}. \tag{1.11}$$

In what follows we shall assume that $\tilde{x}_A(\tau)$, $\tilde{y}_A(\tau)$, and $\tilde{R}(\tau)$ are slowly varying so that $\tilde{x}'_A(\tau)$, $\tilde{y}'_A(\tau)$, $\tilde{R}'(\tau)$ are small. This allows us to approximate

$$g(\tau)\approx-\left(\frac{d\tilde{x}_A(\tau)}{d\tau}\cos\tilde{\theta}(\tau)+\frac{d\tilde{y}_A(\tau)}{d\tau}\sin\tilde{\theta}(\tau)\right)\quad \text{and}\quad h(\tau)\approx 1. \tag{1.12}$$

Then, we may write

$$\hat{\mathbf{t}}_A(\tau)\approx\frac{d\tilde{x}_A(\tau)}{d\tau}\hat{\mathbf{i}}+\frac{d\tilde{y}_A(\tau)}{d\tau}\hat{\mathbf{j}}+\hat{\mathbf{k}}, \tag{1.13}$$

$$\hat{\mathbf{d}}_A(\tau)\approx\cos\tilde{\theta}(\tau)\hat{\mathbf{i}}+\sin\tilde{\theta}(\tau)\hat{\mathbf{j}}-\left[\cos\tilde{\theta}(\tau)\frac{d\tilde{x}(\tau)}{d\tau}+\sin\tilde{\theta}(\tau)\frac{d\tilde{y}(\tau)}{d\tau}\right]\hat{\mathbf{k}}. \tag{1.14}$$

It is useful to define the functions

$$\theta^{(\mu)}(s_\mu(\tau))=\tilde{\theta}(\tau),\; R^{(\mu)}(s_\mu(\tau))=\tilde{R}(\tau),\; x_A^{(\mu)}(s_\mu(\tau))=\tilde{x}_A(\tau),$$

$$y_A^{(\mu)}(s_\mu(\tau))=\tilde{y}_A(\tau),\quad z_A^{(\mu)}(s_\mu(\tau))=\tilde{z}_A(\tau). \tag{1.15}$$

Then, one may write an expression for the tangent vector $\hat{\mathbf{t}}_\mu(s_\mu)$ from Eqs. (1.4), (1.6), (1.14) and (1.15). This reads as

$$\hat{\mathbf{t}}_\mu(s_\mu) \approx \left( \frac{dx_A^{(\mu)}(s_\mu)}{ds_\mu} + \frac{(-1)^\mu}{2} \left( \frac{dR^{(\mu)}(s_\mu)}{ds_\mu} \cos\theta^{(\mu)}(s_\mu) - R^{(\mu)}(s_\mu) \frac{d\theta^{(\mu)}(s_\mu)}{ds_\mu} \sin\theta^{(\mu)}(s_\mu) \right) \right) \hat{\mathbf{i}}$$

$$+ \left( \frac{dy_A^{(\mu)}(s_\mu)}{ds_\mu} + \frac{(-1)^\mu}{2} \left( \frac{dR^{(\mu)}(s_\mu)}{ds_\mu} \sin\theta^{(\mu)}(s_\mu) + R^{(\mu)}(s_\mu) \frac{d\theta^{(\mu)}(s_\mu)}{ds_\mu} \cos\theta^{(\mu)}(s_1) \right) \right) \hat{\mathbf{j}} + \frac{dz_A^{(\mu)}(s_\mu)}{ds_\mu} \hat{\mathbf{k}}.$$

(1.16)

Note that, in deriving Eq. (1.16), we have neglected both the $\tilde{x}'(\tau)$ and $\tilde{y}'(\tau)$ corrections to $\hat{\mathbf{d}}(\tau)$. The unitary nature of the tangent vectors, $\left|\hat{\mathbf{t}}_1(s_1)\right| = \left|\hat{\mathbf{t}}_2(s_2)\right| = 1$, requires that we write

$$\frac{dz_A^{(\mu)}(s_\mu)}{ds_\mu} = \left[ 1 - \left( \frac{dx_A^{(\mu)}(s_1)}{ds_\mu} + \frac{(-1)^\mu}{2} \left( \frac{dR^{(\mu)}(s_\mu)}{ds_\mu} \cos\theta^{(\mu)}(s_\mu) - R^{(\mu)}(s_\mu) \frac{d\theta^{(\mu)}(s_\mu)}{ds_\mu} \sin\theta^{(\mu)}(s_\mu) \right) \right)^2 \right.$$

$$\left. - \left( \frac{dy_A^{(\mu)}(s_\mu)}{ds_\mu} + \frac{(-1)^\mu}{2} \left( \frac{dR^{(\mu)}(s_\mu)}{ds_\mu} \sin\theta^{(\mu)}(s_\mu) + R^{(\mu)}(s_\mu) \frac{d\theta^{(\mu)}(s_\mu)}{ds_\mu} \cos\theta^{(\mu)}(s_\mu) \right) \right)^2 \right]^{1/2}.$$

(1.17)

We now derive an expression for $\frac{ds_\mu}{d\tau}$. To do so we need to consider the derivatives of Eq. (1.4) with respect to $\tau$. Utilizing Eqs. (1.14) , we find that

$$\frac{d\tilde{\mathbf{r}}_\mu(\tau)}{d\tau} \approx \hat{\mathbf{t}}_A(\tau) + \frac{(-1)^\mu}{2} \left( \frac{d\tilde{R}(\tau)}{d\tau} \cos\tilde{\theta}(\tau) - \tilde{R}(\tau) \frac{d\tilde{\theta}(\tau)}{d\tau} \sin\tilde{\theta}(\tau) \right) \hat{\mathbf{i}}$$
$$+ \frac{(-1)^\mu}{2} \left( \frac{d\tilde{R}(\tau)}{d\tau} \sin\tilde{\theta}(\tau) + \tilde{R}(\tau) \frac{d\tilde{\theta}(\tau)}{d\tau} \cos\tilde{\theta}(\tau) \right) \hat{\mathbf{j}}, \quad (1.18)$$

where again, have neglected both the $\tilde{x}'(\tau)$ and $\tilde{y}'(\tau)$ corrections to $\hat{\mathbf{d}}(\tau)$. Thus, from Eqs. (1.2)and (1.18) we find that

$$\frac{ds_\mu}{d\tau} \approx \left[ 1 + (-1)^\mu \frac{d\tilde{x}_A(\tau)}{d\tau} \left( \frac{d\tilde{R}(\tau)}{d\tau} \cos\tilde{\theta}(\tau) - \tilde{R}(\tau) \frac{d\tilde{\theta}(\tau)}{d\tau} \sin\tilde{\theta}(\tau) \right) \right.$$

$$\left. - (-1)^\mu \frac{d\tilde{y}_A(\tau)}{d\tau} \left( \frac{d\tilde{R}(\tau)}{d\tau} \sin\tilde{\theta}(\tau) + \tilde{R}(\tau) \frac{d\tilde{\theta}(\tau)}{d\tau} \cos\tilde{\theta}(\tau) \right) + \frac{1}{4} \left( \frac{d\tilde{R}(\tau)}{d\tau} \right)^2 + \frac{\tilde{R}(\tau)^2}{4} \left( \frac{d\tilde{\theta}(\tau)}{d\tau} \right)^2 \right]^{1/2}.$$

(1.19)

From Eq. (1.19), we can write an expression that relates $L_A$ to $L$

$$L = \frac{1}{2}\left(L_1 + L_2\right) = \frac{1}{2}\left(\int_{-L_1/2}^{L_1/2} ds_1 + \int_{-L_2/2}^{L_2/2} ds_2\right) = \frac{1}{2}\int_{-L_A/2}^{L_A/2} d\tau\left(\frac{ds_1}{d\tau} + \frac{ds_2}{d\tau}\right)$$
$$\approx \int_{-L_A/2}^{L_A/2} d\tau\left(1 + \frac{1}{8}\left(\frac{d\tilde{R}(\tau)}{d\tau}\right)^2 + \frac{\tilde{R}(\tau)^2}{8}\left(\frac{d\theta(\tau)}{d\tau}\right)^2\right). \tag{1.20}$$

In what follows it is useful to define a new coordinate along the braid $s_0$ such that

$$\frac{ds_0}{d\tau} = \left[1 + \frac{1}{4}\left(\frac{d\tilde{R}(\tau)}{d\tau}\right)^2 + \frac{\tilde{R}(\tau)^2}{4}\left(\frac{d\tilde{\theta}(\tau)}{d\tau}\right)^2\right]^{1/2}. \tag{1.21}$$

The coordinate $s_0$ is the unit arc-length along the both molecular centre lines when the braid axis is not fluctuating. This means that when $x_A'(\tau) = y_A'(\tau) = 0$ we have that $s_0 = s_\mu$. We will want to parameterize expressions for the braid is terms $s_0$. From Eqs. (1.19) and (1.21) it is possible to write

$$\frac{ds_\mu}{ds_0} \approx 1 + (-1)^\mu \frac{dx_A(s_0)}{ds_0}\left(\frac{1}{2}\frac{dR(s_0)}{ds_0}\cos\theta(s_0) - \frac{R(s_0)}{2}\frac{d\theta(s_0)}{ds_0}\sin\theta(s_0)\right)$$
$$+(-1)^\mu \frac{dy_A(s_0)}{ds_0}\left(\frac{1}{2}\frac{dR(s_0)}{ds_0}\sin\theta(s_0) + \frac{R(s_0)}{2}\frac{d\theta(s_0)}{ds_0}\cos\theta(s_0)\right). \tag{1.22}$$

Here, we have that $x_A(s_0) = \tilde{x}_A(\tau(s_0))$, $y_A(s_0) = \tilde{y}_A(\tau(s_0))$, $R(s_0) = \tilde{R}(\tau(s_0))$ and $\theta(s_0) = \tilde{\theta}(\tau(s_0))$. Also, we can write

$$L \approx \int_{-L_A/2}^{L_A/2} d\tau\left[\cos\left(\frac{\tilde{\eta}(\tau)}{2}\right)\right]^{-1} \quad \text{and} \quad \frac{ds_0}{d\tau} = \left[\cos\left(\frac{\tilde{\eta}(\tau)}{2}\right)\right]^{-1} = \left[\cos\left(\frac{\eta(s_0)}{2}\right)\right]^{-1}. \tag{1.23}$$

We now obtain an expression for $\cos\eta(s_0)$ ( where $\eta(s_0) = \tilde{\eta}(\tau(s_0))$ ) in terms of $R(s_0)$, $x_A(s_0)$, $y_A(s_0)$ and $\theta(s_0)$. Through Eqs. (1.16) and (1.17), we obtain

$$\cos\eta(s_0) = \hat{\mathbf{t}}_1(s_1(s_0)).\hat{\mathbf{t}}_2(s_2(s_0))$$
$$= \frac{ds_0}{ds_1}\frac{ds_0}{ds_2}\left(\frac{dz_1(s_0)}{ds_0}\frac{dz_2(s_0)}{ds_0} - \frac{1}{4}\left(\frac{dR(s_0)}{ds_0}\right)^2 - \frac{R(s_0)^2}{4}\left(\frac{d\theta(s_0)}{ds_0}\right)^2 + \left(\frac{dx_A(s_0)}{ds_0}\right)^2 + \left(\frac{dy_A(s_0)}{ds_0}\right)^2\right)$$
$$\approx 1 - \frac{1}{2}\left(\frac{dR(s_0)}{ds_0}\right)^2 - \frac{R(s_0)^2}{2}\left(\frac{d\theta(s_0)}{ds_0}\right)^2. \tag{1.24}$$

For small values of $R'(s_0)$ that we consider, we can write from Eq. (1.24)

$$\frac{R(s_0)}{2}\left(\frac{d\theta(s_0)}{ds_0}\right)\approx-\sqrt{\sin^2\left(\frac{\eta(s_0)}{2}\right)-\frac{1}{4}\left(\frac{dR(s_0)}{ds_0}\right)^2}\approx-\sin\left(\frac{\eta(s_0)}{2}\right)+\frac{1}{8\sin\left(\frac{\eta(s_0)}{2}\right)}\left(\frac{dR(s_0)}{ds_0}\right)^2. \tag{1.25}$$

The minus sign in front of the square root Eq. (1.25) comes from defining $\eta(s_0)$ as positive for a left handed braid. When we consider introducing an external moment, $M$ to a braid Eq. (1.25) is useful. It is also important to look at the end to end extension of the braid, when including an external pulling force $F$ acting on the braid. This is defined as

$$\begin{aligned} z &= \int_{-L_A/2}^{-L_A/2} d\tau\left(\frac{d\tilde{z}_A}{d\tau}\right)=\int_{-L_A/2}^{-L_A/2} d\tau\sqrt{1-\left(\frac{d\tilde{x}_A(\tau)}{d\tau}\right)^2-\left(\frac{d\tilde{y}_A(\tau)}{d\tau}\right)^2} \\ &= \int_{-L/2}^{-L/2} ds_0\sqrt{\cos^2\left(\frac{\eta(s_0)}{2}\right)-\left(\frac{dx_A(s_0)}{ds_0}\right)^2-\left(\frac{d\tilde{y}_A(s_0)}{ds_0}\right)^2}. \end{aligned} \tag{1.26}$$

For small $\tilde{x}'_A(\tau)$ and $\tilde{y}'_A(\tau)$ Eq. (1.26) is approximated by

$$z\approx\int_{-L/2}^{L/2} ds_0\left[\cos\left(\frac{\eta(s_0)}{2}\right)-\frac{1}{2\cos\left(\frac{\eta(s_0)}{2}\right)}\left[\left(\frac{dx_A(s_0)}{ds_0}\right)^2+\left(\frac{dy_A(s_0)}{ds_0}\right)^2\right]\right]. \tag{1.27}$$

Last of all, in this section, we can approximate Eq. (1.16) for the tangent vectors as

$$\begin{aligned} \hat{\mathbf{t}}_\mu(s_\mu) &\approx \frac{ds_0}{ds_\mu}\left[\left(\frac{dx_A(s_0)}{ds_0}+(-1)^\mu\left(\frac{1}{2}\frac{dR(s_0)}{ds_0}\cos\theta(s_0)+\sin\left(\frac{\eta(s_0)}{2}\right)\sin\theta(s_0)\right)\right)\hat{\mathbf{i}}\right. \\ &\left.+\left(\frac{dy_A(s_0)}{ds_0}+(-1)^\mu\left(\frac{1}{2}\frac{dR(s_0)}{ds_0}\sin\theta(s_0)-\sin\left(\frac{\eta(s_0)}{2}\right)\cos\theta(s_0)\right)\right)\hat{\mathbf{j}}+\cos\left(\frac{\eta(s_0)}{2}\right)\hat{\mathbf{k}}\right] \\ &\approx\left[\left(\frac{dx_A(s_0)}{ds_0}+(-1)^\mu\left(\frac{1}{2}\frac{dR(s_0)}{ds_0}\cos\theta(s_0)+\sin\left(\frac{\eta(s_0)}{2}\right)\sin\theta(s_0)\right)\right)\hat{\mathbf{i}}\right. \\ &\left.+\left(\frac{dy_A(s_0)}{ds_0}+(-1)^\mu\left(\frac{1}{2}\frac{dR(s_0)}{ds_0}\sin\theta(s_0)-\sin\left(\frac{\eta(s_0)}{2}\right)\cos\theta(s_0)\right)\right)\hat{\mathbf{j}}+\cos\left(\frac{\eta(s_0)}{2}\right)\hat{\mathbf{k}}\right]. \end{aligned} \tag{1.28}$$

## 2. Braid topological quantities

Let us start by defining what we call molecular and braid twist. To describe molecular twist we first need to introduce new unit vectors

$$\hat{\mathbf{v}}_\mu(s_\mu) = \cos\xi_\mu^{(\mu)}(s_\mu)\hat{\mathbf{d}}_\mu(s_\mu) + \sin\xi_\mu^{(\mu)}(s_\mu)\hat{\mathbf{n}}_\mu(s_\mu),$$

$$\hat{\mathbf{u}}_\mu(s_\mu) = -\sin\xi_\mu^{(\mu)}(s_\mu)\hat{\mathbf{d}}_\mu(s_\mu) + \cos\xi_\mu^{(\mu)}(s_\mu)\hat{\mathbf{n}}_\mu(s_\mu). \tag{2.1}$$

The unit vectors $\hat{\mathbf{v}}_\mu(s_\mu)$ may describe the azimuthal orientation of some structural feature of each molecule. For instance, for DNA, it may be chosen to lie on a line bisecting the centre of the minor groove. We then may define a what we call a molecular twist density for molecule $\mu$. This is the rate of precession (angular frequency) of $\hat{\mathbf{v}}_\mu(s)$ about the tangent vector $\hat{\mathbf{t}}_\mu(s)$. This angular frequency is given by

$$g_\mu(s_\mu) = \hat{\mathbf{u}}_\mu(s_\mu).\frac{d\hat{\mathbf{v}}_\mu(s_\mu)}{ds_\mu}. \tag{2.2}$$

The (molecular) twist density can be then be related to topological number called the (molecular) twist

$$Tw_\mu = \frac{1}{2\pi}\int_{-L_\mu/2}^{L_\mu/2} g_\mu(s_\mu)ds_\mu, \tag{2.3}$$

which counts the number of precessions of $\hat{\mathbf{v}}_\mu(s)$ about $\hat{\mathbf{t}}_\mu(s)$ along the length of the braid. We can also define a twist density and twist for the braid

$$g_A(\tau) = \hat{\mathbf{n}}_A(\tau).\frac{d\hat{\mathbf{d}}_A(\tau)}{d\tau} \quad \text{and} \quad Tw_b = \frac{1}{2\pi}\int_{-L_A/2}^{L_A/2} g_A(\tau)d\tau, \tag{2.4}$$

where $\hat{\mathbf{n}}_A(\tau) = \hat{\mathbf{t}}_A(\tau)\times\hat{\mathbf{d}}_A(\tau)$. Through Gauss's integral we can define molecular writhes , $Wr_\mu$ in terms of the molecular centre lines

$$Wr_\mu = \frac{1}{4\pi}\int_{-L/2}^{L/2} ds_\mu \int_{-L/2}^{L/2} ds'_\mu \frac{\left(\mathbf{r}_\mu(s_\mu) - \mathbf{r}_\mu(s'_\mu)\right).\hat{\mathbf{t}}_\mu(s_\mu)\times\hat{\mathbf{t}}_\mu(s'_\mu)}{\left|\mathbf{r}_\mu(s_\mu) - \mathbf{r}_\mu(s'_\mu)\right|^3}. \tag{2.5}$$

We can also define a braid writhe through

$$Wr_b = \frac{1}{4\pi}\int_{-L_A/2}^{L_A/2} d\tau \int_{-L_A/2}^{L_A/2} d\tau' \frac{\left(\mathbf{r}_A(\tau) - \mathbf{r}_A(\tau')\right).\hat{\mathbf{t}}_A(\tau)\times\hat{\mathbf{t}}_A(\tau')}{\left|\mathbf{r}_A(\tau) - \mathbf{r}_A(\tau')\right|^3} \tag{2.6}$$

The writhe in 3-D is an average of the number of crossings a curve makes with itself in all the possible 2-D projections of itself. Through the Fuller -White theorem it is possible to define three different linking numbers for the braid. These are defined as

$$Lk_b = Tw_b + Wr_b,\ \ Lk_\mu = Tw_\mu + Wr_\mu \text{ and } Lk_2 = Tw_2 + Wr_2. \tag{2.7}$$

If all three linking numbers are conserved this may correspond to the situation of an catenane formed by the two molecules or a braid attached to a bead and a substrate where the number of turns of the bead is fixed, with no breaks or nicks in the molecules. If $Lk_1$ or $Lk_2$ are not conserved this corresponds to molecule 1 or 2, respectively, being nicked so that parts about molecule about the nick are able to rotate freely and relieve mechanical stress. In what follows we will look at the situation where both molecules are nicked, so that only the (average) braid linking number is conserved. We will do this by including a Lagrange multiplier $-2\pi M Lk_b$ or work term, where $M$ is an external moment applied to generate a braid with a particular average value of $Lk_b$.

When $\tilde{x}'_A(\tau) = \tilde{y}'_A(\tau) = 0$ from Eqs. (1.5), (1.14), (1.16), (2.1), (2.2) we have derived the following expression for the molecular twisting densities (see [3]).

$$g_\mu(s_\mu) \approx \frac{d\xi_\mu^{(\mu)}(s_\mu)}{ds_\mu} - \frac{\sin\eta^{(\mu)}(s_\mu)}{R} \approx \frac{d\xi_\mu(s_0)}{ds_0} - \frac{\sin\eta(s_0)}{R}. \tag{2.8}$$

At present, we have yet to derive corrections to these expressions when $\tilde{x}'_A(\tau)$ and $\tilde{y}'_A(\tau)$ are not zero, a topic of future work. Now, let's consider the braid twist. We first compute from $\hat{\mathbf{n}}_A(\tau) = \hat{\mathbf{t}}_A(\tau) \times \hat{\mathbf{d}}_A(\tau)$ and Eqs. (1.13) and (1.14) that

$$\begin{aligned}\hat{\mathbf{n}}_A(\tau) \approx &\left\{\left(\frac{1}{2}\left(\frac{dx_A(\tau)}{d\tau}\right)^2 - \frac{1}{2}\left(\frac{dy_A(\tau)}{d\tau}\right)^2 - 1\right)\sin\theta(\tau) - \frac{dx_A(\tau)}{d\tau}\frac{dy_A(\tau)}{d\tau}\cos\theta(\tau)\right\}\hat{\mathbf{i}} \\ &+\left\{\left(1 + \frac{1}{2}\left(\frac{dx_A(\tau)}{d\tau}\right)^2 - \frac{1}{2}\left(\frac{dy_A(\tau)}{d\tau}\right)^2\right)\cos\theta(\tau) + \frac{dx_A(\tau)}{d\tau}\frac{dy_A(\tau)}{d\tau}\sin\theta(\tau)\right\}\hat{\mathbf{j}} \\ &+\left[\sin\theta(\tau)\frac{dx_A(\tau)}{d\tau} - \cos\theta(\tau)\frac{dy_A(\tau)}{d\tau}\right]\hat{\mathbf{k}},\end{aligned} \tag{2.9}$$

$$\begin{aligned}\frac{d\hat{\mathbf{d}}_A(\tau)}{d\tau} = &-\sin\theta(\tau)\frac{d\theta(\tau)}{d\tau}\hat{\mathbf{i}} + \cos\theta(\tau)\frac{d\theta(\tau)}{d\tau}\hat{\mathbf{j}} + \\ &\left\{\left[\sin\theta(\tau)\frac{dx_A(\tau)}{d\tau} - \cos\theta(\tau)\frac{dy_A(\tau)}{d\tau}\right]\frac{d\theta(\tau)}{d\tau} - \left[\frac{d^2x_A(\tau)}{d\tau^2}\cos\theta(\tau) + \frac{d^2y_A(\tau)}{d\tau^2}\sin\theta(\tau)\right]\right\}\hat{\mathbf{k}}.\end{aligned}$$

(2.10)

This yields

$$g_A(\tau) \approx \frac{d\theta(\tau)}{d\tau}\left(1+\frac{1}{2}\left(\frac{dx_A(\tau)}{d\tau}\right)^2+\frac{1}{2}\left(\frac{dy_A(\tau)}{d\tau}\right)^2\right)-\frac{d^2x_A(\tau)}{d\tau^2}\frac{dy_A(\tau)}{d\tau}\cos^2\theta(\tau)+\frac{d^2y_A(\tau)}{d\tau^2}\frac{dx_A(\tau)}{d\tau}\sin^2\theta(\tau)$$
$$+\left[\frac{dx_A(\tau)}{d\tau}\frac{d^2x_A(\tau)}{d\tau^2}-\frac{dy_A(\tau)}{d\tau}\frac{d^2y_A(\tau)}{d\tau^2}\right]\cos\theta(\tau)\sin\theta(\tau). \tag{2.11}$$

Neglecting the higher order terms in Eq. (2.11) and using Eq. (1.25), we derive an expression for the braid twist

$$\begin{aligned} Tw_b &\approx \frac{1}{2\pi}\int_{-L_A/2}^{L_A/2}\frac{d\tilde{\theta}}{d\tau}d\tau = \frac{1}{2\pi}\int_{-L/2}^{L/2}\frac{d\theta}{ds_0}ds_0 \\ &\approx -\frac{1}{\pi}\int_{-L/2}^{L/2}\frac{1}{R(s_0)}\left[\sin\left(\frac{\eta(s_0)}{2}\right)-\frac{1}{8}\left(\frac{dR(s_0)}{ds_0}\right)^2\sin\left(\frac{\eta(s_0)}{2}\right)^{-1}\right]ds_0. \end{aligned} \tag{2.12}$$

## 3. Braid Elastic Energies

If the molecules obey elastic rod theory, we may write down for the bending elastic energy for the two molecules that form the braid

$$\begin{aligned} \frac{E_B}{k_BT} &= \int_{-L_1/2}^{L_1/2} ds_1\left[\frac{l_p}{2}\left(\frac{d\hat{\mathbf{t}}_1(s_1)}{ds_1}\right)^2\right]+\int_{-L_2/2}^{L_2/2} ds_2\left[\frac{l_p}{2}\left(\frac{d\hat{\mathbf{t}}_2(s_2)}{ds_2}\right)^2\right] \\ &= \frac{l_p}{2}\int_{-L/2}^{L/2} ds_0\left(\frac{ds_0}{ds_1}\left(\frac{d\hat{\mathbf{t}}_1(s_0)}{ds_0}\right)^2+\frac{ds_0}{ds_2}\left(\frac{d\hat{\mathbf{t}}_2(s_0)}{ds_0}\right)^2\right) \\ &\approx \frac{l_p}{2}\int_{-L/2}^{L/2} ds_0\left(\left(\frac{d\hat{\mathbf{t}}_1(s_0)}{ds_0}\right)^2+\left(\frac{d\hat{\mathbf{t}}_2(s_0)}{ds_0}\right)^2\right). \end{aligned} \tag{3.1}$$

Here, $l_p$ is the bending persistence length each molecules, which we have supposed to be the same for both molecules. It is defined as $l_p = B/k_BT$ where $B$ is the bending stiffness or rigidity constant. Now, let us find a functional for the bending energy in terms of $x_A$, $y_A$ $R$ and $\eta$. We find that from Eq. (1.28) that we may write

$$\frac{d\hat{\mathbf{t}}_\mu(s_\mu)}{ds_0} \approx \left[\left\{\frac{d^2x_A(s_0)}{ds_0^2} + (-1)^\mu\left(\frac{1}{R(s_0)}\frac{dR(s_0)}{ds_0}\sin\left(\frac{\eta(s_0)}{2}\right) + \frac{1}{2}\frac{d\eta(s)}{ds}\cos\left(\frac{\eta(s_0)}{2}\right)\right)\sin\theta(s_0)\right.\right.$$
$$\left. + (-1)^\mu\left(\frac{1}{2}\frac{d^2R(s_0)}{ds_0^2} - \frac{2}{R(s_0)}\sin^2\left(\frac{\eta(s_0)}{2}\right)\right)\cos\theta(s_0)\right\}\hat{\mathbf{i}}$$
$$+\left\{\frac{d^2y_A(s_0)}{ds_0^2} - (-1)^\mu\left(\frac{1}{R(s_0)}\frac{dR(s_0)}{ds_0}\sin\left(\frac{\eta(s_0)}{2}\right) + \frac{1}{2}\frac{d\eta(s)}{ds}\cos\left(\frac{\eta(s_0)}{2}\right)\right)\cos\theta(s_0)\right.$$
$$\left.\left. + (-1)^\mu\left(\frac{1}{2}\frac{d^2R(s_0)}{ds_0^2} - \frac{2}{R(s_0)}\sin^2\left(\frac{\eta(s_0)}{2}\right)\right)\sin\theta(s_0)\right\}\hat{\mathbf{j}} - \frac{1}{2}\frac{d\eta(s_0)}{ds_0}\sin\left(\frac{\eta(s_0)}{2}\right)\hat{\mathbf{k}}\right].$$

(3.2)

Using Eqs. (3.1) and (3.2) we find that

$$\frac{E_B}{k_BT} = \int_{-L/2}^{L/2} ds_0 l_p\left[\left(\frac{d^2x_A(s_0)}{ds_0^2}\right)^2 + \left(\frac{d^2y_A(s_0)}{ds_0^2}\right)^2 + \frac{1}{4}\left(\frac{d^2R(s_0)}{ds_0^2}\right)^2 + \frac{4}{R(s_0)^2}\sin^4\left(\frac{\eta(s_0)}{2}\right)\right.$$
$$- \frac{2}{R(s_0)}\frac{d^2R(s_0)}{ds_0^2}\sin^2\left(\frac{\eta(s_0)}{2}\right) + \left(\frac{dR(s_0)}{ds_0}\right)^2\frac{1}{R(s_0)}\sin^2\left(\frac{\eta(s_0)}{2}\right) + \frac{1}{4}\left(\frac{d\eta(s_0)}{ds_0}\right)^2 \quad (3.3)$$
$$\left. + \left(\frac{dR(s_0)}{ds_0}\right)\left(\frac{d\eta(s_0)}{ds_0}\right)\frac{1}{R(s_0)}\sin\left(\frac{\eta(s_0)}{2}\right)\cos\left(\frac{\eta(s_0)}{2}\right)\right],$$

our functional for the bending energy. Next, we may consider twisting elastic energies of the molecules in the braid. In the elastic rod model these can be written as

$$\frac{E_T}{k_BT} = \frac{l_c}{2}\int_{-L_1/2}^{L_1/2} ds_1\left(g_1(s_1) - g_1^0(s_1)\right)^2 + \int_{-L_2/2}^{L_2/2} ds_2\left(g_2(s_2) - g_2^0(s_2)\right)^2, \quad (3.4)$$

where $l_c$ is a combined persistence length that takes account of both local stretching and twisting fluctuations effectimg the local twist density. The values $g_1^0(s_1)$ and $g_2^0(s_2)$ are the twist densities of the molecules in their unstressed states. For the case where $g_1^0(s_1) \approx g_2^0(s_2) \approx \bar{g}_0$ where $\bar{g}_0$ is some constant value $l_c = C_tC_s/\left(C_s + \bar{g}^2C_t\right)k_BT$ (see Ref [2]) where $C_t$ is the twisting modulus and $C_s$ is the stretching modulus. Note that (3.4) with $l_c$ is only valid for thermal fluctuations in $g_\mu(s_\mu)$, where the effects of stretching and twisting combine in a simple way. It is important to realize that both $s_\mu(s_0)$ do not include stretching in Eq. (3.4). Particular care must be taken when including externally applied torsional stresses on each molecule, where a more rigorous treatment of thermal stretching and twisting fluctuation is required; $s_\mu(s_0)$ must explicitly take account of stretching fluctuations and in Eq. (3.4) $l_c$ should be replaced with $l_t = C_t / k_bT$ .

Using Eq. (2.8) in Eq. (3.4) in the twisting energy we may approximate

$$\frac{E_{Tw}}{k_B T} \approx \frac{l_c}{2}\int_{-L/2}^{L/2} ds_0 \left[ \left( \frac{d\xi_1(s_0)}{ds_0} - \frac{\sin\eta(s_0)}{R} - g_1^0(s_0) \right)^2 + \left( \frac{d\xi_2(s_0)}{ds_0} - \frac{\sin\eta(s_0)}{R} - g_1^0(s_0) \right)^2 \right]$$

$$= \frac{l_c}{4}\int_{-L/2}^{L/2} ds_0 \left[ \left( \frac{d\Delta\Phi(s_0)}{ds_0} - \Delta g^0(s_0) \right)^2 + \left( \frac{d\bar{\Phi}(s_0)}{ds_0} - \frac{2\sin\eta(s_0)}{R} - \bar{g}^0(s_0) \right)^2 \right] \tag{3.5}$$

$$= \frac{l_c}{4}\int_{-L/2}^{L/2} ds_0 \left[ \left( \frac{d\Delta\Phi(s_0)}{ds_0} - \Delta g^0(s_0) \right)^2 + \left( \bar{g} - \bar{g}^0(s_0) \right)^2 \right]$$

## 4. Including interactions and work terms in the total braid energy functional

As well as the sum of the elastic energies, we may consider three additional terms in the total braid energy functional. So that we may write

$$E_T = E_B + E_{Tw} + E_W + E_{\text{int}} + E_{St}. \tag{4.1}$$

Here, $E_W$ is a work term that includes the pulling force $F$ and the moment $M$, Lagrange multipliers effectively constraining the averages of the end to end distance, $z$ and $Lk_b$. The term $E_{\text{int}}$ is the contribution from interactions between the two molecules, and $E_{St}$ is a steric contribution which we will discuss how to approximate the effect of in the next section. For $E_W$ we may simply write down

$$E_W = -zF - 2\pi Lk_b M. \tag{4.2}$$

Using Eqs.(1.27) and (2.12) we can approximate Eq. (4.2) as

$$E_W \approx -F\int_{-L/2}^{L/2} ds_0 \left[ \cos\left(\frac{\eta(s_0)}{2}\right) - \frac{1}{2\cos\left(\frac{\eta(s_0)}{2}\right)} \left[ \left(\frac{dx_A(s_0)}{ds_0}\right)^2 + \left(\frac{dy_A(s_0)}{ds_0}\right)^2 \right] \right]$$

$$+M\int_{-L/2}^{L/2} \frac{2}{R(s_0)} \left[ \sin\left(\frac{\eta(s_0)}{2}\right) - \frac{1}{8}\left(\frac{dR(s_0)}{ds_0}\right)^2 \sin\left(\frac{\eta(s_0)}{2}\right)^{-1} \right] ds_0 - 2\pi M Wr_b. \tag{4.3}$$

We propose for interaction energy the following general local form

$$E_{\text{int}} = \int ds_0 \varepsilon_{\text{int}}(\eta(s_0), R(s_0), \Delta\Phi(s_0), \eta'(s_0), R'(s_0), x_A'(s_0), y_A'(s_0), g_1(s_0), g_2(s_0)). \tag{4.4}$$

Such a local form should may arise from a derivative expansion of the full interaction between the two molecules in the braid (for instance for molecules described by two helical rods interacting through screened electrostatics [5]). Note that this form does not take account of discrete molecular groups, as we move along the centre lines of the two molecules. However, if these modes are sensitive to thermal fluctuations, their thermal averages may small and not important (See Ref. [6] in regard to DNA).

To simplify things, we'll suppose that twisting/stretching elastic modulii are sufficiently large so that we may approximate the interaction energy as (specific criteria for doing this may be found in [3]).

$$E_{\rm int} \approx \int_{-L/2}^{L/2} ds_0 \varepsilon_{\rm int}(\eta(s_0), R(s_0), \Delta\Phi(s_0), \eta'(s_0), R'(s_0), x'_A(s_0), y'_A(s_0), g_1^0(s_0), g_2^0(s_0)). \tag{4.5}$$

We will consider small fluctuations $\eta(s_0)$ about an average tilt angle that is independent of $s_0$, so that $\eta_0 = \langle \eta(s_0) \rangle$; the brackets, $\langle \ldots \rangle$ correspond to thermal averaging. Therefore, we may write $\eta(s_0) = \eta_0 + \delta\eta(s_0)$, where $\langle \delta\eta(s_0) \rangle = 0$. We assume that $\delta\eta(s_0)$, $x'_A(s_0)$, and $y'_A(s_0)$ can all be expanded up to second order about zero, and that the interaction potential is symmetric under the two transformations $x'_A(s_0) \to -x'_A(s_0)$ and $y'_A(s_0) \to -y'_A(s_0)$.

For the elastic bending energy such an expansion yields

$$\frac{E_B}{k_B T} \approx \frac{E_{A,B}}{k_B T} + \frac{E_{R,B}}{k_B T}, \tag{4.6}$$

where

$$\frac{E_{A,B}}{k_B T} = \int_{-L/2}^{L/2} ds_0 \left( l_p \left[ \left( \frac{dx'_A(s_0)}{ds_0} \right)^2 + \left( \frac{dy'_A(s_0)}{ds_0} \right)^2 \right] \right), \tag{4.7}$$

$$\frac{E_{R,B}}{k_B T} = \int_{-L/2}^{L/2} ds_0 \mathcal{E}_R(R''(s_0), R'(s_0), R(s_0), \delta\eta'(s_0), \delta\eta(s_0)), \tag{4.8}$$

and

$$\mathcal{E}_R(R''(s_0),R'(s_0),R(s_0),\delta\eta'(s_0),\delta\eta(s_0)) =$$
$$\frac{l_p}{4}\left(\frac{d^2R(s_0)}{ds_0^2}\right)^2+\frac{l_p}{4}\left(\frac{d\delta\eta(s_0)}{ds_0}\right)^2-\left(\frac{dR(s_0)}{ds_0}\right)^2\frac{l_p}{R(s_0)^2}\sin^2\left(\frac{\eta_0}{2}\right)$$
$$+\frac{4l_p}{R(s_0)^2}\left[\sin^4\left(\frac{\eta_0}{2}\right)+\frac{\delta\eta(s_0)^2}{2}\left(3\cos^2\left(\frac{\eta_0}{2}\right)\sin^2\left(\frac{\eta_0}{2}\right)-\sin^4\left(\frac{\eta_0}{2}\right)\right)\right] \tag{4.9}$$
$$+\left(\frac{dR(s_0)}{ds_0}\right)\left(\frac{d\delta\eta(s_0)}{ds_0}\right)\frac{3}{R(s_0)}\sin\left(\frac{\eta_0}{2}\right)\cos\left(\frac{\eta_0}{2}\right).$$

For the interaction energy we expand out

$$E_{\text{int}} \approx k_BT\int_{-L/2}^{L/2} ds_0\,\Big(\varepsilon_{\text{int}}^0(\eta_0,R(s_0),\Delta\Phi(s_0),R'(s_0),g_1^0(s_0),g_2^0(s_0))$$
$$+\frac{1}{2}\varepsilon_{\text{int}}^{\eta,\eta}(\eta_0,R(s_0),\Delta\Phi(s_0),R'(s_0),g_1^0(s_0),g_2^0(s_0))\delta\eta(s_0)^2$$
$$+\frac{1}{2}\varepsilon_{\text{int}}^{\eta',\eta}(\eta_0,R(s_0),\Delta\Phi(s_0),R'(s_0),g_1^0(s_0),g_2^0(s_0))\delta\eta'(s_0)^2 \tag{4.10}$$
$$+\varepsilon_{\text{int}}^{\eta',\eta}(\eta_0,R(s_0),\Delta\Phi(s_0),R'(s_0),g_1^0(s_0),g_2^0(s_0))\delta\eta(s_0)\delta\eta'(s_0)$$
$$+\frac{1}{2}\varepsilon_{\text{int}}^{x,x}(\eta_0,R(s_0),\Delta\Phi(s_0),R'(s_0),g_1^0(s_0),g_2^0(s_0))x_A'(s_0)^2$$
$$+\frac{1}{2}\varepsilon_{\text{int}}^{y,y}(\eta_0,R(s_0),\Delta\Phi(s_0),R'(s_0),g_1^0(s_0),g_2^0(s_0))y_A'(s_0)^2\Big),$$

where

$$\varepsilon_{\text{int}}^0(\eta_0,R(s_0),\Delta\Phi(s_0),R'(s_0),g_1^0(s_0),g_2^0(s_0))=\varepsilon_{\text{int}}(\eta_0,R(s_0),\Delta\Phi(s_0),0,R'(s_0),0,0,g_1^0(s_0),g_2^0(s_0)), \tag{4.11}$$

$$\varepsilon_{\text{int}}^{Y,Y}(\eta_0,R(s_0),\Delta\Phi(s_0),R'(s_0),g_1^0(s_0),g_2^0(s_0))$$
$$=\left.\frac{\partial^2\varepsilon_{\text{int}}(\eta_0+\eta,R(s_0),\Delta\Phi(s_0),\eta',R'(s_0),x_A',y_A',g_1^0(s_0),g_2^0(s_0))}{\partial Y^2}\right|_{x_A'=0,y_A=0,\eta=0,\eta'=0}, \tag{4.12}$$

with $Y=\eta,\eta',x_A',y_A'$, and

$$\varepsilon_{\text{int}}^{\eta,\eta'}(\eta_0,R(s_0),\Delta\Phi(s_0),R'(s_0),g_1^0(s_0),g_2^0(s_0))=\left.\frac{\partial^2\varepsilon_{\text{int}}(\eta_0,R(s_0),\Delta\Phi(s_0),\eta',R'(s_0),0,0,g_1^0(s_0),g_2^0(s_0))}{\partial\eta_0\partial\eta'}\right|_{\eta'=0}. \tag{4.13}$$

As the interaction energy between rod like molecules should be invariant under $\xi_\mu(s_0) \to \xi_\mu(s_0) + 2\pi$, hence $\Delta\Phi(s_0) \to \Delta\Phi(s_0) + 2\pi$, it is possible also possible to express the interaction energy in terms of Fourier series, so that

$$
\begin{aligned}
E_{\text{int}} \approx k_B T \sum_{n=-\infty}^{\infty} \int_{-L/2}^{L/2} ds_0 \exp(in\Delta\Phi(s_0)) \Big( & \bar{\varepsilon}^{0}_{\text{int}}(\eta_0, R(s_0), n, R'(s_0), g_1^0(s_0), g_2^0(s_0)) \\
& + \frac{1}{2}\bar{\varepsilon}^{\eta,\eta}_{\text{int}}(\eta_0, R(s_0), n, R'(s_0), g_1^0(s_0), g_2^0(s_0))\delta\eta(s_0)^2 \\
& + \frac{1}{2}\bar{\varepsilon}^{\eta',\eta}_{\text{int}}(\eta_0, R(s_0), n, R'(s_0), g_1^0(s_0), g_2^0(s_0))\delta\eta'(s_0)^2 \\
& + \bar{\varepsilon}^{\eta',\eta}_{\text{int}}(\eta_0, R(s_0), n, R'(s_0), g_1^0(s_0), g_2^0(s_0))\delta\eta(s_0)\delta\eta'(s_0) \\
& + \frac{1}{2}\bar{\varepsilon}^{x,x}_{\text{int}}(\eta_0, R(s_0), n, R'(s_0), g_1^0(s_0), g_2^0(s_0))x_A'(s_0)^2 \\
& + \frac{1}{2}\bar{\varepsilon}^{y,y}_{\text{int}}(\eta_0, R(s_0), n, R'(s_0), g_1^0(s_0), g_2^0(s_0))y_A'(s_0)^2 \Big)
\end{aligned}
\tag{4.14}
$$

The twisting energy is still given by Eq. (3.5). Finally for the work terms we have following expansion

$$
\begin{aligned}
E_W \approx & -F \int_{-L/2}^{L/2} ds_0 \left[ \cos\left(\frac{\eta_0}{2}\right) - \frac{1}{8}\cos\left(\frac{\eta_0}{2}\right)\delta\eta(s_0)^2 - \frac{1}{\cos\left(\frac{\eta_0}{2}\right)} \left[ \left(\frac{dx_A(s_0)}{ds_0}\right)^2 + \left(\frac{dy_A(s_0)}{ds_0}\right)^2 \right] \right] \\
& + M \int_{-L/2}^{L/2} ds_0 \frac{2}{R(s_0)} \left[ \sin\left(\frac{\eta_0}{2}\right) - \frac{1}{8}\sin\left(\frac{\eta_0}{2}\right)\delta\eta(s_0)^2 - \frac{1}{8}\left(\frac{dR(s_0)}{ds_0}\right)^2 \sin\left(\frac{\eta_0}{2}\right)^{-1} \right] - 2\pi M Wr_b .
\end{aligned}
$$

(4.15)

## 5. Estimating steric interactions

We will consider the simplest case for steric interactions, an interaction energy that takes the form

$$
E_{st} = 0 \quad R(s) > 2a,
$$

$$
E_{st} = \infty \quad R(s) \le 2a. \tag{5.1}
$$

Eq. (5.1) supposes that the molecules can be modelled as smooth cylinders. As opposed to the difficulty of dealing with Eq. (5.1) directly in the partition function with the total energy functional given by Eq. (4.1) , we use an approach pioneered by [1]. We introduce a pseudo potential of the form

$$\tilde{E}_{st}=k_BT\int_{-\infty}^{\infty}ds\frac{\alpha_H}{2}(R(s)-R_0)^2=k_BT\int_{-\infty}^{\infty}ds\frac{\alpha_H}{2}\delta R(s)^2, \quad (5.2)$$

where $R_0=\langle R(s_0)\rangle$ and $\delta R(s_0)=R(s_0)-R_0$. As was suggested for the braids [3], and for DNA assembles [2], we now modify all terms in the energy functional that depend on $R(s)$. We do this according to the prescription that for $\delta R(s)>d_{\max}$ we replace $\delta R(s)$ with $d_{\max}$, and for $r(s)<d_{\min}$ we replace $\delta R(s)$ with $d_{\min}$. This prevents an unphysical overestimation (or underestimation) of all the other terms in the energy when we replace the steric interaction term with the harmonic potential. The parameter $\alpha_H$ is chosen so that in the case of when the interaction between molecules is weak

$$\left\langle \delta R(s)^2\right\rangle \approx \frac{\left(d_{\min}-d_{\max}\right)}{4}. \quad (5.3)$$

Following the prescription given above, for the bending energy we write

$$\begin{aligned}\frac{\tilde{E}_{R,B}}{k_BT}=\int_{-L_b/2}^{L_b/2}ds_0\big[&\mathcal{E}_R(R''(s_0),R'(s_0),R(s_0),\delta\eta'(s_0),\delta\eta(s_0))\theta(\delta R(s_0)-d_{\min})\theta(d_{\max}-\delta R(s_0))\\ &+\mathcal{E}_R(R''(s_0),R'(s_0),R_0+d_{\max},\delta\eta'(s_0),\delta\eta(s_0))\theta(\delta R(s_0)-d_{\max})\\ &+\mathcal{E}_R(R''(s_0),R'(s_0),R_0+d_{\max},\delta\eta'(s_0),\delta\eta(s_0))\theta(d_{\min}-\delta R(s_0))\big].\end{aligned}$$

(5.4)

Next, for the interaction energy we have

$$\begin{aligned}\tilde{E}_{\text{int}}\approx k_BT\sum_{n=-\infty}^{\infty}\int_{-L/2}^{L/2}ds_0\exp(in\Delta\Phi(s_0))\Big(&\tilde{\varepsilon}^{0}_{\text{int}}(\eta_0,R(s_0),n,R'(s_0),g_1^0(s_0),g_2^0(s_0))\\ &+\frac{1}{2}\tilde{\varepsilon}^{\eta,\eta}_{\text{int}}(\eta_0,R(s_0),n,R'(s_0),g_1^0(s_0),g_2^0(s_0))\delta\eta(s_0)^2\\ &+\frac{1}{2}\tilde{\varepsilon}^{\eta',\eta}_{\text{int}}(\eta_0,R(s_0),n,R'(s_0),g_1^0(s_0),g_2^0(s_0))\delta\eta'(s_0)^2\\ &+\tilde{\varepsilon}^{\eta',\eta}_{\text{int}}(\eta_0,R(s_0),n,R'(s_0),g_1^0(s_0),g_2^0(s_0))\delta\eta(s_0)\delta\eta'(s_0)\\ &+\frac{1}{2}\tilde{\varepsilon}^{x,x}_{\text{int}}(\eta_0,R(s_0),n,R'(s_0),g_1^0(s_0),g_2^0(s_0))x_A'(s_0)^2\\ &+\frac{1}{2}\tilde{\varepsilon}^{y,y}_{\text{int}}(\eta_0,R(s_0),n,R'(s_0),g_1^0(s_0),g_2^0(s_0))y_A'(s_0)^2\Big),\end{aligned} \quad (5.5)$$

where

$$\tilde{\varepsilon}^0_{\rm int}(\eta_0,R(s_0),n,R'(s_0),g^0_1(s_0),g^0_2(s_0))=\bar{\varepsilon}^0_{\rm int}(\eta_0,R_0+d_{\min},n,R'(s_0),g^0_1(s_0),g^0_2(s_0))\delta(d_{\min}-\delta R(s_0))$$
$$+\bar{\varepsilon}^0_{\rm int}(\eta_0,R(s_0),n,R'(s_0),g^0_1(s_0),g^0_2(s_0))\delta(d_{\max}-\delta R(s_0))\delta(\delta R(s_0)-d_{\min})$$
$$+\bar{\varepsilon}^0_{\rm int}(\eta_0,R_0+d_{\max},n,R'(s_0),g^0_1(s_0),g^0_2(s_0))\delta(\delta R(s_0)-d_{\max}),$$

(5.6)

$$\tilde{\varepsilon}^{Y,Y}_{\rm int}(\eta_0,R(s_0),n,R'(s_0),g^0_1(s_0),g^0_2(s_0))=\bar{\varepsilon}^{Y,Y}_{\rm int}(\eta_0,R_0+d_{\min},n,R'(s_0),g^0_1(s_0),g^0_2(s_0))\delta(d_{\min}-\delta R(s_0))$$
$$+\bar{\varepsilon}^{Y,Y}_{\rm int}(\eta_0,R(s_0),n,R'(s_0),g^0_1(s_0),g^0_2(s_0))\delta(d_{\max}-\delta R(s_0))\delta(\delta R(s_0)-d_{\min})$$
$$+\bar{\varepsilon}^{Y,Y}_{\rm int}(\eta_0,R_0+d_{\max},n,R'(s_0),g^0_1(s_0),g^0_2(s_0))\delta(\delta R(s_0)-d_{\max}),$$

(5.7)

$$\tilde{\varepsilon}^{\eta',\eta}_{\rm int}(\eta_0,R(s_0),n,R'(s_0),g^0_1(s_0),g^0_2(s_0))=\bar{\varepsilon}^{\eta',\eta}_{\rm int}(\eta_0,R_0+d_{\min},n,R'(s_0),g^0_1(s_0),g^0_2(s_0))\delta(d_{\min}-\delta R(s_0))$$
$$+\bar{\varepsilon}^{\eta',\eta}_{\rm int}(\eta_0,R(s_0),n,R'(s_0),g^0_1(s_0),g^0_2(s_0))\delta(d_{\max}-\delta R(s_0))\delta(\delta R(s_0)-d_{\min})$$
$$+\bar{\varepsilon}^{\eta',\eta}_{\rm int}(\eta_0,R_0+d_{\max},n,R'(s_0),g^0_1(s_0),g^0_2(s_0))\delta(\delta R(s_0)-d_{\max}),$$

(5.8)

where $Y=\eta,\eta',x,y$. The work terms are modified to be

$$\tilde{E}_W=-F\int_{-L/2}^{L/2}ds_0\left[\cos\left(\frac{\eta_0}{2}\right)-\frac{1}{8}\cos\left(\frac{\eta_0}{2}\right)\delta\eta(s_0)^2-\frac{1}{2\cos\left(\frac{\eta_0}{2}\right)}\left[\left(\frac{dx_A(s_0)}{ds_0}\right)^2+\left(\frac{dy_A(s_0)}{ds_0}\right)^2\right]\right]$$
$$+M\int_{-L/2}^{L/2}ds_0\left[\sin\left(\frac{\eta_0}{2}\right)-\frac{1}{8}\sin\left(\frac{\eta_0}{2}\right)\delta\eta(s_0)^2-\frac{1}{8}\left(\frac{dR(s_0)}{ds_0}\right)^2\sin\left(\frac{\eta_0}{2}\right)^{-1}\right]\left\{\frac{2}{R(s_0)}\theta(\delta R(s)-d_{\min})\right.$$
$$\left.\theta(d_{\min}-\delta R(s))+\frac{2}{R_0+\delta_{\min}}\theta(\delta R(s)-d_{\min})+\frac{2}{R_0+\delta_{\min}}\theta(d_{\max}-\delta R(s))\right\}-2\pi MWr_b.$$

(5.9)

The total energy functional is now given by

$$\tilde{E}_T=E_{A,B}+\tilde{E}_{R,B}+E_{Tw}+\tilde{E}_W+\tilde{E}_{\rm int}+\tilde{E}_{St}. \quad (5.10)$$

We will consider two possible choices for $d_{\max}$ and $d_{\min}$. The conventional choice is that $d_{\max}=-d_{\min}=R_0-2a$. In making such a choice we assume that the cage formed by one molecule wrapping around the other, in the braided conformation, can be approximated by a hard walled cylinder. However, when the interaction potential is the main contribution to $\langle\delta R(s)^2\rangle$, fluctuations in $\eta(s_0)$ are small, and $P\gg R_0-2a$ it was argued in [3] that a better choice for $d_{\max}$ might be

$$d_{\max} \approx \left( \frac{\pi R_0}{\tan(\eta_0 / 2)} \right)^{3/2} \frac{1}{\left( l_p^b \sqrt{2} \right)^{1/2}}. \tag{5.11}$$

Therefore, for our second choice, which we reserve for the case when interaction between the molecules are strong, and the braid is tight, we have $d_{\max}$ estimated through Eq. (5.11) with $-d_{\min} = R - 2a$. For this second choice, in expressions for the bending energy, interaction energy and work terms, as $d_{\max}$ given by Eq. (5.11) is sufficiently large, we can simply set $d_{\max} = \infty$.

## 6. Constructing the variational free energy function for weakly interacting braids

We have the following free energy to approximate

$$F_{Braid} \approx -k_B T \ln Z_T, \tag{6.1}$$

where

$$\begin{aligned} Z_T = &\int Dx'_A(s_0) \int Dy'_A(s_0) \int D\delta\Phi(s_0) \int D\delta R(s_0) \int D\delta\eta(s_0) \\ &\exp\left( -\frac{\tilde{E}_T[x'_A(s_0), y'_A(s_0), \Delta\Phi_0(s_0) + \delta\Phi(s_0), R_0 + \delta R(s_0), \eta_0 + \delta\eta(s)]}{k_B T} \right), \end{aligned} \tag{6.2}$$

where the total energy functional $\tilde{E}_T$ is given through Eqs.(3.5), (4.7), (5.2), (5.4), (5.5), (5.9) and (5.10). The average braid linking number and the end to end distance may be computed from

$$\begin{aligned} &\left\langle Lk_b[x'_A(s_0), y'_A(s_0), R_0 + \delta R(s_0), \eta_0 + \delta\eta(s)] \right\rangle = \frac{1}{Z_T} \int Dx'_A(s_0) \int Dy'_A(s_0) \int D\delta\Phi(s_0) \int D\delta R(s_0) \int D\delta\eta(s_0) \\ &Lk_b[x'_A(s_0), y'_A(s_0), R_0 + \delta R(s_0), \eta_0 + \delta\eta(s)] \\ &\exp\left( -\frac{\tilde{E}_T[x'_A(s_0), y'_A(s_0), \Delta\Phi_0(s_0) + \delta\Phi(s_0), R_0 + \delta R(s_0), \eta_0 + \delta\eta(s)]}{k_B T} \right), \end{aligned} \tag{6.3}$$

$$\begin{aligned} &\left\langle z[x'_A(s_0), y'_A(s_0), R_0 + \delta R(s_0), \eta_0 + \delta\eta(s)] \right\rangle = \frac{1}{Z_T} \int Dx'_A(s_0) \int Dy'_A(s_0) \int D\delta\Phi(s_0) \int D\delta R(s_0) \int D\delta\eta(s_0) \\ &z[x'_A(s_0), y'_A(s_0), R_0 + \delta R(s_0), \eta_0 + \delta\eta(s)] \\ &\exp\left( -\frac{\tilde{E}_T[x'_A(s_0), y'_A(s_0), \Delta\Phi_0(s_0) + \delta\Phi(s_0), R_0 + \delta R(s_0), \eta_0 + \delta\eta(s)]}{k_B T} \right). \end{aligned} \tag{6.4}$$

Both Eqs. (6.3) and (6.4) can be expressed as derivatives of the free energy with respect to both $F$ and $M$, namely

$$\left\langle Lk_b \right\rangle = -\frac{1}{2\pi}\frac{\partial F_{Braid}}{\partial M} \quad \text{and} \quad \left\langle z \right\rangle = -\frac{\partial F_{Braid}}{\partial F}. \tag{6.5}$$

For weakly interacting braids, we start by constructing the following variational trial functionals

$$\frac{E_{0,R}\left[\delta\eta(s_0),\delta R(s_0)\right]}{k_B T} = \int_{-L/2}^{L/2} ds_0 \left( \frac{l_p}{4}\left(\frac{d^2\delta R(s_0)}{ds_0^2}\right)^2 + \frac{\alpha_H}{2}\delta R(s_0)^2 \right) + \int_{-L/2}^{L/2} ds_0 \left( \frac{\tilde{l}_p}{4}\left(\frac{d\delta\eta(s_0)}{ds_0}\right)^2 + \frac{\alpha_\eta}{2}\delta\eta(s_0)^2 \right), \tag{6.6}$$

and

$$\frac{E_{0,A}[x'_A(s_0), y'_A(s_0)]}{k_B T} = \int_{-L/2}^{L/2} ds_0 \left( l_p \left[ \left(\frac{dx'_A(s_0)}{ds_0}\right)^2 + \left(\frac{dy'_A(s_0)}{ds_0}\right)^2 \right] + \left(\alpha_x x'_A(s_0)^2 + \alpha_y y'_A(s_0)^2\right) \right) - 2\pi M Wr_b, \tag{6.7}$$

where $\alpha_\eta$, $\alpha_x$ and $\alpha_y$ are variational parameters and $\tilde{l}_p$ is a renormalized persistence length, chosen to get rid of singular averages of $\delta\eta'(s)^2$ that occur later in the analysis. We then write down the following variational free energy

$$\begin{aligned} F_T &= -k_B T \ln Z_{0,R} - k_B T \ln Z_{0,A} \\ &+ \left\langle E_T[x'_A(s_0), y'_A(s_0), \Delta\Phi_0(s_0) + \delta\Phi(s_0), R_0 + \delta R(s_0), \eta_0 + \delta\eta(s)] \right\rangle_0 \\ &- \left\langle E_{0,R}[\delta\eta(s), \delta R(s)] + E_{0,A}[x'_A(s_0), y'_A(s_0)] \right\rangle_0, \end{aligned} \tag{6.8}$$

where

$$Z_{0,A} = \int Dx'_A(s_0) \int Dy'_A(s_0) \exp\left( -\frac{E_{0,A}\left[x'_A(s), y'_A(s)\right]}{k_B T} \right), \tag{6.9}$$

$$Z_R = \int D\delta\eta(s) \int D\delta R(s) \exp\left( -\frac{E_{0,R}\left[\delta\eta(s), \delta R(s)\right]}{k_B T} \right), \tag{6.10}$$

and the averaging bracket $\langle \ldots \rangle_0$ corresponds to the path integral

$$\begin{aligned} \langle A \rangle_0 &= \frac{1}{Z_{0,A} Z_{0,R}} \int Dx'_A(s_0) \int Dy'_A(s_0) \int D\delta\eta(s) \int D\delta R(s) \\ & A \exp\left( -\frac{E_{0,A}\left[x'_A(s), y'_A(s)\right]}{k_B T} \right) \exp\left( -\frac{E_{0,R}\left[x'_A(s), y'_A(s)\right]}{k_B T} \right). \end{aligned} \tag{6.11}$$

In the variational approximation we may replace in Eq. (6.5) $F_{Braid}$ with $F_T$.

In writing down Eq. (6.6), we assume that steric interactions primarily determine $\delta R(s_0)$, i.e. $\left\langle \delta R(s_0)^2 \right\rangle \approx (d_{\max} - d_{\min})^2 / 4$ , as well as suppose that suppose in this case the we make our first choice of $d_{\max} \approx -d_{\min} \approx (R_0 - 2a)$. Therefore, to meet this requirement we fix $\alpha_H$ to

$$\alpha_H = \frac{1}{2^{5/3}(R_0 - 2a)^{8/3} l_p^{1/3}}. \tag{6.12}$$

In general we have that

$$\alpha_H = \frac{2}{\left(d_{\max} - d_{\min}\right)^{8/3} \left(l_p\right)^{1/3}}. \tag{6.13}$$

Also, in writing down Eqs. (6.6) and (6.7) we assume that $\Delta\Phi(s_0)$ fluctuates freely. This is only valid when the interactions between the molecules depend weakly on $\Delta\Phi(s_0)$. We may express

$$\begin{aligned}
&\left\langle \tilde{E}_T[x'_A(s_0), y'_A(s_0), \Delta\Phi_0(s_0) + \delta\Phi(s_0), R_0 + \delta R(s_0), \eta_0 + \delta\eta(s)] \right\rangle_0 \\
&- \left\langle E_{0,R}[\delta\eta(s), \delta R(s)] + E_{0,A}[x'_A(s_0), y'_A(s_0)] \right\rangle_0 = \\
&\left\langle \tilde{E}_{R,B}[\eta_0 + \delta\eta(s), R_0 + \delta R(s)] + \tilde{E}_{St}[\delta R(s)] - E_{0,R}[\delta\eta(s), \delta R(s)] \right\rangle \\
&+ \left\langle E_{A,B}[x'_A(s_0), y'_A(s_0)] - E_{0,A}[x'_A(s_0), y'_A(s_0)] \right\rangle_0 + \left\langle E_{Tw}[\Delta\Phi_0(s_0) + \delta\Phi(s_0)] \right\rangle_0 \\
&+ \left\langle \tilde{E}_{\text{int}}[x'_A(s_0), y'_A(s_0), \Delta\Phi_0(s_0) + \delta\Phi(s_0), R_0 + \delta R(s_0), \eta_0 + \delta\eta(s)] \right\rangle_0 \\
&+ \left\langle \tilde{E}_W[x'_A(s_0), y'_A(s_0), R_0 + \delta R(s_0), \eta_0 + \delta\eta(s)] \right\rangle_0 .
\end{aligned} \tag{6.14}$$

Let us first consider the average on the RHS of Eq. (6.14)

$$\frac{\left\langle \tilde{E}_{R,B}[\eta_0+\delta\eta(s),R_0+\delta R(s)]+\tilde{E}_{St}[\delta R(s)]-E_{0,R}[\delta\eta(s),\delta R(s)]\right\rangle_0}{k_BT}=$$

$$\int_{-L/2}^{L/2} ds_0 \left\langle \mathcal{E}_R(R''(s_0),R'(s_0),R(s_0),\delta\eta'(s_0),\delta\eta(s_0))\,\theta(\delta R(s)+R_0-2a)\theta(R_0-2a-\delta R(s))\right.$$
$$+\mathcal{E}_R(R''(s_0),R'(s_0),2a,\delta\eta'(s_0),\delta\eta(s_0))\theta(2a-R_0-\delta R(s))$$
$$+\mathcal{E}_R(R''(s_0),R'(s_0),2R_0-2a,\delta\eta'(s_0),\delta\eta(s_0))\theta(\delta R(s)-R_0+2a)$$
$$\left.-\left(\frac{l_p}{4}\left(\frac{d^2\delta R(s_0)}{ds_0^2}\right)^2\right)-\left(\frac{l_p}{4}\left(\frac{d\delta\eta(s_0)}{ds_0}\right)^2+\frac{\alpha_\eta}{2}\delta\eta(s_0)^2\right)\right\rangle_0$$
$$=\int_{-L/2}^{L/2} ds_0 -\left\langle \Gamma_1(\delta R(s_0))\left(\frac{d\delta R(s_0)}{ds_0}\right)^2\right\rangle_{\delta R} l_p\sin^2\left(\frac{\eta_0}{2}\right)$$
$$+4l_p\left\langle\Gamma_1(\delta R(s_0))\right\rangle_{\delta R}\left[\sin^4\left(\frac{\eta_0}{2}\right)+\frac{\left\langle\delta\eta(s_0)^2\right\rangle_{\delta\eta}}{2}\left(3\cos^2\left(\frac{\eta_0}{2}\right)\sin^2\left(\frac{\eta_0}{2}\right)-\sin^4\left(\frac{\eta_0}{2}\right)\right)\right]$$
$$+3\left\langle\Gamma_2(\delta R(s_0))\left(\frac{dR(s_0)}{ds_0}\right)\right\rangle\left\langle\left(\frac{d\delta\eta(s_0)}{ds_0}\right)\right\rangle_{\delta\eta}\sin\left(\frac{\eta_0}{2}\right)\cos\left(\frac{\eta_0}{2}\right)+\frac{1}{4}\left(l_p-\tilde{l}_p\right)\left\langle\left(\frac{d\delta\eta(s_0)}{ds_0}\right)^2\right\rangle_{\delta\eta}$$
$$-\frac{\alpha_\eta\left\langle\delta\eta(s_0)^2\right\rangle_{\delta\eta}}{2},$$

(6.15)

where

$$\Gamma_1(\delta R(s_0))=\frac{\theta(\delta R(s_0)+R_0-2a)\theta(R_0-2a-\delta R(s_0))}{\left(R_0+\delta R(s_0)\right)^2}+\frac{\theta(2a-R_0-\delta R(s))}{4a^2}$$
$$+\frac{\theta(\delta R(s)-R_0+2a)}{4(R_0-2a)^2}, \qquad (6.16)$$

$$\Gamma_2(\delta R(s_0))=\frac{\theta(\delta R(s_0)+R_0-2a)\theta(R_0-2a-\delta R(s_0))}{\left(R_0+\delta R(s_0)\right)}+\frac{\theta(2a-R_0-\delta R(s))}{2a}$$
$$+\frac{\theta(\delta R(s)-R_0+2a)}{2(R_0-2a)}. \qquad (6.17)$$

The subscripts $\delta\eta$ and $\delta R$ on the averaging brackets in Eq. (6.15) correspond to averaging over fluctuations in $\delta\eta(R_0)$ and $\delta R(R_0)$, respectively, using the variational trail functional. Using the results of Appendix A we may write Eq. (6.15) as

$$\frac{\left\langle \tilde{E}_{R,B}[\eta_0+\delta\eta(s),R_0+\delta R(s)]+\tilde{E}_{St}[\delta R(s)]-E_{0,R}[\delta\eta(s),\delta R(s)]\right\rangle}{k_BT}=-\frac{L\theta_R^2 l_p}{R_0^2}f_1(R_0;a)\sin^2\left(\frac{\eta_0}{2}\right)-\frac{\alpha_\eta d_\eta^2}{2}$$

$$+\frac{4l_p f_1(R_0;a)}{R_0^2}\left[\sin^4\left(\frac{\eta_0}{2}\right)+\frac{d_\eta^2}{2}\left(3\cos^2\left(\frac{\eta_0}{2}\right)\sin^2\left(\frac{\eta_0}{2}\right)-\sin^4\left(\frac{\eta_0}{2}\right)\right)\right]+\frac{1}{4}\left(l_p-\tilde{l}_p\right)\left\langle\left(\frac{d\delta\eta(s_0)}{ds_0}\right)^2\right\rangle_{\delta\eta},$$

(6.18)

where

$$\theta_R^2=\frac{1}{2\pi}\int_{-\infty}^{\infty}\frac{k^2dk}{\frac{l_p}{2}k^4+\alpha_H}=\frac{1}{2^{3/4}}\frac{1}{l_p^{3/4}}\frac{1}{\alpha_H^{1/4}}\approx\frac{1}{2^{1/3}}\left(\frac{R_0-2a}{l_p}\right)^{2/3}, \tag{6.19}$$

$$d_R^2\approx(R-2a)=\frac{1}{2\pi}\int_{-\infty}^{\infty}\frac{dk}{k^4+\alpha_H}=\frac{1}{2^{5/4}}\frac{1}{l_p^{1/4}}\frac{1}{\alpha_H^{3/4}}, \tag{6.20}$$

$$d_\eta^2=\int_{-\infty}^{\infty}\frac{dk}{\frac{\tilde{l}_p}{2}k^2+\alpha_\eta}=\frac{1}{(2\tilde{l}_p\alpha_\eta)^{1/2}}, \tag{6.21}$$

and

$$f_1(R_0;a)=\frac{R_0^2}{(R_0-2a)\sqrt{2\pi}}\int_{2a-R_0}^{R_0-2a}\frac{dx}{(R_0+x)^2}\exp\left(-\frac{x^2}{2(R_0-2a)^2}\right)$$

$$+\frac{R_0^2}{4a^2(R_0-2a)\sqrt{2\pi}}\int_{-\infty}^{2a-R_0}dx\exp\left(-\frac{x^2}{2(R_0-2a)^2}\right)+\frac{R_0^2}{4(R_0-a)^2(R_0-2a)\sqrt{2\pi}}\int_{R_0-2a}^{\infty}dx\exp\left(-\frac{x^2}{2(R_0-2a)^2}\right)$$

$$=\frac{R_0^2}{(R_0-2a)\sqrt{2\pi}}\int_{2a-R_0}^{R_0-2a}\frac{dx}{(R_0+x)^2}\exp\left(-\frac{x^2}{2(R_0-2a)^2}\right)+\frac{1}{2}\left(\frac{R_0^2}{4a^2}+\frac{R_0^2}{4(R_0-a)^2}\right)\left(1-\mathrm{erf}\left(\frac{1}{\sqrt{2}}\right)\right).$$

(6.22)

Now, we consider the second term on the RHS of Eq. (6.14)

$$\frac{\left\langle E_{A,B}[x_A'(s_0),y_A'(s_0)]-E_{0,A}[x_A'(s_0),y_A'(s_0)]\right\rangle_0}{k_BT}=-\int_{-L/2}^{L/2}ds_0\alpha_x\left\langle x_A'(s_0)^2\right\rangle_x+\alpha_y\left\langle y_A'(s_0)^2\right\rangle_y \tag{6.23}$$

$$+2\pi M\left\langle\left\langle Wr_b\right\rangle_x\right\rangle_y,$$

where now $\langle\ldots\rangle_x$ and $\langle\ldots\rangle_y$ corresponds to performing the averages

$$\left\langle A[x'_A(s_0)]\right\rangle_x = \frac{1}{Z_x}\int Dx'_A(s_0) A[x'_A(s_0)]$$
$$\exp\left[-\int_{-L/2}^{L/2} ds_0 \left( l_p^b \left(\frac{dx'_A(s_0)}{ds_0}\right)^2 + \alpha_x x'_A(s_0)^2 \right)\right], \tag{6.24}$$

$$\left\langle A[y'_A(s_0)]\right\rangle_y = \frac{1}{Z_x}\int Dy'_A(s_0) A[y'_A(s_0)]$$
$$\exp\left[-\int_{-L/2}^{L/2} ds_0 \left( l_p^b \left(\frac{dy'_A(s_0)}{ds_0}\right)^2 + \alpha_y y'_A(s_0)^2 \right)\right]. \tag{6.25}$$

In Eqs. (6.24) and (6.25) we have neglected the contribution from $2\pi M Wr_b$, in the averaging, as being small. Using the fact that

$$\left\langle x'_A(s_0)^2\right\rangle_x = \frac{1}{4\pi}\int_{-\infty}^{\infty} dk \frac{1}{l_p k^2 + \alpha_x} = \frac{1}{4}\frac{1}{l_p^{1/2}\alpha_x^{1/2}}, \tag{6.26}$$

$$\left\langle y'_A(s_0)^2\right\rangle_x = \frac{1}{4\pi}\int_{-\infty}^{\infty} dk \frac{1}{l_p k^2 + \alpha_y} = \frac{1}{4}\frac{1}{l_p^{1/2}\alpha_y^{1/2}}, \tag{6.27}$$

we may write Eq. (6.23) as

$$\frac{\left\langle E_{A,B}[x'_A(s_0), y'_A(s_0)] - E_{0,A}[x'_A(s_0), y'_A(s_0)]\right\rangle_0}{k_B T} = -\frac{L}{4}\left[\left(\frac{\alpha_x}{l_p}\right)^{1/2} + \left(\frac{\alpha_y}{l_p}\right)^{1/2}\right] + 2\pi M \left\langle \left\langle Wr_b \right\rangle_x \right\rangle_y . \tag{6.28}$$

Next, we consider the average of the interaction energy appearing in Eq. (6.14). This reads as

$$\begin{aligned}
\left\langle \tilde{E}_{\text{int}} \right\rangle_0 \approx k_B T \sum_{n=-\infty}^{\infty} \int_{-L/2}^{L/2} ds_0 \Big\langle \exp(in\Delta\Phi(s_0)) \Big( & \tilde{\varepsilon}_{\text{int}}^0(\eta_0, R(s_0), n, R'(s_0), g_1^0(s_0), g_2^0(s_0)) \\
& + \frac{1}{2}\tilde{\varepsilon}_{\text{int}}^{\eta,\eta}(\eta_0, R(s_0), n, R'(s_0), g_1^0(s_0), g_2^0(s_0))\delta\eta(s_0)^2 \\
& + \frac{1}{2}\tilde{\varepsilon}_{\text{int}}^{\eta',\eta}(\eta_0, R(s_0), n, R'(s_0), g_1^0(s_0), g_2^0(s_0))\delta\eta'(s_0)^2 \\
& + \tilde{\varepsilon}_{\text{int}}^{\eta',\eta}(\eta_0, R(s_0), n, R'(s_0), g_1^0(s_0), g_2^0(s_0))\delta\eta(s_0)\delta\eta'(s_0) \\
& + \frac{1}{2}\tilde{\varepsilon}_{\text{int}}^{x,x}(\eta_0, R(s_0), n, R'(s_0), g_1^0(s_0), g_2^0(s_0))x'_A(s_0)^2 \\
& + \frac{1}{2}\tilde{\varepsilon}_{\text{int}}^{y,y}(\eta_0, R(s_0), n, R'(s_0), g_1^0(s_0), g_2^0(s_0))y'_A(s_0)^2 \Big)\Big\rangle_0 .
\end{aligned} \tag{6.29}$$

As we allow for free fluctuations in $\Delta\Phi(s_0)$, all terms in the Fourier series average out, except for $n=0$ term. Therefore, we can re-express Eq. (6.29), using Eqs. (6.21), (6.26) and (6.27), as

$$\begin{aligned}\left\langle \tilde{E}_{\text{int}} \right\rangle_0 \approx k_B T \int_{-L/2}^{L/2} ds_0 \Big( &\left\langle \tilde{\varepsilon}^{0}_{\text{int}}(\eta_0, R(s_0), 0, R'(s_0), g_1^0(s_0), g_2^0(s_0)) \right\rangle_{\delta R} \\ &+ \frac{1}{2^{3/2}\alpha_\eta^{1/2} \tilde{l}_p^{1/2}} \left\langle \tilde{\varepsilon}^{\eta,\eta}_{\text{int}}(\eta_0, R(s_0), 0, R'(s_0), g_1^0(s_0), g_2^0(s_0)) \right\rangle_{\delta R} \\ &+ \frac{1}{2} \left\langle \tilde{\varepsilon}^{\eta',\eta'}_{\text{int}}(\eta_0, R(s_0), 0, R'(s_0), g_1^0(s_0), g_2^0(s_0)) \right\rangle_{\delta R} \left\langle \left( \frac{d\delta\eta(s_0)}{ds_0} \right)^2 \right\rangle_{\delta\eta} \\ &+ \frac{1}{8\alpha_x^{1/2} l_p^{1/2}} \left\langle \tilde{\varepsilon}^{x,x}_{\text{int}}(\eta_0, R(s_0), 0, R'(s_0), g_1^0(s_0), g_2^0(s_0)) \right\rangle_{\delta R} \\ &+ \frac{1}{8\alpha_y^{1/2} l_p^{1/2}} \left\langle \tilde{\varepsilon}^{y,y}_{\text{int}}(\eta_0, R(s_0), 0, R'(s_0), g_1^0(s_0), g_2^0(s_0)) \right\rangle_{\delta R} \Big).\end{aligned} \tag{6.30}$$

By performing the averaging over $\delta R(s_0)$, we can express Eq.(6.30) as

$$\begin{aligned}\left\langle \tilde{E}_{\text{int}} \right\rangle_0 \approx k_B T \Big( &\bar{f}^{0}_{\text{int}}(\eta_0, R_0, 0) + \frac{1}{2^{3/2}\alpha_\eta^{1/2} \tilde{l}_p^{1/2}} \bar{f}^{\eta,\eta}_{\text{int}}(\eta_0, R_0, 0) \\ &+ \frac{1}{2} \bar{f}^{\eta',\eta'}_{\text{int}}(\eta_0, R_0, 0) \left\langle \left( \frac{d\delta\eta(s_0)}{ds_0} \right)^2 \right\rangle_{\delta\eta} + \frac{1}{8\alpha_x^{1/2} l_p^{1/2}} \bar{f}^{x,x}_{\text{int}}(\eta_0, R_0, 0) + \frac{1}{8\alpha_y^{1/2} l_p^{1/2}} \bar{f}^{y,y}_{\text{int}}(\eta_0, R_0, 0) \Big),\end{aligned} \tag{6.31}$$

where through the results of Appendix A, we define the functions $\bar{f}^{0}_{\text{int}}(\eta_0, R_0, 0)$, $\bar{f}^{\eta,\eta}_{\text{int}}(\eta_0, R_0, 0)$, and so on, as

$$\begin{aligned}\bar{f}^{0}_{\text{int}}(\eta_0, R_0, n) = \frac{2^{1/6}}{2\pi L} \left( \frac{l_p}{R_0 - 2a} \right)^{1/3} \frac{1}{R_0 - 2a} \int_{-L/2}^{L/2} ds_0 \int_{-\infty}^{\infty} dx \int_{-\infty}^{\infty} dx' \tilde{\varepsilon}^{0}_{\text{int}}(\eta_0, x + R_0, n, x', g_1^0(s_0), g_2^0(s_0)) \\ \exp\left( -\frac{x^2}{2(R_0 - 2a)^2} \right) \exp\left( -\frac{2^{1/3} l_p^{2/3} x'^2}{2(R_0 - 2a)^{2/3}} \right),\end{aligned} \tag{6.32}$$

$$\begin{aligned}\bar{f}^{Y,Y}_{\text{int}}(\eta_0, R_0, n) = \frac{2^{1/6}}{2\pi L} \left( \frac{l_p}{R_0 - 2a} \right)^{1/3} \frac{1}{R_0 - 2a} \int_{-L/2}^{L/2} ds_0 \int_{-\infty}^{\infty} dx \int_{-\infty}^{\infty} dx' \tilde{\varepsilon}^{Y,Y}_{\text{int}}(\eta_0, x + R_0, n, x', g_1^0(s_0), g_2^0(s_0)) \\ \exp\left( -\frac{x^2}{2(R_0 - 2a)^2} \right) \exp\left( -\frac{2^{1/3} l_p^{2/3} x'^2}{2(R_0 - 2a)^{2/3}} \right),\end{aligned}$$

(6.33)

and $Y=\eta,\eta',x,y$.

We move on to consider the average of the work term appearing in the RHS of Eq. (6.14)

$$\left\langle \tilde{E}_W \right\rangle \approx -FL\left[\cos\left(\frac{\eta_0}{2}\right)-\frac{1}{8}\cos\left(\frac{\eta_0}{2}\right)\frac{1}{\left(2\alpha_\eta \tilde{l}_p\right)^{1/2}}-\frac{1}{\cos\left(\frac{\eta_0}{2}\right)}\left[\frac{1}{8\alpha_x^{1/2}l_p^{1/2}}+\frac{1}{8\alpha_y^{1/2}l_p^{1/2}}\right]\right]$$ (6.34)

$$+2ML\left\langle \Gamma_1(\delta R_0)\right\rangle_{\delta R}\left[\sin\left(\frac{\eta_0}{2}\right)-\frac{1}{8}\sin\left(\frac{\eta_0}{2}\right)\frac{1}{\left(2\alpha_\eta l_p\right)^{1/2}}-\frac{1}{8}\theta_R^2\sin\left(\frac{\eta_0}{2}\right)^{-1}\right]-2\pi M\left\langle Wr_b\right\rangle_0.$$

Using Eq. (A.7) of Appendix A, we may write

$$\left\langle \Gamma_1(\delta R_0)\right\rangle_{\delta R}=\frac{f_2(R_0)}{R_0}, \tag{6.35}$$

where

$$f_2(R_0,a)=\frac{R_0}{(R_0-2a)\sqrt{2\pi}}\int_{2a-R_0}^{R_0-2a}\frac{dx}{(R_0+x)^2}\exp\left(-\frac{x^2}{2(R_0-2a)^2}\right)+\frac{1}{2}\left(\frac{R_0}{2a}+\frac{R_0}{2(R_0-a)}\right)\left(1-\mathrm{erf}\left(\frac{1}{\sqrt{2}}\right)\right).$$

(6.36)

As we allow for free fluctuations in $\Delta\Phi(s_0)$, the average elastic twisting energy in Eq. (6.14) becomes

$$\frac{\langle E_{Tw}\rangle}{k_BT}=\frac{l_c}{4}\int_{-L/2}^{L/2}ds_0\left[\left(\frac{d\Delta\Phi_0(s_0)}{ds_0}-\Delta g^0(s_0)\right)^2+\left(\bar{g}-\bar{g}^0(s_0)\right)^2\right], \tag{6.37}$$

where $\Delta\Phi_0(s_0)=\left\langle \Delta\Phi(s_0)\right\rangle_0$, which in this case is ill defined (in the limit $L\to\infty$), although its derivative is not. Eq. (6.37) is trivially minimized by

$$\frac{d\Delta\Phi_0(s_0)}{ds_0}=\Delta g^0(s_0), \qquad \bar{g}=\bar{g}^0(s_0), \tag{6.38}$$

and so $E_{Tw}=0$ is the minimum energy and the term does not contribute. Note that the ill definition of $\Delta\Phi_0(s_0)$ arises from a constant arbitrary constant of integration on integrating the right hand expression in Eq. (6.38).

Now we deal with $\ln Z_{0,A}$ appearing Eq. (6.8). First of all, as in [7] for single molecule twisting, we treat the term $2\pi M Wr_b$ in the energy functional as a perturbation so that for the free energy we may write

$$-k_BT\ln Z_{0,A} \approx -k_BT\ln Z_x - k_BT\ln Z_y - \frac{(2\pi M)^2}{2k_BT}\left\langle\left\langle Wr_b^2\right\rangle_x\right\rangle_y, \tag{6.39}$$

where we have used the fact that $\left\langle\left\langle Wr_b^2\right\rangle_x\right\rangle_y = 0$, when the average braid axis is straight. We will devote the evaluation of $\left\langle\left\langle Wr_b^2\right\rangle_x\right\rangle_y$ to a separate section and focus on $-k_BT\ln Z_x$ and $-k_BT\ln Z_y$. First of all, we note that

$$-\frac{\partial \ln Z_x}{\partial \alpha_x} = L\left\langle x_A'(s_0)^2\right\rangle = \frac{L}{4\alpha_x^{1/2} l_p^{1/2}}, \quad -\frac{\partial \ln Z_y}{\partial \alpha_y} = L\left\langle y_A'(s_0)^2\right\rangle = \frac{L}{4\alpha_y^{1/2} l_p^{1/2}}. \tag{6.40}$$

We can then integrate both expressions in Eq. (6.40) up to obtain

$$-\ln Z_x - \ln Z_y = \frac{L\alpha_x^{1/2}}{2l_p^{1/2}} + \frac{L\alpha_y^{1/2}}{2l_p^{1/2}} + \Theta_{x,y} \tag{6.41}$$

where $\Theta_{x,y}$ is an arbitrary constant of integration that may be discarded. The last term to consider in the expression for the variational Free energy (Eq. (6.8) is $\ln Z_{0,R}$. First of all we may write

$$-\frac{\partial \ln Z_{0,R}}{\partial \alpha_\eta} = \frac{L}{2}\left\langle \delta\eta(s_0)^2\right\rangle = \frac{L}{2^{3/2}\alpha_\eta^{1/2}\tilde{l}_p^{1/2}}, \quad -\frac{\partial \ln Z_{0,R}}{\partial \alpha_H} = \frac{L}{2}\left\langle \delta R(s_0)^2\right\rangle = \frac{L}{2^{9/4}}\frac{1}{l_p^{1/4}}\frac{1}{\alpha_H^{3/4}}. \tag{6.42}$$

Integrating the expression in Eq. (6.42) up yields

$$-\ln Z_{0,R} = \frac{\alpha_\eta^{1/2} L}{2^{1/2}\tilde{l}_p^{1/2}} + \frac{L}{2^{1/4}}\frac{\alpha_H^{1/4}}{l_p^{1/4}}. \tag{6.43}$$

Then, substituting in Eq. (6.12) we obtain

$$-\ln Z_{0,R} = \frac{\alpha_\eta^{1/2} L}{2^{3/2}\tilde{l}_p^{1/2}} + \frac{1}{l_p^{1/3}}\frac{1}{2^{2/3}(R_0-2a)^{2/3}}. \tag{6.44}$$

Combining Eqs. (6.18), (6.28),(6.31), (6.34), (6.41) and (6.44) gives us the following expression for $F_T$ the variational free energy

$$
\begin{aligned}
\frac{F_T}{k_BT} &= \frac{L}{4}\left[\left(\frac{\alpha_x}{l_p}\right)^{1/2}+\left(\frac{\alpha_y}{l_p}\right)^{1/2}\right]+\frac{\alpha_\eta^{1/2}L}{2^{3/2}\tilde{l}_p^{1/2}}+\frac{L}{l_p^{1/3}}\frac{1}{2^{2/3}(R_0-2a)^{2/3}}-\frac{L\theta_R^2}{R_0^2}f_1(R_0;a)l_p\sin^2\left(\frac{\eta_0}{2}\right)\\
&+\frac{4l_pLf_1(R_0;a)}{R_0^2}\sin^4\left(\frac{\eta_0}{2}\right)+L\bar{f}_{\text{int}}^{0}(\eta_0,R_0,0)+\frac{L}{4}\left(l_p+2\bar{f}_{\text{int}}^{\eta',\eta'}(\eta_0,R_0,0)-\tilde{l}_p\right)\left\langle\left(\frac{d\delta\eta(s_0)}{ds_0}\right)^2\right\rangle_{\delta\eta}\\
&+\frac{L}{2^{3/2}\alpha_\eta^{1/2}\tilde{l}_p^{1/2}}\left[\frac{4l_pf_1(R_0;a)}{R_0^2}\left(3\cos^2\left(\frac{\eta_0}{2}\right)\sin^2\left(\frac{\eta_0}{2}\right)-\sin^4\left(\frac{\eta_0}{2}\right)\right)+\bar{f}_{\text{int}}^{\eta,\eta}(\eta_0,R_0,0)+\frac{F_R}{4}\cos\left(\frac{\eta_0}{2}\right)\right.\\
&\left.-\frac{M_Rf_2(R_0;a)}{2R_0}\sin\left(\frac{\eta_0}{2}\right)\right]-F_RL\cos\left(\frac{\eta_0}{2}\right)+\frac{2M_RL}{R_0}\sin\left(\frac{\eta_0}{2}\right)f_2(R_0;a)-\frac{M_RL}{4}\theta_R^2\frac{f_2(R_0;a)}{R_0}\sin\left(\frac{\eta_0}{2}\right)^{-1}\\
&+\frac{L}{8\alpha_x^{1/2}l_p^{1/2}}\left(\bar{f}_{\text{int}}^{x,x}(\eta_0,R_0,0)+\frac{F_R}{\cos\left(\frac{\eta_0}{2}\right)}\right)+\frac{L}{8\alpha_y^{1/2}l_p^{1/2}}\left(\bar{f}_{\text{int}}^{y,y}(\eta_0,R_0,0)+\frac{F_R}{\cos\left(\frac{\eta_0}{2}\right)}\right)-\frac{(2\pi M_R)^2}{2}\left\langle\left\langle Wr_b^2\right\rangle_x\right\rangle_y,
\end{aligned}
\tag{6.45}
$$

where we have introduced the reduced force $F_R = F/k_BT$ and the reduced moment $M_R = M/k_BT$. We may choose

$$
\tilde{l}_p(\eta_0,R_0) = l_p + 2\bar{f}_{\text{int}}^{\eta',\eta'}(\eta_0,R_0,0), \tag{6.46}
$$

so that the singular $\left\langle\left(\frac{d\delta\eta(s_0)}{ds_0}\right)^2\right\rangle$ contribution vanishes. We can also minimize with respect to $\alpha_\eta$ which yields

$$
\begin{aligned}
\alpha_\eta &= \frac{4l_pf_1(R_0;a)}{R_0^2}\left(3\cos^2\left(\frac{\eta_0}{2}\right)\sin^2\left(\frac{\eta_0}{2}\right)-\sin^4\left(\frac{\eta_0}{2}\right)\right)\\
&+\frac{F_R}{4}\cos\left(\frac{\eta_0}{2}\right)-\frac{M_Rf_2(R_0;a)}{2R_0}\sin\left(\frac{\eta_0}{2}\right)+\bar{f}_{\text{int}}^{\eta,\eta}(\eta_0,R_0,0).
\end{aligned}
\tag{6.47}
$$

We can also minimize with respect to $\alpha_x$ and $\alpha_y$, neglecting the contribution from $\frac{(2\pi M_R)^2}{2}\left\langle\left\langle Wr_b^2\right\rangle_x\right\rangle_y$, yielding the simple expressions

$$
\alpha_x = \frac{1}{2}\left(\bar{f}_{\text{int}}^{x,x}(\eta_0,R_0,0)+\frac{F}{\cos\left(\frac{\eta_0}{2}\right)}\right) \quad\text{and}\quad \alpha_y = \frac{1}{2}\left(\bar{f}_{\text{int}}^{y,y}(\eta_0,R_0,0)+\frac{F}{\cos\left(\frac{\eta_0}{2}\right)}\right). \tag{6.48}
$$

Then, we may write from Eqs. (6.45)-(6.48)

$$\frac{F_T}{k_BT}=\frac{L}{2}\left[\left(\frac{\alpha_x}{l_p}\right)^{1/2}+\left(\frac{\alpha_y}{l_p}\right)^{1/2}\right]+\frac{\alpha_\eta^{1/2}L}{2^{1/2}\tilde{l}_p^{1/2}}+\frac{1}{l_p^{1/3}}\frac{L}{2^{2/3}(R_0-2a)^{2/3}}-\frac{L}{R_0^2}\frac{1}{2^{1/3}}\left(\frac{R_0-2a}{l_p}\right)^{2/3}f_1(R_0;a)l_p\sin^2\left(\frac{\eta_0}{2}\right)$$
$$+\frac{4l_pLf_1(R_0;a)}{R_0^2}\sin^4\left(\frac{\eta_0}{2}\right)+L\bar{f}_{\text{int}}^0(\eta_0,R_0,0)-F_RL\cos\left(\frac{\eta_0}{2}\right)+\frac{2M_RL}{R_0}\sin\left(\frac{\eta_0}{2}\right)f_2(R_0;a)$$
$$-\frac{M_RL}{4}\frac{1}{2^{1/3}}\left(\frac{R_0-2a}{l_p}\right)^{2/3}\frac{f_2(R_0;a)}{R_0}\sin\left(\frac{\eta_0}{2}\right)^{-1}-\frac{(2\pi M_R)^2}{2}\left\langle\left\langle Wr_b^2\right\rangle_x\right\rangle_y .$$

(6.49)

## 7. Computing the correction to the free energy due to braid writhing

We devote this section to computing $\left\langle\left\langle Wr_b^2\right\rangle_x\right\rangle_y$. First of all we may write , using Eq. (2.6)

$$\left\langle\left\langle Wr_b^2\right\rangle_x\right\rangle_y=\frac{1}{(4\pi)^2}\int_{-L_A/2}^{L_A/2}d\tau\int_{-L_A/2}^{L_A/2}d\tau'\int_{-L_A/2}^{L_A/2}d\tau''\int_{-L_A/2}^{L_A/2}d\tau'''$$
$$\left\langle\left\langle\frac{\left(\mathbf{r}_A(\tau)-\mathbf{r}_A(\tau')\right).\hat{\mathbf{t}}_A(\tau)\times\hat{\mathbf{t}}_A(\tau')}{\left|\mathbf{r}_A(\tau)-\mathbf{r}_A(\tau')\right|^3}.\frac{\left(\mathbf{r}_A(\tau'')-\mathbf{r}_A(\tau''')\right).\hat{\mathbf{t}}_A(\tau'')\times\hat{\mathbf{t}}_A(\tau''')}{\left|\mathbf{r}_A(\tau'')-\mathbf{r}_A(\tau''')\right|^3}\right\rangle_x\right\rangle_y . \quad (7.1)$$

We can write expressions for the tangent vector and position vectors of the braid axis to lowest order in the derivative expansion (using Eqs.(1.2), (1.13) and (1.23))

$$\hat{\mathbf{t}}_A(\tau)\approx\frac{ds_0}{d\tau}x_A'(s_0)\hat{\mathbf{i}}+\frac{ds_0}{d\tau}y_A'(s_0)\hat{\mathbf{j}}+\hat{\mathbf{k}}\approx\cos\left(\frac{\eta_0}{2}\right)^{-1}x_A'(s_0)\hat{\mathbf{i}}+\cos\left(\frac{\eta_0}{2}\right)^{-1}y_A'(s_0)\hat{\mathbf{j}}+\hat{\mathbf{k}}, \quad (7.2)$$

$$\mathbf{r}_A(\tau)\approx\int_{-L/2}^{s_0(\tau)}ds_0'x_A'(s_0')\hat{\mathbf{i}}+\int_{-L/2}^{s_0(\tau)}ds_0'y_A'(s_0')\hat{\mathbf{j}}+\tau\hat{\mathbf{k}}\equiv\left(x_A(s_0)\hat{\mathbf{i}}+y_A(s_0)\hat{\mathbf{j}}+\cos\left(\frac{\eta_0}{2}\right)s_0\hat{\mathbf{k}}\right). \quad (7.3)$$

Substituting Eqs. (7.2) and (7.3) into Eq. (7.1) allows us to write to lowest order in the derivative expansion

$$\left\langle\left\langle Wr_b^2\right\rangle_x\right\rangle_y=\frac{1}{(4\pi)^2}\frac{1}{\cos\left(\frac{\eta_0}{2}\right)^4}\int_{-L/2}^{L/2}ds_0\int_{-L/2}^{L/2}ds_0'\int_{-L/2}^{L/2}ds_0''\int_{-L/2}^{L/2}ds_0'''\frac{1}{\left|s_0-s_0'\right|^3}\frac{1}{\left|s_0''-s_0'''\right|^3}$$
$$\Big\langle\Big\langle\Big\{\left(x_A'(s_0)y_A'(s_0')-y_A'(s_0)x_A'(s_0')\right)(s_0-s_0')+\left(x_A(s_0)-x_A(s_0')\right)\left(y_A'(s_0)-y_A'(s_0')\right) \quad (7.4)$$
$$-\left(y_A(s_0)-y_A(s_0')\right)\left(x_A'(s_0)-x_A'(s_0')\right)\Big\}\Big\{\left(x_A'(s_0'')y_A'(s_0''')-y_A'(s_0'')x_A'(s_0''')\right)(s_0''-s_0''')$$
$$+\left(x_A(s_0'')-x_A(s_0''')\right)\left(y_A'(s_0'')-y_A'(s_0''')\right)-\left(y_A(s_0'')-y_A(s_0''')\right)\left(x_A'(s_0'')-x_A'(s_0''')\right)\Big\}\Big\rangle_x\Big\rangle_y .$$

We may write

$$\left\langle \left\langle Wr_b^2 \right\rangle_x \right\rangle_y = \mathcal{W}_1 + \mathcal{W}_2 + \mathcal{W}_3 + \mathcal{W}_4, \tag{7.5}$$

where

$$\mathcal{W}_1 = \frac{1}{(4\pi)^2 \cos\left(\frac{\eta_0}{2}\right)^4} \int_{-L/2}^{L/2} ds_0 \int_{-L/2}^{L/2} ds_0' \int_{-L/2}^{L/2} ds_0'' \int_{-L/2}^{L/2} ds_0''' \frac{s_0 - s_0'}{|s_0 - s_0'|^3} \frac{s_0'' - s_0'''}{|s_0'' - s_0'''|^3}$$
$$\Big[\left\langle x_A'(s_0) x_A'(s_0'')\right\rangle_x \left\langle y_A'(s_0') y_A'(s_0''')\right\rangle_y - \left\langle x_A'(s_0) x_A'(s_0''')\right\rangle_x \left\langle y_A'(s_0') y_A'(s_0'')\right\rangle_y$$
$$- \left\langle x_A'(s_0') x_A'(s_0'')\right\rangle_x \left\langle y_A'(s_0) y_A'(s_0''')\right\rangle_y + \left\langle y_A'(s_0) y_A'(s_0'')\right\rangle_y \left\langle x_A'(s_0') x_A'(s_0''')\right\rangle_x \Big]$$
$$= \frac{4}{(4\pi)^2 \cos\left(\frac{\eta_0}{2}\right)^4} \int_{-L/2}^{L/2} ds_0 \int_{-L/2}^{L/2} ds_0' \int_{-L/2}^{L/2} ds_0'' \int_{-L/2}^{L/2} ds_0''' \frac{s_0 - s_0'}{|s_0 - s_0'|^3} \frac{s_0'' - s_0'''}{|s_0'' - s_0'''|^3} \left\langle x_A'(s_0) x_A'(s_0'')\right\rangle_x \left\langle y_A'(s_0') y_A'(s_0''')\right\rangle_y, \tag{7.6}$$

$$\mathcal{W}_2 = \frac{1}{(4\pi)^2 \cos\left(\frac{\eta_0}{2}\right)^4} \int_{-L/2}^{L/2} ds_0 \int_{-L/2}^{L/2} ds_0' \int_{-L/2}^{L/2} ds_0'' \int_{-L/2}^{L/2} ds_0''' \frac{1}{|s_0 - s_0'|^3} \frac{1}{|s_0'' - s_0'''|^3}$$
$$\Big[\left\langle \left(x_A(s_0) - x_A(s_0')\right)\left(x_A(s_0'') - x_A(s_0''')\right)\right\rangle_x \left\langle \left(y_A'(s_0) - y_A'(s_0')\right)\left(y_A'(s_0'') - y_A'(s_0''')\right)\right\rangle_y \tag{7.7}$$
$$+ \left\langle \left(y_A(s_0) - y_A(s_0')\right)\left(y_A(s_0'') - y_A(s_0''')\right)\right\rangle_y \left\langle \left(x_A'(s_0) - x_A'(s_0')\right)\left(x_A'(s_0'') - x_A'(s_0''')\right)\right\rangle_x \Big],$$

$$\mathcal{W}_3 = -\frac{1}{(4\pi)^2} \frac{1}{\cos\left(\frac{\eta_0}{2}\right)^4} \int_{-L/2}^{L/2} ds_0 \int_{-L/2}^{L/2} ds_0' \int_{-L/2}^{L/2} ds_0'' \int_{-L/2}^{L/2} ds_0''' \frac{1}{|s_0 - s_0'|^3} \frac{1}{|s_0'' - s_0'''|^3}$$
$$\Big[\left\langle \left(x_A(s_0) - x_A(s_0')\right)\left(x_A'(s_0'') - x_A'(s_0''')\right)\right\rangle_x \left\langle \left(y_A'(s_0) - y_A'(s_0')\right)\left(y_A(s_0'') - y_A(s_0''')\right)\right\rangle_y$$
$$+ \left\langle \left(x_A(s_0'') - x_A(s_0''')\right)\left(x_A'(s_0) - x_A'(s_0')\right)\right\rangle_x \left\langle \left(y_A'(s_0'') - y_A'(s_0''')\right)\left(y_A(s_0) - y_A(s_0')\right)\right\rangle_y \Big] \tag{7.8}$$
$$= -\frac{2}{(4\pi)^2} \frac{1}{\cos\left(\frac{\eta_0}{2}\right)^4} \int_{-L/2}^{L/2} ds_0 \int_{-L/2}^{L/2} ds_0' \int_{-L/2}^{L/2} ds_0'' \int_{-L/2}^{L/2} ds_0''' \frac{1}{|s_0 - s_0'|^3} \frac{1}{|s_0'' - s_0'''|^3}$$
$$\left\langle \left(x_A(s_0) - x_A(s_0')\right)\left(x_A'(s_0'') - x_A'(s_0''')\right)\right\rangle_x \left\langle \left(y_A(s_0) - y_A(s_0')\right)\left(y_A'(s_0'') - y_A'(s_0''')\right)\right\rangle_y,$$

$$\mathcal{W}_4 = \frac{2}{(2\pi)^2}\frac{1}{\cos\left(\frac{\eta_0}{2}\right)^4}\int_{-L/2}^{L/2} ds_0 \int_{-L/2}^{L/2} ds_0' \int_{-L/2}^{L/2} ds_0'' \int_{-L/2}^{L/2} ds_0''' \frac{s_0 - s_0'}{|s_0 - s_0'|^3}\frac{1}{|s_0'' - s_0'''|^3}$$

$$\Big[\left\langle x_A'(s_0)\left(x_A(s_0'') - x_A(s_0''')\right)\right\rangle_x \left\langle y_A'(s_0')\left(y_A'(s_0'') - y_A'(s_0''')\right)\right\rangle_y$$

$$-\left\langle x_A'(s_0)\left(x_A'(s_0'') - x_A'(s_0''')\right)\right\rangle_x \left\langle y_A'(s_0')\left(y_A(s_0'') - y_A(s_0''')\right)\right\rangle_y$$

$$-\left\langle y_A'(s_0)\left(y_A'(s_0'') - y_A'(s_0''')\right)\right\rangle_y \left\langle x_A'(s_0')\left(x_A(s_0'') - x_A(s_0''')\right)\right\rangle_x$$

$$+\left\langle y_A'(s_0)\left(y_A(s_0'') - y_A(s_0''')\right)\right\rangle_y \left\langle x_A'(s_0')\left(x_A'(s_0'') - x_A'(s_0''')\right)\right\rangle_x \Big]$$

$$= \frac{4}{(2\pi)^2}\frac{1}{\cos\left(\frac{\eta_0}{2}\right)^4}\int_{-L_b/2}^{L_b/2} ds_0 \int_{-L_b/2}^{L_b/2} ds_0' \int_{-L_b/2}^{L_b/2} ds_0'' \int_{-L_b/2}^{L_b/2} ds_0''' \frac{s_0 - s_0'}{|s_0 - s_0'|^3}\frac{1}{|s_0'' - s_0'''|^3}$$

$$\Big[\left\langle x_A'(s_0)\left(x_A(s_0'') - x_A(s_0''')\right)\right\rangle_x \left\langle y_A'(s_0')\left(y_A'(s_0'') - y_A'(s_0''')\right)\right\rangle_y$$

$$+\left\langle y_A'(s_0)\left(y_A(s_0'') - y_A(s_0''')\right)\right\rangle_y \left\langle x_A'(s_0')\left(x_A'(s_0'') - x_A'(s_0''')\right)\right\rangle_x \Big]. \tag{7.9}$$

First we can express the correlation functions in terms of their Fourier transforms. For instance, we have

$$\left\langle x_A'(s) x_A'(s')\right\rangle_x = \frac{1}{2\pi}\int_{-\infty}^{\infty} dk \frac{\exp\left(-(s-s')k\right)}{2(l_p k^2 + \alpha_x)}, \tag{7.10}$$

$$\left\langle y_A'(s) y_A'(s')\right\rangle_y = \frac{1}{2\pi}\int_{-\infty}^{\infty} dk \frac{\exp\left(-(s-s')k\right)}{2(l_p k^2 + \alpha_y)}, \tag{7.11}$$

$$\begin{aligned}\left\langle \left(x_A(s_0) - x_A(s_0')\right)\left(x_A(s_0'') - x_A(s_0''')\right)\right\rangle_x &= \int_{s_0'}^{s_0} d\tilde{s}\int_{s_0'''}^{s_0''} d\tilde{s}' \left\langle x_A'(\tilde{s}) x_A'(\tilde{s}')\right\rangle_x \\ = \frac{1}{2\pi}\int_{-\infty}^{\infty} \frac{dk}{k^2}\frac{\exp\left(-ik\left(s_0 - s_0''\right)\right)}{2l_p k^2 + 2\alpha_x}&\left[1 - \exp\left(ik\left(s_0 - s_0'\right)\right)\right]\left[1 - \exp\left(-ik\left(s_0'' - s_0'''\right)\right)\right],\end{aligned} \tag{7.12}$$

$$\begin{aligned}\left\langle \left(y_A(s_0) - y(s_0')\right)\left(y_A(s_0'') - y_A(s_0''')\right)\right\rangle_y &= \int_{s_0'}^{s_0} d\tilde{s}\int_{s_0'''}^{s_0''} d\tilde{s}' \left\langle y_A'(\tilde{s}) y_A'(\tilde{s}')\right\rangle_y \\ = \frac{1}{2\pi}\int_{-\infty}^{\infty} \frac{dk}{k^2}\frac{\exp\left(-ik\left(s_0 - s_0''\right)\right)}{2l_p k^2 + 2\alpha_y}&\left[1 - \exp\left(ik\left(s_0 - s_0'\right)\right)\right]\left[1 - \exp\left(-ik\left(s_0'' - s_0'''\right)\right)\right],\end{aligned} \tag{7.13}$$

$$\begin{aligned}\left\langle \left(x_A(s_0) - x_A(s_0')\right)\left(x_A'(s_0'') - x_A'(s_0''')\right)\right\rangle_x &= \int_{s_0'}^{s_0} d\tilde{s} \left\langle x_A'(\tilde{s})\left(x_A'(s_0'') - x_A'(s_0''')\right)\right\rangle_x = \\ = \frac{1}{2\pi}\int_{-\infty}^{\infty} \frac{dk}{2ik}\frac{\exp\left(-ik\left(s_0 - s_0''\right)\right)}{l_p k^2 + \alpha_x}&\left[1 - \exp\left(ik\left(s_0 - s_0'\right)\right)\right]\left[1 - \exp\left(-ik\left(s_0'' - s_0'''\right)\right)\right],\end{aligned} \tag{7.14}$$

and

$$\left\langle \left(y_A(s_0)-y_A(s_0')\right)\left(y_A'(s_0'')-y_A'(s_0''')\right)\right\rangle = \int_{s_0'}^{s_0} d\tilde{s}\left\langle y_A'(\tilde{s})\left(y_A'(s_0'')-y_A'(s_0''')\right)\right\rangle$$

$$= \frac{1}{2\pi}\int_{-\infty}^{\infty}\frac{dk}{2ik}\frac{\exp\left(-ik\left(s_0-s_0''\right)\right)}{l_pk^2+\alpha_y}\left[1-\exp\left(ik\left(s_0-s_0'\right)\right)\right]\left[1-\exp\left(-ik\left(s_0''-s_0'''\right)\right)\right]. \tag{7.15}$$

This allows us to write

$$\mathcal{W}_1 = \frac{1}{4(2\pi)^4}\frac{1}{\cos\left(\frac{\eta_0}{2}\right)^4}\int_{-L/2}^{L/2}ds_0\int_{-L/2}^{L/2}ds_0'\int_{-L/2}^{L/2}ds_0''\int_{-L/2}^{L/2}ds_0'''\int_{-\infty}^{\infty}dk\int_{-\infty}^{\infty}dk'\,\frac{s_0-s_0'}{\left|s_0-s_0'\right|^3}\frac{s_0''-s'''}{\left|s_0''-s_0'''\right|^3}$$
$$\frac{\exp\left(-(s_0-s_0'')k\right)\exp\left(-(s_0'-s_0''')k'\right)}{(l_pk^2+\alpha_x)(l_pk'^2+\alpha_y)}, \tag{7.16}$$

$$\mathcal{W}_2 = \frac{1}{4(2\pi)^4}\frac{1}{\cos\left(\frac{\eta_0}{2}\right)^4}\int_{-L/2}^{L/2}ds_0\int_{-L/2}^{L/2}ds_0'\int_{-L/2}^{L/2}ds_0''\int_{-L/2}^{L/2}ds_0'''\frac{1}{\left|s_0-s_0'\right|^3}\frac{1}{\left|s_0''-s_0'''\right|^3}\left(\frac{1}{4k^2}+\frac{1}{4k'^2}\right)$$
$$\frac{\exp\left(-i(k+k')\left(s_0-s_0''\right)\right)}{(l_pk^2+\alpha_x)(l_pk'^2+\alpha_y)}\left[1-\exp\left(ik\left(s_0-s_0'\right)\right)\right]\left[1-\exp\left(-ik\left(s_0''-s_0'''\right)\right)\right] \tag{7.17}$$
$$\left[1-\exp\left(ik'\left(s_0-s_0'\right)\right)\right]\left[1-\exp\left(-ik'\left(s_0''-s_0'''\right)\right)\right],$$

$$\mathcal{W}_3 = -\frac{2}{4(2\pi)^4}\frac{1}{\cos\left(\frac{\eta_0}{2}\right)^4}\int_{-L/2}^{L/2}ds_0\int_{-L/2}^{L/2}ds_0'\int_{-L/2}^{L/2}ds_0''\int_{-L/2}^{L/2}ds_0'''\int_{-\infty}^{\infty}dk\int_{-\infty}^{\infty}dk'\,\frac{1}{\left|s_0-s_0'\right|^3}\frac{1}{\left|s_0''-s_0'''\right|^3}$$
$$\frac{\exp\left(-i\left(k+k'\right)\left(s_0-s_0''\right)\right)\left[1-\exp\left(ik\left(s_0-s_0'\right)\right)\right]\left[1-\exp\left(-ik\left(s_0''-s_0'''\right)\right)\right]}{4kk'\left(2l_pk^2+\alpha_x\right)\left(2l_pk'^2+\alpha_y\right)} \tag{7.18}$$
$$\left[1-\exp\left(ik'\left(s_0-s_0'\right)\right)\right]\left[1-\exp\left(-ik'\left(s_0''-s_0'''\right)\right)\right],$$

$$\mathcal{W}_4 = \frac{1}{4(2\pi)^4}\frac{1}{\cos\left(\frac{\eta_0}{2}\right)^4}\int_{-L/2}^{L/2}ds_0\int_{-L/2}^{L/2}ds_0'\int_{-L/2}^{L/2}ds_0''\int_{-L/2}^{L/2}ds_0'''\int_{-\infty}^{\infty}dk\int_{-\infty}^{\infty}dk'\,\frac{s_0-s_0'}{\left|s_0-s_0'\right|^3}\frac{1}{\left|s_0''-s_0'''\right|^3}$$
$$\frac{\exp\left(-i\left(k+k'\right)\left(s_0-s_0''\right)\right)}{\left(l_pk^2+\alpha_x\right)\left(l_pk'^2+\alpha_y\right)}\left(\frac{\exp(ik\left(s_0-s_0'\right))}{ik'}\left[1-\exp\left(ik\left(s_0''-s_0'''\right)\right)\right]\left[1-\exp\left(-ik'\left(s_0''-s_0'''\right)\right)\right]\right.$$
$$\left.+\frac{\exp(ik'\left(s_0-s_0'\right))}{ik}\left[1-\exp\left(ik'\left(s_0''-s_0'''\right)\right)\right]\left[1-\exp\left(-ik\left(s_0''-s_0'''\right)\right)\right]\right).$$

(7.19)

We make the following variable changes $x_0 = s_0$, $x_1 = s_0 - s_0'$, $x_2 = s_0'' - s_0'''$ and $x_3 = s_0 - s_0''$. Then, we set the limits of integration to infinity on integration of $x_1$, $x_2$ and $x_3$, as we assume that the braid is sufficiently long. The integration over $x_0, x_3$ and $k'$ are easily performed yielding

$$\mathcal{W}_1 \approx \frac{1}{4(2\pi)^3} \frac{L}{\cos\left(\frac{\eta_0}{2}\right)^4} \int_{-\infty}^{\infty} dx_1 \int_{-\infty}^{\infty} dx_2 \int_{-\infty}^{\infty} dk \frac{x_1}{|x_1|^3} \frac{x_2}{|x_2|^3} \frac{\exp\left(ik(x_1 - x_2)\right)}{(l_p k^2 + \alpha_x)(l_p k^2 + \alpha_y)}, \quad (7.20)$$

$$\mathcal{W}_2 + \mathcal{W}_3 \approx \frac{1}{4(2\pi)^3} \frac{L}{\cos\left(\frac{\eta_0}{2}\right)^4} \int_{-\infty}^{\infty} dx_1 \int_{-\infty}^{\infty} dx_2 \int_{-\infty}^{\infty} dk \frac{1}{|x_1|^3} \frac{1}{|x_2|^3} \frac{\left[2 - 2\cos(kx_1)\right]\left[2 - 2\cos(kx_2)\right]}{k^2(l_p k^2 + \alpha_x)(l_p k^2 + \alpha_y)},$$

(7.21)

$$\mathcal{W}_4 \approx \frac{1}{4(2\pi)^3} \frac{2iL_b}{\cos\left(\frac{\eta_0}{2}\right)^4} \int_{-\infty}^{\infty} dx_1 \int_{-\infty}^{\infty} dx_2 \int_{-\infty}^{\infty} dk \frac{\exp\left(ikx_1\right)}{|x_1|^3} \frac{\left[2 - 2\cos(kx_2)\right]}{|x_2|^3} \frac{x_1}{k\left(l_p k^2 + \alpha_x\right)\left(l_p k^2 + \alpha_y\right)}.$$

(7.22)

Summing Eq. (7.20)-(7.22) together and rescaling the $x_1$ and $x_2$ integrals so that $x_1 k = \tilde{x}_1$ and $x_2 k = \tilde{x}_2$, as well as exploiting interchange symmetries of $x_1$ and $x_2$, yields

$$\begin{aligned} \left\langle\left\langle Wr_b^2 \right\rangle_x\right\rangle_y &= \frac{1}{4(2\pi)^3 l_p^2} \frac{L}{\cos\left(\frac{\eta_0}{2}\right)^4} \int_{-\infty}^{\infty} d\tilde{x}_1 \int_{-\infty}^{\infty} d\tilde{x}_2 \int_{-\infty}^{\infty} dk \frac{1}{|\tilde{x}_1|^3} \frac{1}{|\tilde{x}_2|^3} \frac{k^2}{\left(k^2 + \alpha_x / l_p\right)\left(k^2 + \alpha_y / l_p\right)} \\ &\left[\left(2 - 2\cos\left(\tilde{x}_1\right)\right) - i\tilde{x}_1 \exp\left(-i\tilde{x}_1\right)\right]\left[\left(2 - 2\cos\left(\tilde{x}_2\right)\right) + i\tilde{x}_2 \exp\left(i\tilde{x}_2\right)\right] \\ &= \frac{1}{4(2\pi) l_p^2} \frac{LI^2}{\cos\left(\frac{\eta_0}{2}\right)^4} \int_{-\infty}^{\infty} dk \frac{k^2}{\left(k^2 + \alpha_x / l_p\right)\left(k^2 + \alpha_y / l_p\right)}, \end{aligned} \quad (7.23)$$

where

$$\begin{aligned} I &= \frac{1}{2\pi} \int_0^{\infty} dx \frac{2}{x^3}\left[2 - 2\cos x - x\sin x\right] = -\frac{1}{2\pi} \int_0^{\infty} dx 2 \frac{d}{dx}\left(\frac{1 - \cos x}{x^2}\right) \\ &= -\frac{1}{\pi} \lim_{x \to 0}\left\{\frac{1 - \cos x}{x^2}\right\} = -\frac{1}{2\pi}. \end{aligned} \quad (7.24)$$

It then follows from Eq. (7.23) that

$$\left\langle \left\langle Wr_b^2 \right\rangle_x \right\rangle_y = \frac{1}{4(2\pi)^3 l_p^2} \frac{L}{\cos\left(\frac{\eta_0}{2}\right)^4} \int_{-\infty}^{\infty} dk \frac{1}{(\alpha_y - \alpha_x)} \left( \frac{\alpha_y}{k^2 + \alpha_y / l_p} - \frac{\alpha_x}{k^2 + \alpha_x / l_p} \right)$$

$$= \frac{1}{8(2\pi)^2 l_p^{3/2}} \frac{L}{\cos\left(\frac{\eta_0}{2}\right)^4} \frac{1}{\left(\alpha_y^{1/2} + \alpha_x^{1/2}\right)}. \qquad (7.25)$$

Therefore, we have as our final result for the Free energy for weakly interacting braids

$$\frac{F_T}{k_B T} = \frac{L}{2}\left[\left(\frac{\alpha_x}{l_p}\right)^{1/2} + \left(\frac{\alpha_y}{l_p}\right)^{1/2}\right] + \frac{\alpha_\eta^{1/2} L}{2^{1/2} \tilde{l}_p^{1/2}} + \frac{1}{l_p^{1/3}} \frac{L}{2^{2/3}(R_0 - 2a)^{2/3}} - \frac{L}{R_0^2} \frac{1}{2^{1/3}} \left(\frac{R_0 - 2a}{l_p}\right)^{2/3} f_1(R_0;a) l_p \sin^2\left(\frac{\eta_0}{2}\right)$$

$$+ \frac{4 l_p L f_1(R_0;a)}{R_0^2} \sin^4\left(\frac{\eta_0}{2}\right) + L\bar{f}_{\text{int}}^0(\eta_0, R_0, 0) - F_R L \cos\left(\frac{\eta_0}{2}\right) + \frac{2 M_R L}{R_0} \sin\left(\frac{\eta_0}{2}\right) f_2(R_0;a)$$

$$- \frac{M_R L}{4} \frac{1}{2^{1/3}} \left(\frac{R_0 - 2a}{l_p}\right)^{2/3} \frac{f_2(R_0;a)}{R_0} \sin\left(\frac{\eta_0}{2}\right)^{-1} - \frac{M_R^2}{16 l_p^{3/2}} \frac{L}{\cos\left(\frac{\eta_0}{2}\right)^4} \frac{1}{\left(\alpha_y^{1/2} + \alpha_x^{1/2}\right)}. \qquad (7.26)$$

## 8. Equations that minimizes the free energy for weakly interacting braids, the average extension and braid linking number

We can now minimize Eq. (7.26) with respect to the geometric parameters $R_0$ and $\eta_0$ obtain the following Equations. Minimization with respect to $\eta_0$ yields

$$
\begin{aligned}
0 &= \frac{1}{16}\left[\left(\frac{1}{\alpha_x l_p}\right)^{1/2}+\left(\frac{1}{\alpha_y l_p}\right)^{1/2}\right]\frac{F_R\sin\left(\frac{\eta_0}{2}\right)}{\cos\left(\frac{\eta_0}{2}\right)^2}+\frac{1}{8}\left[\left(\frac{1}{\alpha_x l_p}\right)^{1/2}\frac{\partial \bar{f}_{\text{int}}^{x,x}(\eta_0,R_0,0)}{\partial\eta_0}+\left(\frac{1}{\alpha_y l}\right)^{1/2}\frac{\partial \bar{f}_{\text{int}}^{y,y}(\eta_0,R_0,0)}{\partial\eta_0}\right] \\
&+\frac{1}{2^{3/2}\tilde{l}_p^{1/2}\alpha_\eta^{1/2}}\frac{\partial\alpha_\eta}{\partial\eta_0}-\frac{\alpha_\eta^{1/2}}{2^{1/2}\tilde{l}_p^{3/2}}\frac{\partial \bar{f}_{\text{int}}^{\eta',\eta'}(\eta_0,R_0,0)}{\partial\eta_0}-\frac{1}{R_0^2}\frac{1}{2^{1/3}}\left(\frac{R_0-2a}{l_p}\right)^{2/3} f_1(R_0;a)l_p\cos\left(\frac{\eta_0}{2}\right)\sin\left(\frac{\eta_0}{2}\right) \\
&+\frac{8l_p f_1(R_0;a)}{R_0^2}\cos\left(\frac{\eta_0}{2}\right)\sin^3\left(\frac{\eta_0}{2}\right)+\frac{\partial \bar{f}_{\text{int}}^{0}(\eta_0,R_0,0)}{\partial\eta_0}+\frac{F_R}{2}\sin\left(\frac{\eta_0}{2}\right)+\frac{M_R}{R_0}\cos\left(\frac{\eta_0}{2}\right)f_2(R_0;a) \\
&+\frac{M_R}{8}\frac{1}{2^{1/3}}\left(\frac{R_0-2a}{l_p}\right)^{2/3}\frac{f_2(R_0;a)}{R_0}\frac{\cos\left(\frac{\eta_0}{2}\right)}{\sin^2\left(\frac{\eta_0}{2}\right)}+\frac{M_R^2}{128 l_p^{3/2}}\frac{L}{\cos\left(\frac{\eta_0}{2}\right)^4}\frac{1}{\left(\alpha_y^{1/2}+\alpha_x^{1/2}\right)\alpha_y^{1/2}\alpha_x^{1/2}}\frac{F_R\sin\left(\frac{\eta_0}{2}\right)}{\cos^2\left(\frac{\eta_0}{2}\right)} \\
&+\frac{M_R^2}{64 l_p^{3/2}}\frac{1}{\cos\left(\frac{\eta_0}{2}\right)^4}\frac{1}{\left(\alpha_y^{1/2}+\alpha_x^{1/2}\right)^2}\left(\frac{1}{\alpha_y^{1/2}}\frac{\partial \bar{f}_{\text{int}}^{y,y}(\eta_0,R_0,0)}{\partial\eta_0}+\frac{1}{\alpha_x^{1/2}}\frac{\partial \bar{f}_{\text{int}}^{x,x}(\eta_0,R_0,0)}{\partial\eta_0}\right) \\
&-\frac{M_R^2}{8 l_p^{3/2}}\frac{\sin\left(\frac{\eta_0}{2}\right)}{\cos\left(\frac{\eta_0}{2}\right)^5}\frac{1}{\left(\alpha_y^{1/2}+\alpha_x^{1/2}\right)},
\end{aligned}
\tag{8.1}
$$

where

$$
\begin{aligned}
\frac{\partial\alpha_\eta}{\partial\eta_0} &= \frac{4l_p f_1(R_0;a)}{R_0^2}\left(3\cos^3\left(\frac{\eta_0}{2}\right)\sin^2\left(\frac{\eta_0}{2}\right)-5\sin^3\left(\frac{\eta_0}{2}\right)\cos\left(\frac{\eta_0}{2}\right)\right)+\frac{\partial \bar{f}_{\text{int}}^{\eta,\eta}(\eta_0,R_0,0)}{\partial\eta_0} \\
&-\frac{F_R}{8}\sin\left(\frac{\eta_0}{2}\right)-\frac{M_R f_2(R;a)}{4R_0}\cos\left(\frac{\eta_0}{2}\right).
\end{aligned}
\tag{8.2}
$$

Minimization of Eq. (7.26) with respect to $R_0$ yields

$$0=\frac{1}{8}\left[\left(\frac{1}{\alpha_x l_p}\right)^{1/2}\frac{\partial\bar{f}_{\text{int}}^{x,x}(\eta_0,R_0,0)}{\partial R_0}+\left(\frac{1}{\alpha_y l_p}\right)^{1/2}\frac{\partial\bar{f}_{\text{int}}^{y,y}(\eta_0,R_0,0)}{\partial R_0}\right]+\frac{1}{2^{3/2}\tilde{l}_p^{1/2}\alpha_\eta^{1/2}}\frac{\partial\alpha_\eta}{\partial R_0}-\frac{\alpha_\eta^{1/2}}{2^{1/2}\tilde{l}_p^{3/2}}\frac{\partial\bar{f}_{\text{int}}^{\eta',\eta'}(\eta_0,R_0,0)}{\partial R_0}$$
$$-\frac{2}{3}\frac{1}{l_p^{1/3}}\frac{1}{2^{2/3}(R_0-2a)^{5/3}}+\frac{2l_p}{R_0^3}\frac{1}{2^{1/3}}\left(\frac{R_0-2a}{l_p}\right)^{2/3}h_1(R_0;a)\sin^2\left(\frac{\eta_0}{2}\right)-\frac{2}{3R_0^2}\frac{1}{2^{1/3}}\left(\frac{l_p}{(R_0-2a)}\right)^{1/3}f_1(R_0;a)\sin^2\left(\frac{\eta_0}{2}\right)$$
$$-\frac{8l_p h_1(R_0;a)}{R_0^3}\sin^4\left(\frac{\eta_0}{2}\right)+\frac{\partial\bar{f}_{\text{int}}^{0}(\eta_0,R_0,0)}{\partial R_0}-\frac{2M_R}{R_0^2}\sin\left(\frac{\eta_0}{2}\right)h_2(R_0;a)$$
$$+\frac{M_R}{4}\frac{1}{2^{1/3}}\left(\frac{R_0-2a}{l_p}\right)^{2/3}\frac{h_2(R_0;a)}{R_0^2}\sin\left(\frac{\eta_0}{2}\right)^{-1}-\frac{M_R}{6}\frac{1}{2^{1/3}}\left(\frac{l_p}{R_0-2a}\right)^{1/3}\frac{f_2(R_0;a)}{l_pR_0}\sin\left(\frac{\eta_0}{2}\right)^{-1}$$
$$+\frac{M_R^2}{64l_p^{3/2}}\frac{1}{\cos\left(\frac{\eta_0}{2}\right)^4}\frac{1}{\left(\alpha_y^{1/2}+\alpha_x^{1/2}\right)^2}\left(\frac{1}{\alpha_y^{1/2}}\frac{\partial\bar{f}_{\text{int}}^{y,y}(\eta_0,R_0,0)}{\partial R_0}+\frac{1}{\alpha_x^{1/2}}\frac{\partial\bar{f}_{\text{int}}^{x,x}(\eta_0,R_0,0)}{\partial R_0}\right),$$

(8.3)

where

$$h_1(R_0;a)=\frac{R_0^3}{\sqrt{2\pi}}\int_{-1}^{1}dy\frac{(1+y)\exp\left(-y^2/2\right)}{\left(R_0+(R_0-2a)y\right)^3}+\frac{R_0^3}{8\left(R_0-a\right)^3}\left(1-\text{erf}\left(\frac{1}{\sqrt{2}}\right)\right),\qquad(8.4)$$

$$h_2(R_0;a)=\frac{R_0^2}{\sqrt{2\pi}}\int_{-1}^{1}dy\frac{(1+y)\exp\left(-y^2/2\right)}{\left(R_0+(R_0-2a)y\right)^2}+\frac{R_0^2}{4\left(R_0-a\right)^2}\left(1-\text{erf}\left(\frac{1}{\sqrt{2}}\right)\right),\qquad(8.5)$$

and

$$\frac{\partial\alpha_\eta}{\partial R_0}=-\frac{8l_p h_1(R_0;a)}{R_0^3}\left(3\cos^2\left(\frac{\eta_0}{2}\right)\sin^2\left(\frac{\eta_0}{2}\right)-\sin^4\left(\frac{\eta_0}{2}\right)\right)+\frac{\partial\bar{f}_{\text{int}}^{\eta,\eta}(\eta_0,R_0,0)}{\partial R_0}+\frac{M_R h_2(R_0;a)}{2R_0^2}\sin\left(\frac{\eta_0}{2}\right).$$

(8.6)

By taking the derivative of our expression for the free energy, Eq. (7.26) with respect to $M$ and $F$ (c.f. Eq. (6.5)) we obtain expressions for the average braid linking number and braid extension

$$-\frac{\langle Lk_b\rangle}{L}=\frac{\sin\left(\frac{\eta_0}{2}\right)}{\pi R_0}f_2(R_0;a)\left(1-\frac{1}{8\tilde{l}_p^{1/2}\left(2\alpha_\eta\right)^{1/2}}\right)$$
$$-\frac{1}{8\pi}\frac{1}{2^{1/3}}\left(\frac{R_0-2a}{l_p}\right)^{2/3}\frac{f_2(R_0;a)}{R_0}\sin\left(\frac{\eta_0}{2}\right)^{-1}-\frac{M_R}{16\pi l_p^{3/2}}\frac{1}{\cos\left(\frac{\eta_0}{2}\right)^4}\frac{1}{\left(\alpha_y^{1/2}+\alpha_x^{1/2}\right)},\qquad(8.7)$$

$$\frac{\langle z\rangle}{L}=\cos\left(\frac{\eta_0}{2}\right)\left(1-\frac{1}{2^{1/2}\tilde{l}_p^{1/2}\alpha_\eta^{1/2}}\frac{1}{8}\right)-\frac{1}{8}\left[\left(\frac{1}{\alpha_x l_p}\right)^{1/2}+\left(\frac{1}{\alpha_y l_p}\right)^{1/2}\right]\frac{1}{\cos\left(\frac{\eta_0}{2}\right)}$$
$$-\frac{M_R^2}{64 l_p^{3/2}}\frac{1}{\cos\left(\frac{\eta_0}{2}\right)^5}\frac{1}{\left(\alpha_y^{1/2}+\alpha_x^{1/2}\right)\alpha_x^{1/2}\alpha_y^{1/2}}. \quad (8.8)$$

If $R-2a$ is sufficiently small we may approximate in Eqs. (7.26), (8.1), (8.2), (8.3), (8.6) and (8.7)

$$f_1(R_0)\approx 1, \quad f_2(R_0)\approx 1, \quad h_1(R_0)\approx 1, \qquad h_1(R_0)\approx 1, \quad (8.9)$$

$$\overline{f}_{\text{int}}^{0}(\eta_0,R_0,n)\approx\frac{1}{L}\int_{-L/2}^{L/2}ds_0\overline{\varepsilon}_{\text{int}}^{0}(\eta_0,R_0,n,x',g_1^0(s_0),g_2^0(s_0)), \quad (8.10)$$

$$\overline{f}_{\text{int}}^{Y,Y}(\eta_0,R_0,n)\approx\frac{1}{L}\int_{-L/2}^{L/2}ds_0\overline{\varepsilon}_{\text{int}}^{Y,Y}(\eta_0,R_0,n,x',g_1^0(s_0),g_2^0(s_0)), \quad (8.11)$$

where $Y=\eta,\eta',x,y$. We can further suppose that we can neglect $\overline{f}_{\text{int}}^{\eta,\eta}(\eta_0,R_0,n)$, $\overline{f}_{\text{int}}^{\eta',\eta'}(\eta_0,R_0,n)$, $\overline{f}_{\text{int}}^{x,x}(\eta_0,R_0,n)$ and $\overline{f}_{\text{int}}^{y,y}(\eta_0,R_0,n)$, valid when braid interactions are sufficiently weak. This is precisely the approach that was used in [8], describing the mechanical braiding of two DNA molecules, where $g_1^0(s_0), g_2^0(s_0)$ have been approximated as being constant (see section 14 below for a justification of this for DNA), to derive the undulation free energy contribution (Eq. (19), therein). This approximation is in the spirit of [9], first to tackle the statistical mechanics of the braiding of two molecules.

## 9. Non interacting braids

In this case we simply set

$$\overline{f}_{\text{int}}^{0}(\eta_0,R_0,n)=\overline{f}_{\text{int}}^{\eta,\eta}(\eta_0,R_0,n)=\overline{f}_{\text{int}}^{\eta',\eta'}(\eta_0,R_0,n)=\overline{f}_{\text{int}}^{x,x}(\eta_0,R_0,n)=\overline{f}_{\text{int}}^{y,y}(\eta_0,R_0,n)=0. \quad (9.1)$$

We can then make the following rescaling $F_R=\tilde{F}_R/l_p$, $a=l_p\tilde{a}$, $R_0=l_p\tilde{R}_0$ and $L=l_p\tilde{L}$ to remove all $l_p$ dependence. Therefore, we may write Eqs. (7.26), (8.1), (8.2), (8.3) and (8.6), with Eq. (9.1), as

$$\frac{F_T}{k_BT}=\tilde{L}\left(\frac{\tilde{F}_R}{2\cos\left(\frac{\eta_0}{2}\right)}\right)^{1/2}+\frac{\tilde{\alpha}_\eta^{1/2}\tilde{L}}{2^{1/2}}+\frac{\tilde{L}}{2^{2/3}(\tilde{R}_0-2\tilde{a})^{2/3}}-\frac{\tilde{L}}{\tilde{R}_0^2}\frac{1}{2^{1/3}}\left(\tilde{R}_0-2\tilde{a}\right)^{2/3}f_1(\tilde{R}_0;\tilde{a})\sin^2\left(\frac{\eta_0}{2}\right)$$
$$+\frac{4\tilde{L}f_1(\tilde{R}_0;\tilde{a})}{\tilde{R}_0^2}\sin^4\left(\frac{\eta_0}{2}\right)-\tilde{F}_R\tilde{L}\cos\left(\frac{\eta_0}{2}\right)+\frac{2M_R\tilde{L}}{\tilde{R}_0}\sin\left(\frac{\eta_0}{2}\right)f_2(\tilde{R}_0;\tilde{a}) \quad (9.2)$$
$$-\frac{M_R\tilde{L}}{4}\frac{1}{2^{1/3}}\left(\tilde{R}_0-2\tilde{a}\right)^{2/3}\frac{f_2(\tilde{R}_0;\tilde{a})}{\tilde{R}_0}\sin\left(\frac{\eta_0}{2}\right)^{-1}-\frac{M_R^2}{16\left(2\tilde{F}_R\right)^{1/2}}\frac{\tilde{L}}{\cos\left(\frac{\eta_0}{2}\right)^{7/2}},$$

with

$$\tilde{\alpha}_\eta=\left[\frac{4f_1(\tilde{R}_0;\tilde{a})}{\tilde{R}_0^2}\left(3\cos^2\left(\frac{\eta_0}{2}\right)\sin^2\left(\frac{\eta_0}{2}\right)-\sin^4\left(\frac{\eta_0}{2}\right)\right)+\frac{\tilde{F}_R}{4}\cos\left(\frac{\eta_0}{2}\right)-\frac{M_Rf_2(\tilde{R}_0;\tilde{a})}{2\tilde{R}_0}\sin\left(\frac{\eta_0}{2}\right)\right], \quad (9.3)$$

and

$$0=\frac{1}{4}\left(\frac{\tilde{F}_R}{2\cos\left(\frac{\eta_0}{2}\right)}\right)^{1/2}\tan\left(\frac{\eta_0}{2}\right)+\frac{1}{2^{3/2}\tilde{\alpha}_\eta^{1/2}}\frac{\partial\tilde{\alpha}_\eta}{\partial\eta_0}-\frac{1}{\tilde{R}_0^2}\frac{1}{2^{1/3}}\left(\tilde{R}_0-2\tilde{a}\right)^{2/3}f_1(\tilde{R}_0;\tilde{a})\cos\left(\frac{\eta_0}{2}\right)\sin\left(\frac{\eta_0}{2}\right)$$
$$+\frac{8f_1(\tilde{R}_0;\tilde{a})}{\tilde{R}_0^2}\cos\left(\frac{\eta_0}{2}\right)\sin^3\left(\frac{\eta_0}{2}\right)+\frac{\tilde{F}_R}{2}\sin\left(\frac{\eta_0}{2}\right)+\frac{M_R}{\tilde{R}_0}\cos\left(\frac{\eta_0}{2}\right)f_2(\tilde{R}_0;\tilde{a})$$
$$+\frac{M_R}{8}\frac{1}{2^{1/3}}\left(\tilde{R}_0-2\tilde{a}\right)^{2/3}\frac{f_2(\tilde{R}_0;\tilde{a})}{\tilde{R}_0}\frac{\cos\left(\frac{\eta_0}{2}\right)}{\sin^2\left(\frac{\eta_0}{2}\right)}-\frac{7M_R^2}{64(2\tilde{F}_R)^{1/2}}\frac{\sin\left(\frac{\eta_0}{2}\right)}{\cos\left(\frac{\eta_0}{2}\right)^{9/2}}, \quad (9.4)$$

$$\frac{\partial\tilde{\alpha}_\eta}{\partial\eta_0}=\frac{4f_1(\tilde{R}_0;\tilde{a})}{\tilde{R}_0^2}\left(3\cos^3\left(\frac{\eta_0}{2}\right)\sin^2\left(\frac{\eta_0}{2}\right)-5\sin^3\left(\frac{\eta_0}{2}\right)\cos\left(\frac{\eta_0}{2}\right)\right)-\frac{M_R}{4}\frac{f_2(\tilde{R}_0;\tilde{a})}{\tilde{R}_0}\cos\left(\frac{\eta_0}{2}\right) \quad (9.5)$$
$$-\frac{\tilde{F}_R}{8}\sin\left(\frac{\eta_0}{2}\right),$$

$$0=\frac{1}{2^{3/2}\tilde{\alpha}_\eta^{1/2}}\frac{\partial\tilde{\alpha}_\eta}{\partial\tilde{R}_0}-\frac{2^{1/3}}{3}\frac{1}{(\tilde{R}_0-2\tilde{a})^{5/3}}+\frac{2^{2/3}}{\tilde{R}_0^3}\left(\tilde{R}_0-2\tilde{a}\right)^{2/3}h_1(\tilde{R}_0;\tilde{a})\sin^2\left(\frac{\eta_0}{2}\right)$$
$$-\frac{2^{2/3}}{3\tilde{R}_0^2}\left(\frac{1}{\tilde{R}_0-2\tilde{a}}\right)^{1/3}f_1(\tilde{R}_0;\tilde{a})\sin^2\left(\frac{\eta_0}{2}\right)-\frac{8h_1(\tilde{R}_0;\tilde{a})}{\tilde{R}_0^3}\sin^4\left(\frac{\eta_0}{2}\right) \quad (9.6)$$
$$-\frac{2M_R}{\tilde{R}_0^2}\sin\left(\frac{\eta_0}{2}\right)h_2(\tilde{R}_0;\tilde{a})+\frac{M_R}{4}\frac{1}{2^{1/3}}\left(\tilde{R}_0-2\tilde{a}\right)^{2/3}\frac{h_2(\tilde{R}_0;\tilde{a})}{\tilde{R}_0^2}\sin\left(\frac{\eta_0}{2}\right)^{-1}$$
$$-\frac{M_R}{6}\frac{1}{2^{1/3}}\left(\frac{1}{\tilde{R}_0-2\tilde{a}}\right)^{1/3}\frac{f_2(\tilde{R}_0;\tilde{a})}{\tilde{R}_0}\sin\left(\frac{\eta_0}{2}\right)^{-1},$$

$$\frac{\partial\tilde{\alpha}_\eta}{\partial\tilde{R}_0}=-\frac{8h_1(\tilde{R}_0;\tilde{a})}{\tilde{R}_0^3}\left(3\cos^2\left(\frac{\eta_0}{2}\right)\sin^2\left(\frac{\eta_0}{2}\right)-\sin^4\left(\frac{\eta_0}{2}\right)\right)+\frac{Mh_2(\tilde{R}_0;\tilde{a})}{2\tilde{R}_0^2}\sin\left(\frac{\eta_0}{2}\right). \tag{9.7}$$

On rescaling and application of Eq. (9.1), Eqs. (8.7) and (8.8) are modified to be

$$\begin{aligned}-\frac{\langle Lk_b\rangle}{\tilde{L}}=&\frac{1}{\pi\tilde{R}_0}f_2(\tilde{R}_0;\tilde{a})\left(1-\frac{1}{8\left(2\tilde{\alpha}_\eta\right)^{1/2}}\right)\sin\left(\frac{\eta_0}{2}\right)-\frac{1}{8\pi}\frac{1}{2^{1/3}}\left(\tilde{R}_0-2\tilde{a}\right)^{2/3}\frac{f_2(\tilde{R}_0;\tilde{a})}{\tilde{R}_0}\sin\left(\frac{\eta_0}{2}\right)^{-1}\\&-\frac{M_R}{16\pi(2\tilde{F}_R)^{1/2}}\frac{1}{\cos\left(\frac{\eta_0}{2}\right)^4},\end{aligned} \tag{9.8}$$

$$\frac{\langle z\rangle}{\tilde{L}}=\cos\left(\frac{\eta_0}{2}\right)\left(1-\frac{1}{2^{1/2}\tilde{\alpha}_\eta^{1/2}}\frac{1}{8}\right)-\frac{1}{2}\frac{1}{\left(2\tilde{F}_R\cos\left(\frac{\eta_0}{2}\right)\right)^{1/2}}-\frac{M_R^2}{16\left(2\tilde{F}_R\right)^{3/2}}\frac{1}{\cos\left(\frac{\eta_0}{2}\right)^{7/2}}. \tag{9.9}$$

Eqs. (9.4), (9.6) are numerically solved, and the braiding free energy, average braid linking number and end to end distance are evaluated using Eqs. (9.2), (9.8) and (9.9), respectively, These numerical results are used in a model to describe mechanical braiding of two molecules, with no interaction between them, attached to a bead and a substrate in [10], similar to that presented in [8].

## 10. Variational approximation for strongly interacting Braids

For strongly interacting Braids we change the form of the variational trial functional $E_{0,R}$. Here, we consider two possible choices for $d_{\max}$. For a loosely braided state the choice $d_{\max}\approx R_0-2a$ may be appropriate. However, as was argued previously in [3] and in Section 5, a better choice of $d_{\max}$ for a tightly braided state, where the $\delta R(s_0)$ fluctuations are primarily limited by the interactions, may be that given by Eq. (5.11).

Now we write

$$\begin{aligned}\frac{E_{0,R}\left[\delta\eta(s_0),\delta R(s_0),\delta\Phi(s_0)\right]}{k_BT}=&\int_{-L/2}^{L/2}ds_0\left(\frac{l_p}{4}\left(\frac{d^2\delta R(s_0)}{ds_0^2}\right)^2+\frac{\beta_R}{2}\left(\frac{d\delta R(s_0)}{ds_0}\right)^2+\frac{\alpha_R}{2}\delta R(s_0)^2\right)\\&+\int_{-L/2}^{L/2}ds_0\left(\frac{l_p}{4}\left(\frac{d\delta\eta(s_0)}{ds_0}\right)^2+\frac{\alpha_\eta}{2}\delta\eta(s_0)^2\right)+\int_{-L/2}^{L/2}ds_0\left(\frac{l_p}{4}\left(\frac{d\delta\Phi(s_0)}{ds_0}\right)^2+\frac{\alpha_\Phi}{2}\delta\Phi(s_0)^2\right),\end{aligned} \tag{10.1}$$

where now we have that $\Delta\Phi(s_0)=\Delta\Phi_0(s_0)+\delta\Phi(s_0)$ and the $\delta\Phi(s_0)$-fluctuations are now limited through the variational parameter $\alpha_\Phi$. Instead of $\alpha_H$ , we now have variational parameters $\beta_R$ and $\alpha_R$ that determine both $\left\langle\delta R(s)^2\right\rangle$ and $\left\langle\delta R'(s)^2\right\rangle$. These parameters take account that interactions now also contribute to $\left\langle\delta R(s)^2\right\rangle$ and $\left\langle\delta R'(s)^2\right\rangle$. The general form for the variational trail function, given by Eqs. (6.8) and (6.14), still applies. We also use similar analysis to that contained in Section 6; but there are some changes. Firstly, we replace Eq. (6.18) with

$$
\begin{aligned}
&\frac{\left\langle \tilde{E}_{R,B}[\eta_0+\delta\eta(s),R_0+\delta R(s)]+\tilde{E}_{St}[\delta R(s)]-E_{0,R}[\delta\eta(s),\delta R(s)]\right\rangle_0}{k_BT}= \\
&-\frac{L\theta_R^2 l_p}{R_0^2}\tilde{f}_1(R_0,d_R,d_{\max},d_{\min})\sin^2\left(\frac{\eta_0}{2}\right)+\frac{1}{4}\left(l_p-\tilde{l}_p\right)\left\langle\left(\frac{d\delta\eta(s_0)}{ds_0}\right)^2\right\rangle_{\delta\eta} \\
&+4l_p^b\frac{\tilde{f}_1(R_0,d_R,d_{\max},d_{\min})}{R_0^2}\left[\sin^4\left(\frac{\eta_0}{2}\right)+\frac{d_\eta^2}{2}\left(3\cos^2\left(\frac{\eta_0}{2}\right)\sin^2\left(\frac{\eta_0}{2}\right)-\sin^4\left(\frac{\eta_0}{2}\right)\right)\right] \\
&-\frac{\alpha_\eta d_\eta^2}{2}+\frac{(\alpha_H-\alpha_R)}{2}d_R^2-\frac{\beta_R}{2}\theta_R^2-\frac{\alpha_\Phi d_\Phi^2}{2}-\frac{l_c}{4}\left\langle\left(\frac{d\delta\Phi(s_0)}{ds_0}\right)^2\right\rangle_{\delta\Phi},
\end{aligned}
\tag{10.2}
$$

where now

$$
\theta_R^2=\frac{1}{2\pi}\int_{-\infty}^{\infty}\frac{k^2dk}{\frac{l_p}{2}k^4+\beta_R k^2+\alpha_R}=\frac{1}{2\pi\alpha_R}\left(\frac{2\alpha_R}{l_p}\right)^{3/4}\int_{-\infty}^{\infty}\frac{k^2dk}{k^4+\gamma k^2+1},\quad \text{where } \gamma=\beta_R\left(\frac{2}{\alpha_R l_p}\right)^{1/2},
\tag{10.3}
$$

$$
d_R^2=\frac{1}{2\pi}\int_{-\infty}^{\infty}\frac{dk}{\frac{l_p}{2}k^4+\beta_R k^2+\alpha_R}=\frac{1}{2\pi\alpha_R}\left(\frac{2\alpha_R}{l_p}\right)^{1/4}\int_{-\infty}^{\infty}\frac{dk}{k^4+\gamma k^2+1},
\tag{10.4}
$$

as well as

$$
d_\Phi=\frac{1}{2\pi}\int_{-\infty}^{\infty}dk\frac{1}{\frac{l_c}{2}k^2+\alpha_\Phi}=\frac{2}{l_c}\frac{1}{2\pi}\int_{-\infty}^{\infty}dk\frac{1}{k^2+\frac{1}{\lambda_h^2}}=\frac{\lambda_h}{l_c},\quad \text{where } \lambda_h=\left(\frac{l_c}{2\alpha_\Phi}\right)^{1/2}.
\tag{10.5}
$$

The relationship between $d_\eta$ and $\alpha_\eta$ remains unchanged from Eq. (6.21). In Appendix B we show how one can evaluate the integrals in Eqs. (10.3) and (10.4). This yields

$$
\theta_R^2=\frac{1}{2\alpha_R}\left(\frac{2\alpha_R}{l_p}\right)^{3/4}\frac{1}{\sqrt{\gamma+2}}\quad\text{and}\quad d_R^2=\frac{1}{2\alpha_R}\left(\frac{2\alpha_R}{l_p}\right)^{1/4}\frac{1}{\sqrt{\gamma+2}}.
\tag{10.6}
$$

In Eq. (10.2) the functions $\tilde{f}_1(R_0,d_R,d_{\max},d_{\min})$ are defined as

$$\tilde{f}_1(R_0,d_R,d_{\max},d_{\min}) = \frac{R_0^2}{d_R\sqrt{2\pi}}\int_{d_{\min}}^{d_{\max}} \frac{dx}{(R_0+x)^2}\exp\left(-\frac{x^2}{2d_R^2}\right)$$
$$+\frac{1}{2}\left(\frac{R_0^2}{(R_0+d_{\min})^2}\left(1-\text{erf}\left(-\frac{d_{\min}}{d_R\sqrt{2}}\right)\right)+\frac{R_0^2}{(R_0+d_{\max})}\left(1-\text{erf}\left(\frac{d_{\max}}{d_R\sqrt{2}}\right)\right)\right). \quad (10.7)$$

Secondly, the average of the interaction energy is now modified to be

$$\frac{\langle \tilde{E}_{\text{int}}\rangle_0}{k_BT} \approx \sum_{n=-\infty}^{\infty}\int_{-L/2}^{L/2} ds_0 \left\langle \exp(in\Delta\Phi(s_0))\right\rangle_{\delta\Phi}\Big(\left\langle \bar{\varepsilon}_{\text{int}}^0(\eta_0,R(s_0),n,R'(s_0),g_1^0(s_0),g_2^0(s_0))\right\rangle_{\delta R}$$
$$+\frac{1}{2^{3/2}\alpha_\eta^{1/2}\tilde{l}_p^{1/2}}\left\langle \bar{\varepsilon}_{\text{int}}^{\eta,\eta}(\eta_0,R(s_0),n,R'(s_0),g_1^0(s_0),g_2^0(s_0))\right\rangle_{\delta R}$$
$$+\frac{1}{2}\left\langle \bar{\varepsilon}_{\text{int}}^{\eta',\eta'}(\eta_0,R(s_0),n,R'(s_0),g_1^0(s_0),g_2^0(s_0))\right\rangle_{\delta R}\left\langle\left(\frac{d\delta\eta(s_0)}{ds_0}\right)^2\right\rangle_{\delta\eta} \quad (10.8)$$
$$+\frac{1}{8\alpha_x^{1/2}l_p^{1/2}}\left\langle \bar{\varepsilon}_{\text{int}}^{x,x}(\eta_0,R(s_0),n,R'(s_0),g_1^0(s_0),g_2^0(s_0))\right\rangle_{\delta R}$$
$$+\frac{1}{8\alpha_y^{1/2}l_p^{1/2}}\left\langle \bar{\varepsilon}_{\text{int}}^{y,y}(\eta_0,R(s_0),n,R'(s_0),g_1^0(s_0),g_2^0(s_0))\right\rangle_{\delta R}\Big).$$

This is due to the fact that the fluctuations in $\Delta\Phi(s_0)$ are now limited and forms for the $\delta R$ averages will be different. Using the result Eq. (A.7) of Appendix A we may write

$$\left\langle \exp(in\Delta\Phi(s_0))\right\rangle_{\delta\Phi} = \frac{1}{d_\Phi\sqrt{2\pi}}\int_{-\infty}^{\infty} dx\exp(in\Delta\Phi_0(s_0))\exp(inx)\exp\left(-\frac{x^2}{2d_\Phi}\right)$$
$$=\exp(in\Delta\Phi_0(s_0))\exp\left(-\frac{n^2}{2d_\Phi}\right). \quad (10.9)$$

Then, using Eqs. (10.5), (10.8) and (10.9) we may write

$$\frac{\langle \tilde{E}_{\text{int}}\rangle_0}{k_BT} \approx \sum_{n=-\infty}^{\infty} L\exp\left(-\frac{n^2\lambda_h}{2l_c}\right)\left[\tilde{f}_{\text{int}}^0(\eta_0,R_0,n,d_R,\theta_R)+\frac{1}{2^{3/2}\alpha_\eta^{1/2}\tilde{l}_p^{1/2}}\tilde{f}_{\text{int}}^{\eta,\eta}(\eta_0,R_0,n,d_R,\theta_R)\right.$$
$$+\frac{1}{2}\tilde{f}_{\text{int}}^{\eta',\eta'}(\eta_0,R_0,n,,d_R,\theta_R)\left\langle\left(\frac{d\delta\eta(s_0)}{ds_0}\right)^2\right\rangle_{\delta\eta}+\frac{1}{8\alpha_x^{1/2}l_p^{1/2}}\tilde{f}_{\text{int}}^{x,x}(\eta_0,R_0,n,d_R,\theta_R) \quad (10.10)$$
$$\left.+\frac{1}{8\alpha_y^{1/2}l_p^{1/2}}\tilde{f}_{\text{int}}^{y,y}(\eta_0,R_0,n,d_R,\theta_R)\right],$$

where

$$\tilde{f}^{0}_{\rm int}(\eta_0,R_0,n,,d_R,\theta_R)=\frac{1}{2\pi\theta_R d_R L}\int_{-L/2}^{L/2}ds_0\int_{-\infty}^{\infty}dx\int_{-\infty}^{\infty}dx'\exp(in\Delta\Phi_0(s_0))\tilde{\varepsilon}^{0}_{\rm int}(\eta_0,x,n,x',g^0_1(s_0),g^0_2(s_0))$$
$$\exp\left(-\frac{x^2}{2d_R^2}\right)\exp\left(-\frac{x'^2}{2\theta_R^2}\right),$$

(10.11)

$$\tilde{f}^{Y,Y}_{\rm int}(\eta_0,R_0,n,d_R,\theta_R)=\frac{1}{2\pi\theta_R d_R L}\int_{-L/2}^{L/2}ds_0\int_{-\infty}^{\infty}dx\int_{-\infty}^{\infty}dx'\exp(in\Delta\Phi_0(s_0))\tilde{\varepsilon}^{Y,Y}_{\rm int}(\eta_0,x,n,x',g^0_1(s_0),g^0_2(s_0))$$
$$\exp\left(-\frac{x^2}{2d_R^2}\right)\exp\left(-\frac{x'^2}{2\theta_R^2}\right),$$

(10.12)

where $Y=\eta,\eta',x,y$. Thirdly, the average of the work term now becomes

$$\left\langle\tilde{E}_W\right\rangle\approx -FL\left[\cos\left(\frac{\eta_0}{2}\right)-\frac{1}{8}\cos\left(\frac{\eta_0}{2}\right)\frac{1}{\left(2\alpha_\eta\tilde{l}_p\right)^{1/2}}-\frac{1}{\cos\left(\frac{\eta_0}{2}\right)}\left[\frac{1}{8\alpha_x^{1/2}l_p^{1/2}}+\frac{1}{8\alpha_y^{1/2}l_p^{1/2}}\right]\right]$$
$$+2ML\frac{\tilde{f}_2(R_0,d_R,d_{\max},d_{\min})}{R_0}\left[\sin\left(\frac{\eta_0}{2}\right)-\frac{1}{8}\sin\left(\frac{\eta_0}{2}\right)\frac{1}{\left(2\alpha_\eta l_p\right)^{1/2}}-\frac{1}{8}\theta_R^2\sin\left(\frac{\eta_0}{2}\right)^{-1}\right]-2\pi M\left\langle Wr_b\right\rangle_0,$$

(10.13)

where

$$\tilde{f}_2(R_0,d_R,d_{\max},d_{\min})=\frac{R_0^2}{d_R\sqrt{2\pi}}\int_{d_{\min}}^{d_{\max}}\frac{dx}{(R_0+x)}\exp\left(-\frac{x^2}{2d_R^2}\right)$$
$$+\frac{1}{2}\left(\frac{R_0}{\left(R_0+d_{\min}\right)}\left(1-\mathrm{erf}\left(-\frac{d_{\min}}{d_R\sqrt{2}}\right)\right)+\frac{R_0}{\left(R_0+d_{\max}\right)}\left(1-\mathrm{erf}\left(\frac{d_{\max}}{d_R\sqrt{2}}\right)\right)\right).\qquad(10.14)$$

The evaluation of $\ln Z_{0,A}$ remains the same as before with

$$-\ln Z_{0,A}\approx\frac{L\alpha_x^{1/2}}{2l_p^{1/2}}+\frac{L\alpha_y^{1/2}}{2l_p^{1/2}}-\frac{M_R^2}{8l_p^{3/2}}\frac{L}{\cos\left(\frac{\eta_0}{2}\right)^4}\frac{1}{\left(\alpha_y^{1/2}+\alpha_x^{1/2}\right)}.\qquad(10.15)$$

The evaluation of $\ln Z_{0,R}$ is now more complicated. From the form of Eq. (10.1) we see that we may write

$$-\frac{\partial\ln Z_{0,R}}{\partial\alpha_\eta}=\frac{L}{2}\left\langle\delta\eta(s_0)^2\right\rangle_{\delta\eta}=\frac{L}{2^{3/2}\alpha_\eta^{1/2}\tilde{l}_p^{1/2}},\qquad-\left(\frac{\partial\ln Z_{0,R}}{\partial\alpha_R}\right)_{\beta_R}=\frac{L}{2}\left\langle\delta R(s_0)^2\right\rangle_{\delta R}=\frac{d_R^2L}{2}$$

$$-\left(\frac{\partial \ln Z_{0,R}}{\partial \beta_R}\right)_{\alpha_R} = \frac{L}{2}\left\langle\left(\frac{d\delta R(s_0)}{ds_0}\right)^2\right\rangle = \frac{\theta_R^2 L}{2}, \quad -\frac{\partial \ln Z_{0,R}}{\partial \alpha_\Phi} = \frac{L}{2}\left\langle \delta\Phi(s_0)^2\right\rangle = \frac{L}{2^{3/2}\alpha_\Phi^{1/2} l_c^{1/2}}. \quad (10.16)$$

Though differentiation with respect to $\alpha_R$ and $\beta_R$ is useful in generating the initial expressions in Eq. (10.16), a better choice of variable to enable integrating up $\ln Z_{0,R}$ is $\gamma$ and $\alpha_R$. Utilizing the chain rule we can write

$$\left(\frac{\partial \ln Z_{0,R}}{\partial \alpha_R}\right)_{\gamma} = \left(\frac{\partial \ln Z_{0,R}}{\partial \alpha_R}\right)_{\beta_R} + \left(\frac{\partial \ln Z_{0,R}}{\partial \beta_R}\right)_{\alpha_R}\left(\frac{\partial \beta_R}{\partial \alpha_R}\right)_{\gamma} = -\frac{Ld_R}{2} - \frac{L\theta_R^2}{2}\frac{\gamma l_p^{1/2}}{2\sqrt{2}\alpha_R^{1/2}},$$

$$\left(\frac{\partial \ln Z_{0,R}}{\partial \gamma}\right)_{\alpha_R} = \left(\frac{\partial \ln Z_{0,R}}{\partial \beta_R}\right)_{\alpha_R}\left(\frac{\partial \beta_R}{\partial \gamma}\right)_{\beta_R} = -\frac{L\theta_R^2 l_p^{1/2}\alpha_R^{1/2}}{2\sqrt{2}}. \quad (10.17)$$

Then, through substitution of expressions in Eq. (10.16) into Eq. (10.17) we have that

$$-\left(\frac{\partial \ln Z_{0,R}}{\partial \alpha_R}\right)_{\gamma} = \frac{L}{42^{3/4}\alpha_R^{3/4} l_p^{1/4}}\sqrt{\gamma+2}, \qquad -\left(\frac{\partial \ln Z_{0,R}}{\partial \gamma}\right)_{\alpha_R} = \frac{1}{2^{7/4}}\frac{L\alpha_R^{1/4}}{l_p^{1/4}\sqrt{\gamma+2}},$$

$$-\frac{\partial \ln Z_{0,R}}{\partial \alpha_\eta} = \frac{L}{2^{3/2}\alpha_\eta^{1/2}\tilde{l}_p^{1/2}}, \qquad -\frac{\partial \ln Z_{0,R}}{\partial \alpha_\Phi} = \frac{L}{2^{3/2}\alpha_\Phi^{1/2} l_c^{1/2}}. \quad (10.18)$$

The partial differentials in Eq. (10.18) can now easily be integrated up to yield

$$-\ln Z_{0,R} = \frac{\alpha_\eta^{1/2}L}{2^{1/2}\tilde{l}_p^{1/2}} + \frac{L}{2^{3/4}}\frac{\alpha_R^{1/4}}{l_p^{1/4}}\sqrt{\gamma+2} + \frac{\alpha_\Phi^{1/2}L}{2^{1/2}l_c^{1/2}} = \frac{\alpha_\eta^{1/2}L}{2^{1/2}\tilde{l}_p^{1/2}} + \frac{L}{2l_p\theta_R^2} + \frac{L}{2\lambda_h}. \quad (10.19)$$

Lastly, the average twisting energy is now given by

$$\frac{\langle E_{Tw}\rangle}{k_BT} = \frac{l_c}{4}\int_{-L/2}^{L/2} ds_0\left[\left(\frac{d\Delta\Phi_0(s_0)}{ds_0} - \Delta g^0(s_0)\right)^2 + \left(\bar{g} - \bar{g}^0(s_0)\right)^2 + \left\langle\left(\frac{d\delta\Phi(s_0)}{ds_0}\right)^2\right\rangle_{\delta\Phi}\right]. \quad (10.20)$$

Using the expressions to eliminate $\alpha_R$ and $\beta_R$ in favour of $\theta_R$ and $d_R$, which arise from rearrangement of Eq. (10.6), which are

$$\alpha_R = \frac{l_p}{2}\left(\frac{\theta_R}{d_R}\right)^4 \quad \text{and} \quad \beta_R = \frac{l_p}{2}\left(\frac{1}{l_p^2\theta_R^4} - 2\left(\frac{\theta_R}{d_R}\right)^2\right), \quad (10.21)$$

and adding up all the contributions (Eqs.(6.28), (10.2), (10.10), (10.13), (10.15), (10.19) and (10.20)) , yields for the free energy

$$
\begin{aligned}
\frac{F_T}{k_BT} &= \frac{L}{4}\left[\left(\frac{\alpha_x}{l_p}\right)^{1/2}+\left(\frac{\alpha_y}{l_p}\right)^{1/2}\right]+\frac{\alpha_\eta^{1/2}L}{2^{3/2}\tilde{l}_p^{1/2}}+\frac{L}{4l_p\theta_R^2}+\frac{L}{4\lambda_h}+\frac{l_p\theta_R^4 L}{4d_R^2}+\frac{d_R^2\alpha_H L}{2} \\
&-\frac{L\theta_R^2 l_p}{R_0^2}\tilde{f}_1(R_0,d_R,d_{\max},d_{\min})\sin^2\left(\frac{\eta_0}{2}\right)+\frac{l_c}{4}\int_{-L/2}^{L/2}ds_0\left(\frac{d\Delta\Phi_0(s_0)}{ds_0}-\Delta g^0(s_0)\right)^2 \\
&+\frac{4l_pL\tilde{f}_1(R_0,d_R,d_{\max},d_{\min})}{R_0^2}\sin^4\left(\frac{\eta_0}{2}\right)-F_RL\cos\left(\frac{\eta_0}{2}\right) \\
&+\frac{L}{2^{3/2}\alpha_\eta^{1/2}\tilde{l}_p^{1/2}}\left[\frac{4l_p\tilde{f}_1(R_0,d_R,d_{\max},d_{\min})}{R_0^2}\left(3\cos^2\left(\frac{\eta_0}{2}\right)\sin^2\left(\frac{\eta_0}{2}\right)-\sin^4\left(\frac{\eta_0}{2}\right)\right)\right. \\
&\left.+\sum_{n=-\infty}^{\infty}\tilde{f}_{\rm int}^{\eta,\eta}(\eta_0,R_0,n,d_R,\theta_R)\exp\left(-\frac{n^2\lambda_h}{2l_c}\right)+\frac{F_R}{4}\cos\left(\frac{\eta_0}{2}\right)-\frac{M_R}{2}\sin\left(\frac{\eta_0}{2}\right)\right] \\
&+L\sum_{n=-\infty}^{\infty}\tilde{f}_{\rm int}^{0}(\eta_0,R_0,n,d_R,\theta_R)\exp\left(-\frac{n^2\lambda_h}{2l_c}\right)-\frac{M_R^2}{16l_p^{3/2}}\frac{L}{\cos\left(\frac{\eta_0}{2}\right)^4}\frac{1}{\left(\alpha_y^{1/2}+\alpha_x^{1/2}\right)} \\
&+\frac{L}{4}\left(l_p+2\sum_{n=-\infty}^{\infty}\tilde{f}_{\rm int}^{\eta',\eta'}(\eta_0,R_0,n,d_R,\theta_R)\exp\left(-\frac{n^2\lambda_h}{2l_c}\right)-\tilde{l}_p\right)\left\langle\left(\frac{d\delta\eta(s_0)}{ds_0}\right)^2\right\rangle_{\delta\eta} \\
&+\frac{L}{8\alpha_x^{1/2}l_p^{1/2}}\left(\sum_{n=-\infty}^{\infty}\tilde{f}_{\rm int}^{x,x}(\eta_0,R_0,n,d_R,\theta_R)\exp\left(-\frac{n^2\lambda_h}{2l_c}\right)+\frac{F_R}{\cos\left(\frac{\eta_0}{2}\right)}\right) \\
&+\frac{L}{8\alpha_y^{1/2}l_p^{1/2}}\left(\sum_{n=-\infty}^{\infty}\tilde{f}_{\rm int}^{y,y}(\eta_0,R_0,n,d_R,\theta_R)\exp\left(-\frac{n^2\lambda_h}{2l_c}\right)+\frac{F_R}{\cos\left(\frac{\eta_0}{2}\right)}\right) \\
&+\frac{2M_RL}{R_0}\sin\left(\frac{\eta_0}{2}\right)\tilde{f}_2(R_0,d_R,d_{\max},d_{\min})-\frac{M_RL}{4}\theta_R^2\frac{\tilde{f}_2(R_0,d_R,d_{\max},d_{\min})}{R_0}\sin\left(\frac{\eta_0}{2}\right)^{-1}.
\end{aligned}
\tag{10.22}
$$

Recollect that $M_R = M/k_BT$ and $F_R = F/k_BT$. Here, we may choose

$$
\tilde{l}_p(R_0,\eta_0,d_R,\theta_R)=l_p+2\sum_{n=-\infty}^{\infty}\tilde{f}_{\rm int}^{\eta',\eta'}(\eta_0,R_0,n,d_R,\theta_R)\exp\left(-\frac{n^2\lambda_h}{2l_c}\right) \tag{10.23}
$$

so that singular $\left\langle\left(\frac{d\delta\eta(s_0)}{ds_0}\right)^2\right\rangle_{\delta R}$ contribution vanishes and also we minimize with respect to $\alpha_\eta$. The latter operation yields

$$
\begin{aligned}
\alpha_\eta &= \left[\frac{4l_p\tilde{f}_1(R_0,d_R,d_{\max},d_{\min})}{R_0^2}\left(3\cos^2\left(\frac{\eta_0}{2}\right)\sin^2\left(\frac{\eta_0}{2}\right)-\sin^4\left(\frac{\eta_0}{2}\right)\right)\right. \\
&\left.+\sum_{n=-\infty}^{\infty}\tilde{f}_{\rm int}^{\eta,\eta}(\eta_0,R_0,n,d_R,\theta_R)\exp\left(-\frac{n^2\lambda_h}{2l_c}\right)+\frac{F_R}{4}\cos\left(\frac{\eta_0}{2}\right)-\frac{M_R\tilde{f}_2(R_0,d_R,d_{\max},d_{\min})}{2R_0}\sin\left(\frac{\eta_0}{2}\right)\right].
\end{aligned}
$$

(10.24)

Also, if we neglect the writhe correction in the minimization of Eq. (10.22) with respect to $\alpha_x$ and $\alpha_y$, we obtain

$$\alpha_x \approx \frac{1}{2}\left( \sum_{n=-\infty}^{\infty} \tilde{f}_{\rm int}^{x,x}(\eta_0, R_0, n, d_R, \theta_R) \exp\left( -\frac{n^2 \lambda_h}{2 l_c} \right) + \frac{F_R}{\cos\left(\frac{\eta_0}{2}\right)} \right) \quad \text{and}$$

$$\alpha_y \approx \frac{1}{2}\left( \sum_{n=-\infty}^{\infty} \tilde{f}_{\rm int}^{y,y}(\eta_0, R_0, n, d_R, \theta_R) \exp\left( -\frac{n^2 \lambda_h}{2 l_c} \right) + \frac{F_R}{\cos\left(\frac{\eta_0}{2}\right)} \right). \tag{10.25}$$

Eq. (10.22) for the free energy is then simplified to

$$\frac{F_T}{k_B T} = \frac{L}{2}\left[ \left(\frac{\alpha_x}{l_p}\right)^{1/2} + \left(\frac{\alpha_y}{l_p}\right)^{1/2} \right] + \frac{\alpha_\eta^{1/2} L}{2^{1/2} \tilde{l}_p^{1/2}} + \frac{d_R^2 \alpha_H L}{2} + \frac{L}{4 l_p \theta_R^2} + \frac{L}{4\lambda_h} + \frac{l_p \theta_R^4 L}{4 d_R^2} - \frac{M_R^2}{16 l_p^{3/2}} \frac{L}{\cos\left(\frac{\eta_0}{2}\right)^4} \frac{1}{\left(\alpha_y^{1/2} + \alpha_x^{1/2}\right)}$$
$$-\frac{L\theta_R^2}{R_0^2} \tilde{f}_1(R_0, d_R, d_{\max}, d_{\min}) l_p \sin^2\left(\frac{\eta_0}{2}\right) + \frac{l_c}{4} \int_{-L/2}^{L/2} ds_0 \left( \frac{d\Delta\Phi_0(s_0)}{ds_0} - \Delta g^0(s_0) \right)^2$$
$$+\frac{4 l_p L \tilde{f}_1(R_0, d_R, d_{\max}, d_{\min})}{R_0^2} \sin^4\left(\frac{\eta_0}{2}\right) + L \sum_{n=-\infty}^{\infty} \tilde{f}_{\rm int}^{0}(\eta_0, R_0, n, d_R, \theta_R) \exp\left( -\frac{n^2 \lambda_h}{2 l_c} \right)$$
$$-F_R L \cos\left(\frac{\eta_0}{2}\right) + \frac{2 M_R L}{R_0} \sin\left(\frac{\eta_0}{2}\right) \tilde{f}_2(R_0, d_R, d_{\max}, d_{\min}) - \frac{M_R L}{4} \theta_R^2 \frac{\tilde{f}_2(R_0, d_R, d_{\max}, d_{\min})}{R_0} \sin\left(\frac{\eta_0}{2}\right)^{-1}.$$

(10.26)

By differentiating the Free energy given by Eq. (10.26) with respect to $M_R$ and $F_R$ we obtain expressions for $\langle Lk_b \rangle$ and $\langle z \rangle$, respectively, which read as

$$-\frac{\langle Lk_b \rangle}{L} = \frac{1}{\pi R_0}\left( 1 - \frac{1}{8 \tilde{l}_p^{1/2} (2\alpha_\eta)^{1/2}} \right) \sin\left(\frac{\eta_0}{2}\right) \tilde{f}_2(R_0, d_R, d_{\max}, d_{\min}) - \frac{M_R}{16\pi l_p^{3/2}} \frac{1}{\cos\left(\frac{\eta_0}{2}\right)^4} \frac{1}{\left(\alpha_y^{1/2} + \alpha_x^{1/2}\right)}$$
$$-\frac{1}{8\pi} \theta_R^2 \frac{\tilde{f}_2(R_0, d_R, d_{\max}, d_{\min})}{R_0} \sin\left(\frac{\eta_0}{2}\right)^{-1},$$

(10.27)

$$\frac{\langle z\rangle}{L}=\left(1-\frac{1}{8\left(2\alpha_\eta\right)^{1/2}\tilde{l}_p^{1/2}}\right)\cos\left(\frac{\eta_0}{2}\right)-\frac{1}{8}\left[\left(\frac{1}{\alpha_x l_p}\right)^{1/2}+\left(\frac{1}{\alpha_x l_p}\right)^{1/2}\right]\frac{1}{\cos\left(\frac{\eta_0}{2}\right)}-\frac{M_R^2}{64 l_p^{3/2}}\frac{1}{\cos\left(\frac{\eta_0}{2}\right)^5}\frac{1}{\left(\alpha_y^{1/2}+\alpha_x^{1/2}\right)\alpha_y^{1/2}\alpha_x^{1/2}}.$$

(10.28)

## 11. Further minimization of the Free Energy for Strong Interactions

We now have the following additional requirements to minimize the Free energy

$$\frac{\delta F_T}{\delta\Delta\Phi_0(s)}=0,\quad \frac{\partial F_T}{\partial R_0}=0,\quad \frac{\partial F_T}{\partial\eta_0}=0,\quad \frac{\partial F_T}{\partial\theta_R}=0,\quad \frac{\partial F_T}{\partial d_R}=0\ \text{ and }\ \frac{\partial F_T}{\partial\lambda_h}=0, \tag{11.1}$$

where $\frac{\delta}{\delta\Delta\Phi_0(s)}$ is the functional differentiation with respect to $\Delta\Phi_0(s)$. Let us consider each in turn. First, we may write

$$\begin{aligned}
&\frac{\delta F_T}{\delta\Delta\Phi_0(s)}=-\frac{l_c}{2}\left(\frac{d\Delta\Phi_0(s)}{ds^2}-\frac{d\Delta g^0(s)}{ds}\right)\\
&+L\sum_{n=-\infty}^{\infty}\exp\left(-\frac{n^2\lambda_h}{2l_c}\right)\left[\frac{1}{8}\left(\frac{1}{l_p\alpha_x}\right)^{1/2}\frac{\delta\tilde{f}_{\text{int}}^{x,x}(\eta_0,R_0,n,d_R,\theta_R)}{\delta\Delta\Phi_0(s)}+\frac{1}{8}\left(\frac{1}{l_p\alpha_y}\right)^{1/2}\frac{\delta\tilde{f}_{\text{int}}^{y,y}(\eta_0,R_0,n,d_R,\theta_R)}{\delta\Delta\Phi_0(s)}\right.\\
&\left.\frac{1}{2^{3/2}\tilde{l}_p^{1/2}\alpha_\eta^{1/2}}\frac{\delta\tilde{f}_{\text{int}}^{\eta,\eta}(\eta_0,R_0,n,d_R,\theta_R)}{\delta\Delta\Phi_0(s)}-\frac{\alpha_\eta^{1/2}}{2^{1/2}\tilde{l}_p^{3/2}}\frac{\delta\tilde{f}_{\text{int}}^{\eta',\eta'}(\eta_0,R_0,n,d_R,\theta_R)}{\delta\Delta\Phi_0(s)}+\frac{\delta\tilde{f}_{\text{int}}^{0}(\eta_0,R_0,n,d_R,\theta_R)}{\delta\Delta\Phi_0(s)}\right]\\
&+\frac{M_R^2}{64l_p^{3/2}}\frac{L}{\cos\left(\frac{\eta_0}{2}\right)^4}\sum_{n=-\infty}^{\infty}\frac{\exp\left(-\frac{n^2\lambda_h}{2l_c}\right)}{\left(\alpha_y^{1/2}+\alpha_x^{1/2}\right)^2}\left[\frac{1}{\alpha_x^{1/2}}\frac{\delta\tilde{f}_{\text{int}}^{x,x}(\eta_0,R_0,n,d_R,\theta_R)}{\delta\Delta\Phi_0(s)}+\frac{1}{\alpha_y^{1/2}}\frac{\delta\tilde{f}_{\text{int}}^{y,y}(\eta_0,R_0,n,d_R,\theta_R)}{\delta\Delta\Phi_0(s)}\right]=0.
\end{aligned}$$

(11.2)

This equation can be recast as

$$\begin{aligned}
&\frac{l_c}{2}\left(\frac{d^2\Delta\Phi_0(s)}{ds^2}-\frac{d\Delta g^0(s)}{ds}\right)=\sum_{n=-\infty}^{\infty}in\exp\left(in\Delta\Phi_0(s)\right)\exp\left(-\frac{n^2\lambda_h}{2l_c}\right)\\
&\left[\frac{1}{8}\left(\frac{1}{l_p\alpha_x}\right)^{1/2}\tilde{p}_{\text{int}}^{x,x}(\eta_0,R_0,n,d_R,\theta_R)+\frac{1}{8}\left(\frac{1}{l_p\alpha_y}\right)^{1/2}\tilde{p}_{\text{int}}^{y,y}(\eta_0,R_0,n,d_R,\theta_R)\right.\\
&+\frac{1}{2^{3/2}\tilde{l}_p^{1/2}\alpha_\eta^{1/2}}\tilde{p}_{\text{int}}^{\eta,\eta}(\eta_0,R_0,n,d_R,\theta_R)-\frac{\alpha_\eta^{1/2}}{2^{1/2}\tilde{l}_p^{3/2}}\tilde{p}_{\text{int}}^{\eta',\eta'}(\eta_0,R_0,n,d_R,\theta_R)+\tilde{p}_{\text{int}}^{0}(\eta_0,R_0,n,d_R,\theta_R)
\end{aligned}$$

$$+\frac{M_R^2}{64 l_p^{3/2}} \frac{1}{\cos\left(\frac{\eta_0}{2}\right)^4 \left(\alpha_y^{1/2}+\alpha_x^{1/2}\right)^2} \left[ \frac{1}{\alpha_x^{1/2}} \tilde{p}_{\text{int}}^{x,x}(\eta_0, R_0, n, d_R, \theta_R) + \frac{1}{\alpha_y^{1/2}} \tilde{p}_{\text{int}}^{y,y}(\eta_0, R_0, n, d_R, \theta_R) \right] \Bigg],$$

(11.3)

where

$$\tilde{p}_{\text{int}}^{0}(\eta_0, R_0, n,, d_R, \theta_R) = \frac{1}{2\pi} \frac{1}{\theta_R d_R} \int_{-\infty}^{\infty} dx \int_{-\infty}^{\infty} dx' \overline{\varepsilon}_{\text{int}}^{0}(\eta_0, x, n, x', g_1^0(s_0), g_2^0(s_0)) \exp\left(-\frac{x^2}{2d_R^2}\right) \exp\left(-\frac{x'^2}{2\theta_R^2}\right),$$

(11.4)

$$\tilde{p}_{\text{int}}^{Y,Y}(\eta_0, R_0, n,, d_R, \theta_R) = \frac{1}{2\pi} \frac{1}{\theta_R d_R} \int_{-\infty}^{\infty} dx \int_{-\infty}^{\infty} dx' \overline{\varepsilon}_{\text{int}}^{Y,Y}(\eta_0, x, n, x', g_1^0(s_0), g_2^0(s_0)) \exp\left(-\frac{x^2}{2d_R^2}\right) \exp\left(-\frac{x'^2}{2\theta_R^2}\right),$$

(11.5)

and $Y = \eta, \eta', x, y$. If the corrections of $\tilde{f}_{\text{int}}^{\eta,\eta}(\eta_0, R_0, n,, d_R, \theta_R)$, $\tilde{f}_{\text{int}}^{\eta',\eta'}(\eta_0, R_0, n,, d_R, \theta_R)$ $\tilde{f}_{\text{int}}^{x,x}(\eta_0, R_0, n,, d_R, \theta_R)$ and $\tilde{f}_{\text{int}}^{y,y}(\eta_0, R_0, n,, d_R, \theta_R)$ to $\alpha_\eta$, $\tilde{l}_p$, $\alpha_x$ and $\alpha_y$ are small, we can ignore the latter's functional dependence.

Let us examine Eq. (11.3) in two situations, one where $\Delta g^0(s) = 0$ and the other where $\Delta g^0(s) = c$, where $c$ is a constant. In the first case, the lowest energy solution to Eq.(11.3) is $\Delta\Phi_0(s) = \Delta\overline{\Phi}$, a constant solution that satisfies

$$0 = \sum_{n=-\infty}^{\infty} in \exp\left(in\Delta\overline{\Phi}\right) \exp\left(-\frac{n^2 \lambda_h}{2l_c}\right) \left[ \frac{1}{8}\left(\frac{1}{l_p \alpha_x}\right)^{1/2} \tilde{p}_{\text{int}}^{x,x}(\eta_0, R_0, n, d_R, \theta_R) \right.$$

$$+\frac{1}{8}\left(\frac{1}{l_p \alpha_y}\right)^{1/2} \tilde{p}_{\text{int}}^{y,y}(\eta_0, R_0, n, d_R, \theta_R) + \frac{L}{2^{3/2} \tilde{l}_p^{1/2} \alpha_\eta^{1/2}} \tilde{p}_{\text{int}}^{\eta,\eta}(\eta_0, R_0, n, d_R, \theta_R)$$

$$-\frac{\alpha_\eta^{1/2} L}{2^{1/2} \tilde{l}_p^{3/2}} \tilde{p}_{\text{int}}^{\eta',\eta'}(\eta_0, R_0, n, d_R, \theta_R) + \tilde{p}_{\text{int}}^{0}(\eta_0, R_0, n, d_R, \theta_R)$$

$$\left. +\frac{M_R^2}{64 l_p^{3/2}} \frac{1}{\cos\left(\frac{\eta_0}{2}\right)^4 \left(\alpha_y^{1/2}+\alpha_x^{1/2}\right)^2} \left[ \frac{1}{\alpha_x^{1/2}} \tilde{p}_{\text{int}}^{x,x}(\eta_0, R_0, n, d_R, \theta_R) + \frac{1}{\alpha_y^{1/2}} \tilde{p}_{\text{int}}^{y,y}(\eta_0, R_0, n, d_R, \theta_R) \right] \right].$$

(11.6)

In the second case, Eq. (11.6) no longer yields the solution that minimizes the free energy. Instead, we may multiply Eq.(11.3) by $\left(\frac{d\Delta\Phi_0}{ds}\right)$ and integrate with respect to $s$ yielding

$$\left(\frac{d\Delta\Phi_0(s)}{ds}\right)^2 = \sum_{n=-\infty}^{\infty} \exp\left(in\Delta\Phi_0(s)\right)\chi_n(\eta_0, R_0, d_R, \theta_R) + k_1, \tag{11.7}$$

or

$$\int \frac{d\Delta\Phi_0}{k_1 + \sum_{n=-\infty}^{\infty} \exp\left(in\Delta\Phi_0\right)\chi_n(\eta_0, R_0, d_R, \theta_R)} = s + k_2, \tag{11.8}$$

where

$$\chi(\eta_0, R_0, n, d_R, \theta_R) = \frac{4}{l_c}\exp\left(-\frac{n^2\lambda_h}{2l_c}\right)\left[\frac{1}{8}\left(\frac{1}{l_p\alpha_x}\right)^{1/2} \tilde{p}_{\text{int}}^{x,x}(\eta_0, R_0, n, d_R, \theta_R)\right.$$
$$+\frac{1}{8}\left(\frac{1}{l_p\alpha_y}\right)^{1/2} \tilde{p}_{\text{int}}^{y,y}(\eta_0, R_0, n, d_R, \theta_R) + \frac{1}{2^{3/2}\tilde{l}_p^{1/2}\alpha_\eta^{1/2}} \tilde{p}_{\text{int}}^{\eta,\eta}(\eta_0, R_0, n, d_R, \theta_R)$$
$$-\frac{\alpha_\eta^{1/2}}{2^{1/2}\tilde{l}_p^{3/2}} \tilde{p}_{\text{int}}^{\eta',\eta'}(\eta_0, R_0, n, d_R, \theta_R) + \tilde{p}_{\text{int}}^{0}(\eta_0, R_0, n, d_R, \theta_R)$$
$$\left.+\frac{M_R^2}{64 l_p^{3/2}} \frac{1}{\cos\left(\frac{\eta_0}{2}\right)^4 \left(\alpha_y^{1/2}+\alpha_x^{1/2}\right)^2}\left[\frac{1}{\alpha_x^{1/2}} \tilde{p}_{\text{int}}^{x,x}(\eta_0, R_0, n, d_R, \theta_R) + \frac{1}{\alpha_y^{1/2}} \tilde{p}_{\text{int}}^{y,y}(\eta_0, R_0, n, d_R, \theta_R)\right]\right],$$

(11.9)

and both $k_1$ and $k_2$ are constants of integration, which the free energy needs to be minimized over if the boundary conditions on $\Delta\Phi_0(s)$ are free. Eq. (11.8) may correspond to the situation in which a braid is formed of molecules that form regular helices with different pitches, but roughly the same elasticity. If we suppose that that only $n = 0, -1, 1$ modes survive and $\tilde{p}_{\text{int}}^{x,x}(\eta_0, R_0, 1, d_R, \theta_R) = \tilde{p}_{\text{int}}^{x,x}(\eta_0, R_0, -1, d_R, \theta_R)$, Eq. (11.8) then becomes the general solution to a Sine-Gordon equation, and one can perform the integration on the LHS of Eq. (11.8) analytically in terms of Elliptic integrals.

Now partial differentiation of Eq. (10.26) w.r.t $R_0$ yields

$$0=\frac{1}{8}\sum_{n=-\infty}^{\infty}\exp\left(-\frac{n^2\lambda_h}{2l_c}\right)\left[\left(\frac{1}{\alpha_x l_p}\right)^{1/2}\frac{\partial\tilde{f}_{\text{int}}^{x,x}(\eta_0,R_0,n,d_R,\theta_R)}{\partial R_0}+\left(\frac{1}{\alpha_y l_p}\right)^{1/2}\frac{\partial\tilde{f}_{\text{int}}^{y,y}(\eta_0,R_0,n,d_R,\theta_R)}{\partial R_0}\right]$$

$$+\frac{1}{\alpha_\eta^{1/2}2^{3/2}\tilde{l}_p^{1/2}}\frac{\partial\alpha_\eta}{\partial R_0}+\frac{d_R^2}{2}\frac{\partial\alpha_H}{\partial R_0}-\frac{\alpha_\eta^{1/2}}{2^{1/2}\tilde{l}_p^{3/2}}\sum_{n=-\infty}^{\infty}\frac{\partial\tilde{f}_{\text{int}}^{\eta',\eta'}(\eta_0,R_0,n,d_R,\theta_R)}{\partial R_0}\exp\left(-\frac{n^2\lambda_h}{2l_c}\right)$$

$$+\frac{\theta_R^2 l_p}{R_0^3}\tilde{h}_1(R_0,d_R,d_{\max},d_{\min})\sin^2\left(\frac{\eta_0}{2}\right)-\frac{4l_p\tilde{h}_1(R_0,d_R,d_{\max},d_{\min})}{R_0^3}\sin^4\left(\frac{\eta_0}{2}\right)+\sum_{n=-\infty}^{\infty}\frac{\partial\tilde{f}_{\text{int}}^{0}(\eta_0,R_0,n,d_R,\theta_R)}{\partial R_0}\exp\left(-\frac{n^2\lambda_h}{2l_c}\right)$$

$$-\frac{2M_R}{R_0^2}\sin\left(\frac{\eta_0}{2}\right)\tilde{h}_2(R_0,d_R,d_{\max},d_{\min})+\frac{M_R}{4}\theta_R^2\frac{\tilde{h}_2(R_0,d_R,d_{\max},d_{\min})}{R_0^2}\sin\left(\frac{\eta_0}{2}\right)^{-1}$$

$$+\frac{M_R^2}{64l_p^{3/2}}\frac{1}{\cos\left(\frac{\eta_0}{2}\right)^4}\frac{1}{\left(\alpha_y^{1/2}+\alpha_x^{1/2}\right)^2}\sum_{n=-\infty}^{\infty}\exp\left(-\frac{n^2\lambda_h}{2l_c}\right)\left(\frac{1}{\alpha_y^{1/2}}\frac{\partial\tilde{f}_{\text{int}}^{y,y}(\eta_0,R_0,n,d_R,\theta_R)}{\partial R_0}+\frac{1}{\alpha_x^{1/2}}\sum_{n=-\infty}^{\infty}\frac{\partial\tilde{f}_{\text{int}}^{x,x}(\eta_0,R_0,n,d_R,\theta_R)}{\partial R_0}\right),$$

(11.10)

whilst minimization with respect to $\eta_0$ yields

$$0=\frac{1}{16}\left[\left(\frac{1}{\alpha_x l_p}\right)^{1/2}+\left(\frac{1}{\alpha_y l}\right)^{1/2}\right]\frac{F_R\sin\left(\frac{\eta_0}{2}\right)}{\cos\left(\frac{\eta_0}{2}\right)^2}-\frac{M_R^2}{8l_p^{3/2}}\frac{\sin\left(\frac{\eta_0}{2}\right)}{\cos\left(\frac{\eta_0}{2}\right)^5}\frac{1}{\left(\alpha_y^{1/2}+\alpha_x^{1/2}\right)}$$

$$+\frac{1}{8}\sum_{n=-\infty}^{\infty}\exp\left(-\frac{n^2\lambda_h}{2l_c}\right)\left[\left(\frac{1}{\alpha_x l_p}\right)^{1/2}\frac{\partial\tilde{f}_{\text{int}}^{x,x}(\eta_0,R_0,n,d_R,\theta_R)}{\partial\eta_0}+\left(\frac{1}{\alpha_y l_p}\right)^{1/2}\frac{\partial\tilde{f}_{\text{int}}^{y,y}(\eta_0,R_0,n,d_R,\theta_R)}{\partial\eta_0}\right]$$

$$+\frac{d_R^2}{2}\frac{\partial\alpha_H}{\partial\eta_0}+\frac{1}{\alpha_\eta^{1/2}2^{3/2}\tilde{l}_p^{1/2}}\frac{\partial\alpha_\eta}{\partial\eta_0}-\frac{\alpha_\eta^{1/2}}{2^{1/2}\tilde{l}_p^{3/2}}\sum_{n=-\infty}^{\infty}\frac{\partial\tilde{f}_{\text{int}}^{\eta',\eta'}(\eta_0,R_0,n,d_R,\theta_R)}{\partial\eta_0}\exp\left(-\frac{n^2\lambda_h}{2l_c}\right)$$

$$-\frac{\theta_R^2 l_p}{R_0^2}\tilde{f}_1(R_0,d_R,d_{\max},d_{\min})\sin\left(\frac{\eta_0}{2}\right)\cos\left(\frac{\eta_0}{2}\right)+\frac{8l_pL\tilde{f}_1(R_0,d_R,d_{\max},d_{\min})}{R_0^2}\cos\left(\frac{\eta_0}{2}\right)\sin^3\left(\frac{\eta_0}{2}\right)$$

$$+\sum_{n=-\infty}^{\infty}\frac{\partial\tilde{f}_{\text{int}}^{0}(\eta_0,R_0,n,d_R,\theta_R)}{\partial\eta_0}\exp\left(-\frac{n^2\lambda_h}{2l_c}\right)+\frac{M_R}{R_0}\cos\left(\frac{\eta_0}{2}\right)\tilde{f}_2(R_0,d_R,d_{\max},d_{\min})+\frac{F_R}{2}\sin\left(\frac{\eta_0}{2}\right)$$

$$+\frac{M_R}{8}\theta_R^2\frac{\tilde{f}_2(R_0,d_R,d_{\max},d_{\min})}{R_0}\cos\left(\frac{\eta_0}{2}\right)\sin\left(\frac{\eta_0}{2}\right)^{-2}+\frac{M_R^2}{128l_p^{3/2}}\frac{1}{\cos\left(\frac{\eta_0}{2}\right)^4}\frac{1}{\left(\alpha_y^{1/2}+\alpha_x^{1/2}\right)\alpha_y^{1/2}\alpha_x^{1/2}}\frac{F_R\sin\left(\frac{\eta_0}{2}\right)}{\cos^2\left(\frac{\eta_0}{2}\right)}$$

$$+\frac{M_R^2}{64l_p^{3/2}}\frac{1}{\cos\left(\frac{\eta_0}{2}\right)^4}\frac{1}{\left(\alpha_y^{1/2}+\alpha_x^{1/2}\right)^2}\sum_{n=-\infty}^{\infty}\exp\left(-\frac{n^2\lambda_h}{2l_c}\right)\left(\frac{1}{\alpha_y^{1/2}}\frac{\partial\tilde{f}_{\text{int}}^{y,y}(\eta_0,R_0,n,d_R,\theta_R)}{\partial\eta_0}+\frac{1}{\alpha_x^{1/2}}\frac{\partial\tilde{f}_{\text{int}}^{x,x}(\eta_0,R_0,n,d_R,\theta_R)}{\partial\eta_0}\right),$$

(11.11)

where

$$\frac{\partial \alpha_\eta}{\partial R_0} = \left[ -\frac{4 l_p \tilde{h}_1(R_0, d_R, d_{\max}, d_{\min})}{R_0^3} \left( 3\cos^2\left(\frac{\eta_0}{2}\right)\sin^2\left(\frac{\eta_0}{2}\right) - \sin^4\left(\frac{\eta_0}{2}\right) \right) \right.$$

$$\left. + \sum_{n=-\infty}^{\infty} \frac{\partial \tilde{f}_{\text{int}}^{\eta,\eta}(\eta_0, R_0, n, d_R, \theta_R)}{\partial R_0} \exp\left( -\frac{n^2 \lambda_h}{2 l_c} \right) + \frac{M_R \tilde{h}_2(R_0, d_R, d_{\max}, d_{\min})}{2 R_0^2} \sin\left(\frac{\eta_0}{2}\right) \right], \quad (11.12)$$

and

$$\frac{\partial \alpha_\eta}{\partial \eta_0} = \left[ \frac{4 l_p \tilde{f}_1(R_0, d_R, d_{\max}, d_{\min})}{R_0^2} \left( 3\cos^3\left(\frac{\eta_0}{2}\right)\sin^2\left(\frac{\eta_0}{2}\right) - 5\sin^3\left(\frac{\eta_0}{2}\right)\cos\left(\frac{\eta_0}{2}\right) \right) \right.$$

$$\left. + \sum_{n=-\infty}^{\infty} \frac{\partial \tilde{f}_{\text{int}}^{\eta,\eta}(\eta_0, R_0, n, d_R, \theta_R)}{\partial \eta_0} \exp\left( -\frac{n^2 \lambda_h}{2 l_c} \right) - \frac{M_R \tilde{f}_2(R_0, d_R, d_{\max}, d_{\min})}{4 R_0} \cos\left(\frac{\eta_0}{2}\right) - \frac{F_R}{8} \sin\left(\frac{\eta_0}{2}\right) \right].$$

(11.13)

For our choice of $d_{\max} = -d_{\min} = (R_0 - 2a)$, we have that the functions $\tilde{h}_1$ and $\tilde{h}_2$ are defined by

$$\tilde{h}_1(R_0, d_R, R_0 - 2a, 2a - R_0) = -\frac{R_0^3}{\sqrt{2\pi} d_R} \int_{-1}^{1} dx \frac{1}{\left(x(R_0 - 2a) + R_0\right)^2} \exp\left( -\frac{(R_0 - 2a)^2 x^2}{2 d_R^2} \right)$$

$$+ \frac{2(R_0 - 2a) R_0^3}{\sqrt{2\pi} d_R} \int_{-1}^{1} dx \frac{(x+1)}{\left(x(R_0 - 2a) + R_0\right)^3} \exp\left( -\frac{(R_0 - 2a)^2 x^2}{2 d_R^2} \right)$$

$$+ \frac{(R_0 - 2a)^2 R_0^3}{\sqrt{2\pi} d_R^3} \int_{-1}^{1} dx \frac{x^2}{\left(x(R_0 - 2a) + R_0\right)^2} \exp\left( -\frac{(R_0 - 2a)^2 x^2}{2 d_R^2} \right)$$

$$+ \frac{2 R_0^3}{(2R_0 - 2a)^3} \left( 1 - \text{erf}\left( \frac{R_0 - 2a}{d_R \sqrt{2}} \right) \right) + \frac{R_0^3}{\sqrt{2\pi} d_R} \left[ \frac{1}{(2a)^2} + \frac{1}{(2R_0 - 2a)^2} \right] \exp\left( -\frac{(R_0 - 2a)^2}{2 d_R^2} \right),$$

(11.14)

and

$$\tilde{h}_2(R_0,d_R,R_0-2a,2a-R_0) = -\frac{R_0^2}{\sqrt{2\pi}d_R}\int_{-1}^{1}dx\frac{1}{\left(x(R_0-2a)+R_0\right)}\exp\left(-\frac{\left(R_0-2a\right)^2x^2}{2d_R^2}\right)$$
$$+\frac{(R_0-2a)R_0^2}{\sqrt{2\pi}d_R}\int_{-1}^{1}dx\frac{(x+1)}{\left(x(R_0-2a)+R_0\right)^2}\exp\left(-\frac{\left(R_0-2a\right)^2x^2}{2d_R^2}\right)$$
$$+\frac{\left(R_0-2a\right)^2R_0^2}{\sqrt{2\pi}d_R^3}\int_{-1}^{1}dx\frac{x^2}{\left(x(R_0-2a)+R_0\right)}\exp\left(-\frac{\left(R_0-2a\right)^2x^2}{2d_R^2}\right) \tag{11.15}$$
$$+\frac{R_0^2}{(2R_0-2a)^2}\left(1-\operatorname{erf}\left(\frac{R_0-2a}{d_R\sqrt{2}}\right)\right)+\frac{R_0^2}{\sqrt{2\pi}d_R}\left[\frac{1}{(2a)}+\frac{1}{(2R_0-2a)}\right]\exp\left(-\frac{\left(R_0-2a\right)^2}{2d_R^2}\right).$$

For our second choice of $d_{\min}$ and $d_{\max}$ we may effectively set $d_{\max}=\infty$ in the integrals for $\tilde{f}_1$ and $\tilde{f}_2$ with $d_{\max}=2a-R_0$. Doing this, we also obtain for $\tilde{h}_1$ and $\tilde{h}_2$

$$\tilde{h}_1(R_0,d_R,\infty,2a-R_0) = -\frac{R_0^3}{\sqrt{2\pi}d_R}\int_{-1}^{\infty}dx\frac{1}{\left(x(R_0-2a)+R_0\right)^2}\exp\left(-\frac{\left(R_0-2a\right)^2x^2}{2d_R^2}\right)$$
$$+\frac{2(R_0-2a)R_0^3}{\sqrt{2\pi}d_R}\int_{-1}^{\infty}dx\frac{(x+1)}{\left(x(R_0-2a)+R_0\right)^3}\exp\left(-\frac{\left(R_0-2a\right)^2x^2}{2d_R^2}\right)$$
$$+\frac{\left(R_0-2a\right)^2R_0^3}{\sqrt{2\pi}d_R^3}\int_{-1}^{\infty}dx\frac{x^2}{\left(x(R_0-2a)+R_0\right)^2}\exp\left(-\frac{\left(R_0-2a\right)^2x^2}{2d_R^2}\right)+\frac{R_0^3}{\sqrt{2\pi}d_R}\frac{1}{(2a)^2}\exp\left(-\frac{\left(R_0-2a\right)^2}{2d_R^2}\right), \tag{11.16}$$

and

$$\tilde{h}_2(R_0,d_R,\infty,2a-R_0) = -\frac{R_0^2}{\sqrt{2\pi}d_R}\int_{-1}^{\infty}dx\frac{1}{\left(x(R_0-2a)+R_0\right)}\exp\left(-\frac{\left(R_0-2a\right)^2x^2}{2d_R^2}\right)$$
$$+\frac{(R_0-2a)R_0^2}{\sqrt{2\pi}d_R}\int_{-1}^{\infty}dx\frac{(x+1)}{\left(x(R_0-2a)+R_0\right)^2}\exp\left(-\frac{\left(R_0-2a\right)^2x^2}{2d_R^2}\right)$$
$$+\frac{\left(R_0-2a\right)^2R_0^2}{\sqrt{2\pi}d_R^3}\int_{-1}^{\infty}dx\frac{x^2}{\left(x(R_0-2a)+R_0\right)}\exp\left(-\frac{\left(R_0-2a\right)^2x^2}{2d_R^2}\right)+\frac{R_0^2}{\sqrt{2\pi}d_R}\frac{1}{2a}\exp\left(-\frac{\left(R_0-2a\right)^2}{2d_R^2}\right). \tag{11.17}$$

In general $\tilde{h}_1$ and $\tilde{h}_2$ are defined through the relations

$$\tilde{h}_1(R_0,d_R,d_{\max},d_{\min}) = 2\tilde{f}_1(R_0,d_R,d_{\max},d_{\min})-R_0\frac{\partial\tilde{f}_1(R_0,d_R,d_{\max},d_{\min})}{\partial R_0}, \tag{11.18}$$

$$\tilde{h}_2(R_0, d_R, d_{\max}, d_{\min}) = \tilde{f}_2(R_0, d_R, d_{\max}, d_{\min}) - R_0 \frac{\partial \tilde{f}_2(R_0, d_R, d_{\max}, d_{\min})}{\partial R_0}. \quad (11.19)$$

On minimization with respect to $\theta_R$ of Eq.(10.26) , we obtain

$$\begin{aligned}
0 &= \frac{l_p \theta_R^3}{d_R^2} + \sum_{n=-\infty}^{\infty} \exp\left(-\frac{n^2 \lambda_h}{2l_c}\right)\left(\frac{1}{\alpha_\eta^{1/2} 2^{3/2} \tilde{l}_p^{1/2}} \frac{\partial \tilde{f}_{\text{int}}^{\eta,\eta}(\eta_0, R_0, n, d_R, \theta_R)}{\partial \theta_R} - \frac{\alpha_\eta^{1/2}}{2^{1/2} \tilde{l}_p^{3/2}} \frac{\partial \tilde{f}_{\text{int}}^{\eta',\eta'}(\eta_0, R_0, n, d_R, \theta_R)}{\partial \theta_R}\right) \\
&+ \frac{1}{8} \sum_{n=-\infty}^{\infty} \exp\left(-\frac{n^2 \lambda_h}{2l_c}\right)\left[\left(\frac{1}{\alpha_x l_p}\right)^{1/2} \frac{\partial \tilde{f}_{\text{int}}^{x,x}(\eta_0, R_0, n, d_R, \theta_R)}{\partial \theta_R} + \left(\frac{1}{\alpha_y l_p}\right)^{1/2} \frac{\partial \tilde{f}_{\text{int}}^{y,y}(\eta_0, R_0, n, d_R, \theta_R)}{\partial \theta_R}\right] - \frac{1}{2 l_p \theta_R^3} \\
&- \frac{2\theta_R l_p}{R_0^2} \tilde{f}_1(R_0, d_R, d_{\max}, d_{\min}) \sin^2\left(\frac{\eta_0}{2}\right) + \sum_{n=-\infty}^{\infty} \frac{\partial \tilde{f}_{\text{int}}^{0}(\eta_0, R_0, n, d_R, \theta_R)}{\partial \theta_R} \exp\left(-\frac{n^2 \lambda_h}{2l_c}\right) - \frac{M_R}{2} \theta_R \frac{\tilde{f}_2(R_0, d_R, d_{\max}, d_{\min})}{R_0} \sin\left(\frac{\eta_0}{2}\right)^{-1} \\
&+ \frac{M_R^2}{64 l_p^{3/2}} \frac{1}{\cos\left(\frac{\eta_0}{2}\right)^4} \frac{1}{\left(\alpha_y^{1/2} + \alpha_x^{1/2}\right)^2} \sum_{n=-\infty}^{\infty} \exp\left(-\frac{n^2 \lambda_h}{2l_c}\right)\left(\frac{1}{\alpha_y^{1/2}} \frac{\partial \tilde{f}_{\text{int}}^{y,y}(\eta_0, R_0, n, d_R, \theta_R)}{\partial \theta_R} + \frac{1}{\alpha_x^{1/2}} \frac{\partial \tilde{f}_{\text{int}}^{x,x}(\eta_0, R_0, n, d_R, \theta_R)}{\partial \theta_R}\right),
\end{aligned}$$

(11.20)

and from minimization with respect to $d_R$

$$\begin{aligned}
0 &= \frac{1}{\alpha_\eta^{1/2} 2^{3/2} \tilde{l}_p^{1/2}} \frac{\partial \alpha_\eta}{\partial d_R} - \frac{\alpha_\eta^{1/2}}{2^{1/2} \tilde{l}_p^{3/2}} \sum_{n=-\infty}^{\infty} \frac{\partial \tilde{f}_{\text{int}}^{\eta',\eta'}(\eta_0, R_0, n, d_R, \theta_R)}{\partial d_R} \exp\left(-\frac{n^2 \lambda_h}{2l_c}\right) \\
&+ \frac{1}{8} \sum_{n=-\infty}^{\infty} \exp\left(-\frac{n^2 \lambda_h}{2l_c}\right)\left[\left(\frac{1}{\alpha_x l_p}\right)^{1/2} \frac{\partial \tilde{f}_{\text{int}}^{x,x}(\eta_0, R_0, n, d_R, \theta_R)}{\partial d_R} + \left(\frac{1}{\alpha_y l_p}\right)^{1/2} \frac{\partial \tilde{f}_{\text{int}}^{y,y}(\eta_0, R_0, n, d_R, \theta_R)}{\partial d_R}\right] \\
&- \frac{l_p \theta_R^2}{d_R R_0^2} \tilde{l}_1(R_0, d_R, d_{\max}, d_{\min}) \sin^2\left(\frac{\eta_0}{2}\right) + d_R \alpha_H - \frac{l_p \theta_R^4}{2 d_R^3} \\
&+ \frac{4 l_p}{d_R R_0^2} \tilde{l}_1(R_0, d_R, d_{\max}, d_{\min}) \sin^4\left(\frac{\eta_0}{2}\right) + \sum_{n=-\infty}^{\infty} \frac{\partial \tilde{f}_{\text{int}}^{0}(\eta_0, R_0, n, d_R, \theta_R)}{\partial d_R} \exp\left(-\frac{n^2 \lambda_h}{2l_c}\right) \\
&+ \frac{2 M_R}{d_R R_0} \tilde{l}_2(R_0, d_R, d_{\max}, d_{\min}) \sin\left(\frac{\eta_0}{2}\right) - \frac{M_R \theta_R^2}{4 d_R R_0} \tilde{l}_2(R_0, d_R, d_{\max}, d_{\min}) \sin\left(\frac{\eta_0}{2}\right)^{-1} \\
&+ \frac{M_R^2}{64 l_p^{3/2}} \frac{1}{\cos\left(\frac{\eta_0}{2}\right)^4} \frac{1}{\left(\alpha_y^{1/2} + \alpha_x^{1/2}\right)^2} \sum_{n=-\infty}^{\infty} \exp\left(-\frac{n^2 \lambda_h}{2l_c}\right)\left(\frac{1}{\alpha_y^{1/2}} \frac{\partial \tilde{f}_{\text{int}}^{y,y}(\eta_0, R_0, n, d_R, \theta_R)}{\partial d_R} + \frac{1}{\alpha_x^{1/2}} \frac{\partial \tilde{f}_{\text{int}}^{x,x}(\eta_0, R_0, n, d_R, \theta_R)}{\partial d_R}\right),
\end{aligned}$$

(11.21)

where

$$\begin{aligned}
\frac{\partial \alpha_\eta}{\partial d_R} &= \left[\frac{4 l_p \tilde{l}_1(R_0, d_R, d_{\max}, d_{\min})}{d_R R_0^2}\left(3\cos^2\left(\frac{\eta_0}{2}\right)\sin^2\left(\frac{\eta_0}{2}\right) - \sin^4\left(\frac{\eta_0}{2}\right)\right)\right. \\
&\left.+ \sum_{n=-\infty}^{\infty} \frac{\partial \tilde{f}_{\text{int}}^{\eta,\eta}(\eta_0, R_0, n, d_R, \theta_R)}{\partial d_R} \exp\left(-\frac{n^2 \lambda_h}{2l_c}\right) - \frac{M_R \tilde{l}_2(R_0, d_R, d_{\max}, d_{\min})}{2 d_R R_0} \sin\left(\frac{\eta_0}{2}\right)\right],
\end{aligned} \quad (11.22)$$

$$\tilde{l}_1(R_0,d_R,d_{\max},d_{\min})=\frac{R_0^2}{\sqrt{2\pi}}\int_{\frac{d_{\min}}{d_R}}^{\frac{d_{\max}}{d_R}}dx\frac{(x^2-1)}{(R_0+d_Rx)^2}\exp\left(-\frac{x^2}{2}\right)$$

$$+\frac{R_0^2}{\sqrt{2\pi}}\left(\frac{d_{\max}}{d_R}\frac{\exp\left(-\frac{1}{2}\left(\frac{d_{\max}}{d_R}\right)^2\right)}{(R_0+d_{\max})^2}-\frac{d_{\min}}{d_R}\frac{\exp\left(-\frac{1}{2}\left(\frac{d_{\min}}{d_R}\right)^2\right)}{(R_0+d_{\min})^2}\right),\qquad(11.23)$$

and

$$\tilde{l}_2(R_0,d_R,d_{\max},d_{\min})=\frac{R_0}{\sqrt{2\pi}}\int_{\frac{d_{\min}}{d_R}}^{\frac{d_{\max}}{d_R}}dx\frac{(x^2-1)}{(R_0+d_Rx)}\exp\left(-\frac{x^2}{2}\right)$$

$$+\frac{R_0}{\sqrt{2\pi}}\left(\frac{d_{\max}}{d_R}\frac{\exp\left(-\frac{1}{2}\left(\frac{d_{\max}}{d_R}\right)^2\right)}{R_0+d_{\max}}-\frac{d_{\min}}{d_R}\frac{\exp\left(-\frac{1}{2}\left(\frac{d_{\min}}{d_R}\right)^2\right)}{R_0+d_{\min}}\right).\qquad(11.24)$$

Last of all, we obtain from minimization of Eq. (10.26) with respect to $\lambda_h$ the equation

$$0=-\frac{1}{4\lambda_h^2}-\sum_{n=-\infty}^{\infty}\frac{n^2}{2\lambda_c}\exp\left(-\frac{n^2\lambda_h}{2l_c}\right)\left[\frac{1}{\alpha_\eta^{1/2}2^{3/2}\tilde{l}_p^{1/2}}\tilde{f}_{\text{int}}^{\eta,\eta}(\eta_0,R_0,n,d_R,\theta_R)-\frac{\alpha_\eta^{1/2}}{2^{1/2}\tilde{l}_p^{3/2}}\tilde{f}_{\text{int}}^{\eta',\eta'}(\eta_0,R_0,n,d_R,\theta_R)\right.$$
$$\left.+\tilde{f}_{\text{int}}^{0}(\eta_0,R_0,n,d_R,\theta_R)+\frac{1}{8}\left(\frac{1}{\alpha_xl_p}\right)^{1/2}\tilde{f}_{\text{int}}^{x,x}(\eta_0,R_0,n,d_R,\theta_R)+\frac{1}{8}\left(\frac{1}{\alpha_yl_p}\right)^{1/2}\tilde{f}_{\text{int}}^{y,y}(\eta_0,R_0,n,d_R,\theta_R)\right]$$
$$-\frac{M_R^2}{64l_p^{3/2}}\frac{1}{\cos\left(\frac{\eta_0}{2}\right)^4}\frac{1}{\left(\alpha_y^{1/2}+\alpha_x^{1/2}\right)^2}\sum_{n=-\infty}^{\infty}\frac{n^2}{2\lambda_c}\exp\left(-\frac{n^2\lambda_h}{2\lambda_c}\right)\left(\frac{1}{\alpha_y^{1/2}}\tilde{f}_{\text{int}}^{y,y}(\eta_0,R_0,n,d_R,\theta_R)+\frac{1}{\alpha_x^{1/2}}\tilde{f}_{\text{int}}^{x,x}(\eta_0,R_0,n,d_R,\theta_R)\right).$$

(11.25)

In our first choice of $d_{\min}$ and $d_{\max}$ the derivatives of $\alpha_H$ are computed to be

$$\frac{\partial\alpha_H}{\partial R_0}=-\frac{8}{3}\frac{1}{2^{5/3}(R_0-2a)^{11/3}l_p^{1/3}},\qquad\frac{\partial\alpha_H}{\partial\eta_0}=0,\qquad(11.26)$$

whereas, in our second choice (c.f Eq.(5.11)), the derivatives of $\alpha_H$ are computed to be

$$\frac{\partial \alpha_H}{\partial R_0} \approx -\frac{16}{3(d_{\max}-d_{\min})^{11/3} l_p^{1/3}} \left( 1+\frac{3}{2}\left(\frac{\pi}{\tan(\eta_0/2)}\right)^{3/2} \frac{R_0^{1/2}}{\left(l_p^b\sqrt{2}\right)^{1/2}} \right), \tag{11.27}$$

$$\frac{\partial \alpha_H}{\partial \eta_0} \approx \frac{4}{(d_{\max}-d_{\min})^{11/3} l_p^{1/3}} \frac{(\pi R_0)^{3/2}}{\left(l_p^b\sqrt{2}\right)^{1/2} \cos^2(\eta_0/2)\tan^{5/2}(\eta_0/2)}, \tag{11.28}$$

where $d_{\max}$ is approximately given by Eq. (5.11).

We now have a set of general equations that can be applied to braids formed of different rod like molecules and different models of intermolecular interaction. In last two sections we will deal with special cases of Eqs. (10.26), (11.10), (11.11),(11.20),(11.21) and (11.25) .

## 12. A braid formed of molecules which may be described by uniformly charged elastic rods

For uniformly charged rods the interaction does not depend on $\Delta\Phi(s)$. Furthermore we can suppose $\eta(s),\ x_A'(s),\ y_A'(s),$ and $R'(s)$ are small so that

$$E_{\rm int} \approx \int_{-L/2}^{L/2} ds_0 \tilde{\varepsilon}_{\rm int}^0(0,R(s_0),0,0,0,0). \tag{12.1}$$

For molecules described by uniformly charged rods in salt solution, interacting through electrostatics, we may further write

$$\begin{aligned}\bar{\varepsilon}_{\rm int}^0(0,R(s_0),0,0,0,0) &= \frac{2l_B(1-\theta_c)^2}{l_e^2}\frac{K_0(\kappa_D R(s_0))}{(a\kappa_D)^2 K_1(a\kappa_D)^2} - \theta_{cor}^2 K_0(\kappa_c R(s_0)) \\ &+ \frac{2l_B(1-\theta_c)^2}{l_e^2(a\kappa_D)^2 K_1(a\kappa_D)^2}\sum_{j=-\infty}^{\infty} K_j(\kappa_D R(s_0))K_j(\kappa_D R(s_0))\frac{I_j'(a\kappa_D)}{K_j'(a\kappa_D)}.\end{aligned} \tag{12.2}$$

The first term in Eq. (12.2) represents the repulsive contribution from mean field electrostatics, where $\theta_c$ is fraction of the bare charge that is compensated by condensed or adsorbed ions near the molecular surface. The length $l_e$ is defined as $l_e = e/(\pi a|\sigma|)$, where $\sigma$ is the surface charge density of the molecular surface and $l_B$ is the Bjerrum length. The range of this interaction is determined through the inverse Debye screening length $\kappa_D$. The second term is an attractive contribution from correlation forces [6,11] (or some other mechanism), where the empirical parameters $\theta_{cor}$ and $\kappa_c$ determine the strength and range of this interaction. The last term in Eq. (12.2) is a mean field electrostatic term due to image charge repulsion where the dielectric constant of the molecular core is assumed to be much smaller than that of the solvent, a more general expression may be found in [12]. The functions $I_n(x)$ and $K_n(x)$ are modified Bessel functions of the first and second kind of order $n$; $I_n'(x)$ and $K_n'(x)$ are their derivatives with respect to argument.

The parameter $\theta_c$ may be calculated through application of the non-linear Poisson Boltzmann equation; however near the molecular surface, the effects of discrete solvent molecules and chemi-adsorption effects may be important factors in determining $\theta_c$. The parameter $\theta_{cor}$ depends on the number density of ions near the surface of the molecule as well as the correlation parameter $\Xi = (2\pi)^2 q^3 l_B^2 |\sigma|$. For mono-valent salt $\theta_{cor}$ will usually be quite small and this term can be neglected. For multivalent ions, the magnitude of $\theta_{cor}$ may depend on whether ions near the surface molecule are free to move or are adsorbed specifically at sites at which they are pinned and cannot move, for a further discussion see Ref. [13].

Utilizing Eqs. (12.1) and (12.2), in the general expression for the Free Energy, Eq. (10.26), simplifies to

$$
\begin{aligned}
\frac{F_T}{k_B T} &= L\left(\frac{F_R}{2l_p\cos(\eta_0/2)}\right)^{1/2} + \frac{\alpha_\eta^{1/2} L}{2^{1/2}\tilde{l}_p^{1/2}} + \frac{\alpha_H d_R^2 L}{2} + \frac{L}{4l_p\theta_R^2} + \frac{L}{4\lambda_h} + \frac{l_p\theta_R^4 L}{4d_R^2} - \frac{M_R^2}{16 l_p^{3/2}(2F_R)^{1/2}}\frac{L}{\cos\left(\frac{\eta_0}{2}\right)^{7/2}} \\
&- \frac{L\theta_R^2 l_p}{R_0^2}\tilde{f}_1(R_0,d_R,d_{\max},d_{\min})\sin^2\left(\frac{\eta_0}{2}\right) + \frac{2l_B L(1-\theta_c)^2}{l_e^2(a\kappa_D)^2 K_1(a\kappa_D)^2} g_0(\kappa_D R_0,\kappa_D d_R, d_{\max}/d_R, d_{\min}/d_R) \\
&+ \frac{2l_B L}{l_e^2}\frac{2l_B L(1-\theta_c)^2}{(\kappa_n a K_n'(\kappa_n a))^2} g_{img}(0,\kappa_D R_0,\kappa_D d_R, d_{\max}/d_R, d_{\min}/d_R; a) - \theta_{cor}^2 L g_0(\kappa_c R_0,\kappa_c d_R, d_{\max}/d_R, d_{\min}/d_R) \\
&+ \frac{4l_p L\tilde{f}_1(R_0,d_R,d_{\max},d_{\min})}{R_0^2}\sin^4\left(\frac{\eta_0}{2}\right) - F_R L\cos\left(\frac{\eta_0}{2}\right) + \frac{2M_R L}{R_0}\sin\left(\frac{\eta_0}{2}\right)\tilde{f}_2(R_0,d_R,d_{\max},d_{\min}) \\
&- \frac{M_R L\theta_R^2}{4}\frac{\tilde{f}_2(R_0,d_R,d_{\max},d_{\min})}{R_0}\sin\left(\frac{\eta_0}{2}\right)^{-1},
\end{aligned}
\tag{12.3}
$$

and we define the functions

$$
\begin{aligned}
g_j(\kappa R_0,\kappa d_R, d_{\max}/d_R, d_{\min}/d_R) &= \frac{1}{\sqrt{2\pi}}\int_{d_{\min}/d_R}^{d_{\max}/d_R} dy K_j(\kappa R_0 + y\kappa d_R)\exp\left(-\frac{y^2}{2}\right) \\
&+ \frac{1}{2}K_j(\kappa(R_0+d_{\min}))\left[1-\operatorname{erf}\left(-\frac{1}{\sqrt{2}}\frac{d_{\min}}{d_R}\right)\right] + \frac{1}{2}K_j(\kappa(R_0+d_{\max}))\left[1-\operatorname{erf}\left(\frac{1}{\sqrt{2}}\frac{d_{\max}}{d_R}\right)\right],
\end{aligned}
\tag{12.4}
$$

and

$$g_{img}(n,\kappa R_0,\kappa d_r,d_{\max}/d_R,d_{\min}/d_R;a)=\frac{1}{\sqrt{2\pi}}\sum_{j=-\infty}^{\infty}\int_{d_{\min}/d_R}^{d_{\max}/d_R}dyK_{n-j}(\kappa R_0+y\kappa d_R)K_{n-j}(\kappa R_0+y\kappa d_R)\frac{I'_j(\kappa a)}{K'_j(\kappa a)}$$

$$\exp\left(-\frac{y^2}{2}\right)+\frac{1}{2}\sum_{j=-\infty}^{\infty}K_{n-j}(\kappa(R_0+d_{\min}))K_{n-j}(\kappa(R_0+d_{\min}))\frac{I'_j(\kappa a)}{K'_j(\kappa a)}\left[1-\mathrm{erf}\left(-\frac{1}{\sqrt{2}}\frac{d_{\min}}{d_R}\right)\right]$$

$$+\frac{1}{2}\sum_{j=-\infty}^{\infty}K_{n-j}(\kappa(R_0+d_{\max}))K_{n-j}(\kappa(R_0+d_{\max}))\frac{I'_j(\kappa a)}{K'_j(\kappa a)}\left[1-\mathrm{erf}\left(\frac{1}{\sqrt{2}}\frac{d_{\max}}{d_R}\right)\right].$$

(12.5)

Here, $\alpha_\eta$ is given by the expression

$$\alpha_\eta=\left[\frac{4l_p\tilde{f}_1(R_0,d_R,d_{\max},d_{\min})}{R_0^2}\left(3\cos^2\left(\frac{\eta_0}{2}\right)\sin^2\left(\frac{\eta_0}{2}\right)-\sin^4\left(\frac{\eta_0}{2}\right)\right)\right.$$
$$\left.-\frac{M_R\tilde{f}_2(R_0,d_R,d_{\max},d_{\min})}{2R_0}\sin\left(\frac{\eta_0}{2}\right)+\frac{F_R}{4}\cos\left(\frac{\eta_0}{2}\right)\right]. \qquad (12.6)$$

In the case of uniform charged rods, Eqs (11.10), (11.11), (11.20) and (11.21) simplify to

$$0=\frac{\theta_R^2 l_p}{R_0^3}\tilde{h}_1(R_0,d_R,d_{\max},d_{\min})\sin^2\left(\frac{\eta_0}{2}\right)-\frac{4l_p\tilde{h}_1(R_0,d_R,d_{\max},d_{\min})}{R_0^3}\sin^4\left(\frac{\eta_0}{2}\right)-\frac{2M_R}{R_0^2}\sin\left(\frac{\eta_0}{2}\right)\tilde{h}_2(R_0,d_R,d_{\max},d_{\min})$$
$$+\frac{2l_B(1-\theta_c)^2}{l_e^2(a\kappa_D)^2R_0K_1(a\kappa_D)^2}\left[q_0(\kappa_DR_0,\kappa_Dd_R,d_{\max}/d_R,d_{\min}/d_R)+q_{img}(0,\kappa_DR_0,\kappa_Dd_R,d_{\max}/d_R,d_{\min}/d_R)\right]$$
$$-\frac{\theta_{cor}^2}{R_0}g_0(\kappa_cR_0,\kappa_cd_R,d_{\max}/d_R,d_{\min}/d_R)+\frac{M_R\theta_R^2}{4}\frac{\tilde{h}_2(R_0,d_R,d_{\max},d_{\min})}{R_0^2}\sin\left(\frac{\eta_0}{2}\right)^{-1}+\frac{d_R^2}{2}\frac{d\alpha_H}{dR_0}+\frac{1}{\alpha_\eta^{1/2}2^{3/2}\tilde{l}_p^{1/2}}\frac{\partial\alpha_\eta}{\partial R_0},$$

(12.7)

$$0=\frac{1}{4}\left(\frac{F_R}{2l_p}\right)^{1/2}\frac{\sin\left(\frac{\eta_0}{2}\right)}{\cos\left(\frac{\eta_0}{2}\right)^{3/2}}+\frac{1}{\alpha_\eta^{1/2}2^{3/2}\tilde{l}_p^{1/2}}\frac{\partial\alpha_\eta}{\partial\eta_0}-\frac{\theta_R^2l_p}{R_0^2}\tilde{f}_1(R_0,d_R,d_{\max},d_{\min})\sin\left(\frac{\eta_0}{2}\right)\cos\left(\frac{\eta_0}{2}\right)$$
$$+\frac{8l_p\tilde{f}_1(R_0,d_R,d_{\max},d_{\min})}{R_0^2}\cos\left(\frac{\eta_0}{2}\right)\sin^3\left(\frac{\eta_0}{2}\right)+\frac{M_R}{R_0}\cos\left(\frac{\eta_0}{2}\right)\tilde{f}_2(R_0,d_R,d_{\max}d_{\min})+\frac{F_R}{2}\sin\left(\frac{\eta_0}{2}\right) \qquad (12.8)$$
$$+\frac{M_R\theta_R^2}{8}\frac{\tilde{f}_2(R_0,d_R,d_{\max}d_{\min})}{R_0}\cos\left(\frac{\eta_0}{2}\right)\sin\left(\frac{\eta_0}{2}\right)^{-2}-\frac{7M_R^2}{64l_p^{3/2}\left(2F_R\right)^{1/2}}\frac{\sin\left(\frac{\eta_0}{2}\right)}{\cos\left(\frac{\eta_0}{2}\right)^{9/5}},$$

$$0=-\frac{1}{2l_p\theta_R^3}+\frac{l_p\theta_R^3}{d_R^2}-\frac{2\theta_Rl_p}{R_0^2}\tilde{f}_1(R_0,d_R,d_{\max},d_{\min})\sin^2\left(\frac{\eta_0}{2}\right)-\frac{M_R\theta_R}{2}\frac{\tilde{f}_2(R_0,d_R,d_{\max},d_{\min})}{R_0}\sin\left(\frac{\eta_0}{2}\right)^{-1},$$

(12.9)

$$0 = \frac{1}{\alpha_\eta^{1/2} 2^{3/2} \tilde{l}_p^{1/2}} \frac{\partial \alpha_\eta}{\partial d_R} + d_R \alpha_H - \frac{l_p \theta_R^4}{8 d_R^3} - \frac{l_p \theta_R^2}{d_R R_0^2} \tilde{l}_1(R_0, d_R, d_{\max}, d_{\min}) \sin^2\left(\frac{\eta_0}{2}\right)$$
$$+ \frac{4 l_p}{d_R R_0^2} \tilde{l}_1(R_0, d_R, d_{\max}, d_{\min}) \sin^4\left(\frac{\eta_0}{2}\right) - \frac{M_R \theta_R^2}{4 d_R R_0} \tilde{l}_2(R_0, d_R, d_{\max}, d_{\min})) \sin\left(\frac{\eta_0}{2}\right)^{-1}$$
$$+ \frac{2 l_B (1-\theta_c)^2}{d_R l_e^2 (a\kappa_D)^2 K_1(a\kappa_D)^2} \Big( m_0(\kappa_D R_0, \kappa_D d_R, d_{\max}/d_R, d_{\min}/d_R) + m_{img}(0, \kappa_D R_0, \kappa_D d_R, d_{\max}/d_R, d_{\min}/d_R) \Big)$$
$$- \frac{\theta_{cor}^2}{d_R} m_0(\kappa_c R_0, \kappa_c d_R, d_{\max}/d_R, d_{\min}/d_R) + \frac{2 M_R}{d_R R_0} \tilde{l}_2(R_0, d_R, d_{\max}, d_{\min})) \sin\left(\frac{\eta_0}{2}\right). \qquad (12.10)$$

The partial derivatives of $\alpha_\eta$ are now given by

$$\frac{\partial \alpha_\eta}{\partial R_0} = \Bigg[ -\frac{4 l_p \tilde{h}_1(R_0, d_R, d_{\max}, d_{\min})}{R_0^3} \left( 3\cos^2\left(\frac{\eta_0}{2}\right) \sin^2\left(\frac{\eta_0}{2}\right) - \sin^4\left(\frac{\eta_0}{2}\right) \right)$$
$$+ \frac{M_R \tilde{h}_2(R_0, d_R, d_{\max}, d_{\min})}{2 R_0^2} \sin\left(\frac{\eta_0}{2}\right) \Bigg], \qquad (12.11)$$

$$\frac{\partial \alpha_\eta}{\partial \eta_0} = \Bigg[ \frac{4 l_p \tilde{f}_1(R_0, d_R, d_{\max}, d_{\min})}{R_0^2} \left( 3\cos^3\left(\frac{\eta_0}{2}\right) \sin^2\left(\frac{\eta_0}{2}\right) - 5\sin^3\left(\frac{\eta_0}{2}\right) \cos\left(\frac{\eta_0}{2}\right) \right)$$
$$- \frac{M_R \tilde{f}_2(R_0, d_R, d_{\max}, d_{\min})}{4 R_0} \cos\left(\frac{\eta_0}{2}\right) - \frac{F_R}{8} \sin\left(\frac{\eta_0}{2}\right) \Bigg], \qquad (12.12)$$

$$\frac{\partial \alpha_\eta}{\partial d_R} = \Bigg[ \frac{4 l_p \tilde{l}_1(R_0, d_R, d_{\max}, d_{\min})}{d_R R_0^2} \left( 3\cos^2\left(\frac{\eta_0}{2}\right) \sin^2\left(\frac{\eta_0}{2}\right) - \sin^4\left(\frac{\eta_0}{2}\right) \right)$$
$$- \frac{M_R \tilde{l}_2(R_0, d_R, d_{\max}, d_{\min})}{2 d_R R_0} \sin\left(\frac{\eta_0}{2}\right) \Bigg]. \qquad (12.13)$$

The derivatives of $\alpha_H$ are given by either Eq. (11.26) or Eqs. (11.27) and(11.28) for our two choices of $d_{\max}$ and $d_{\min}$. For our first choice of $d_{\max} = -d_{\min} = R_0 - 2a$ we have that

$$q_j(\kappa R_0,\kappa d_R,(R_0-2a)/d_R,-(R_0-2a)/d_R)$$
$$=\frac{1}{\sqrt{2\pi}}\frac{R_0}{d_R}\int_{-1}^{1}dxK_j(\kappa R_0+x(R_0-2a)\kappa)\exp\left(-\frac{(R_0-2a)^2x^2}{2d_R^2}\right)$$
$$+\frac{1}{\sqrt{2\pi}}\frac{\left(R_0-2a\right)\kappa R_0}{d_R}\int_{-1}^{1}dx(1+x)K_j'(\kappa R_0+x(R_0-2a)\kappa)\exp\left(-\frac{(R_0-2a)^2x^2}{2d_R^2}\right)$$
$$-\frac{1}{\sqrt{2\pi}}\frac{\left(R_0-2a\right)^2R_0}{d_R^3}\int_{-1}^{1}dxx^2K_j(\kappa R_0+x(R_0-2a)\kappa)\exp\left(-\frac{(R_0-2a)^2x^2}{2d_R^2}\right) \qquad (12.14)$$
$$-\frac{1}{\sqrt{2\pi}}\frac{R_0}{d_R}\left[K_j(2\kappa a)+K_j(2(R_0-a)\kappa)\right]\exp\left(-\frac{(R_0-2a)^2}{2d_R^2}\right)$$
$$+\kappa R_0K_j'(2(R_0-a)\kappa)\left[1-\mathrm{erf}\left(\frac{R_0-2a}{d_R\sqrt{2}}\right)\right],$$

$$q_{img}(n,\kappa R_0,\kappa d_R,(R_0-2a)/d_R,-(R_0-2a)/d_R)$$
$$=\sum_{j=-\infty}^{\infty}\frac{I_j'(\kappa a)}{K_j'(\kappa a)}\left[\frac{R_0}{\sqrt{2\pi}d_R}\int_{-1}^{1}dxK_{j-n}(\kappa R_0+x(R_0-2a)\kappa)K_{j-n}(\kappa R_0+x(R_0-2a)\kappa)\exp\left(-\frac{(R_0-2a)^2x^2}{2d_R^2}\right)\right.$$
$$+\frac{2\left(R_0-2a\right)\kappa R_0}{\sqrt{2\pi}d_R}\int_{-1}^{1}dx(1+x)K_{j-n}'(\kappa R_0+x(R_0-2a)\kappa)K_{j-n}(\kappa R_0+x(R_0-2a)\kappa)\exp\left(-\frac{(R_0-2a)^2x^2}{2d_R^2}\right)$$
$$-\frac{\left(R_0-2a\right)^2R_0}{\sqrt{2\pi}d_R^3}\int_{-1}^{1}dxx^2K_{j-n}(\kappa R_0+x(R_0-2a)\kappa)K_{j-n}(\kappa R_0+x(R_0-2a)\kappa)\exp\left(-\frac{(R_0-2a)^2x^2}{2d_R^2}\right)$$
$$-\frac{1}{\sqrt{2\pi}}\frac{R_0}{d_R}\left[K_{j-n}(2\kappa a)K_{j-n}(2\kappa a)+K_{j-n}(2(R_0-a)\kappa)K_{j-n}(2(R_0-a)\kappa)\right]\exp\left(-\frac{(R_0-2a)^2}{2d_R^2}\right)$$
$$\left.+2\kappa R_0K_{j-n}'(2(R_0-a)\kappa)K_{j-n}(2(R_0-2)\kappa)\left[1-\mathrm{erf}\left(\frac{R_0-2a}{d_R\sqrt{2}}\right)\right]\right],$$

(12.15)

$$m_j(\kappa R_0,\kappa d_R,(R_0-2a)/d_R,-(R_0-2a)/d_R)=\frac{1}{\sqrt{2\pi}}\int_{(2a-R_0)/d_R}^{(R_0-2a)/d_R}dy\left(y^2-1\right)\exp\left(-\frac{y^2}{2}\right)K_j\left(\kappa\left(R_0+yd_R\right)\right)$$
$$+\frac{(R_0-2a)}{d_R\sqrt{2\pi}}\left(K_j(2\kappa a)+K_j(2\kappa(R_0-2a))\right)\exp\left(-\frac{(R_0-2a)^2}{2d_R^2}\right),$$

(12.16)

$$m_{img}(n,\kappa R_0,\kappa d_R,(R_0-2a)/d_R,-(R_0-2a)/d_R)=\sum_{j=-\infty}^{\infty}\frac{I_j'(\kappa a)}{K_j'(\kappa a)}$$

$$\left[\frac{1}{\sqrt{2\pi}}\int_{(2a-R_0)/d_R}^{(R_0-2a)/d_R}dy\left(y^2-1\right)\exp\left(-\frac{y^2}{2}\right)K_{n-j}\left(\kappa\left(R_0+yd_R\right)\right)K_{n-j}\left(\kappa\left(R_0+yd_R\right)\right)\right.$$

$$\left.+\frac{(R_0-2a)}{d_R\sqrt{2\pi}}\left(K_{n-j}(2\kappa a)K_{n-j}(2\kappa a)+K_{n-j}(2\kappa(R_0-a))K_{n-j}(2\kappa(R_0-a))\right)\exp\left(-\frac{(R_0-2a)^2}{2d_R^2}\right)\right].$$

(12.17)

In general, the functions $m_j(\kappa R_0,\kappa d_R,d_{\max}/d_R,d_{\min}/d_R)$, $m_{img}(n,\kappa R_0,\kappa d_R,d_{\max}/d_R,d_{\min}/d_R)$, $q_j(\kappa R_0,\kappa d_R,d_{\max}/d_R,d_{\min}/d_R)$ and $q_{img}(\kappa R_0,\kappa d_R,d_{\max}/d_R,d_{\min}/d_R)$ are given by

$$q_j(\kappa R_0,\kappa d_R,d_{\max}/d_R,d_{\min}/d_R)=R_0\frac{\partial g_j(\kappa R_0,\kappa d_R,d_{\max}/d_R,d_{\min}/d_R)}{\partial R_0}, \quad (12.18)$$

$$q_{img}(n,\kappa R_0,\kappa d_R,d_{\max}/d_R,d_{\min}/d_R)=R_0\frac{\partial g_{img}(n,\kappa R_0,\kappa d_R,d_{\max}/d_R,d_{\min}/d_R)}{\partial R_0}, \quad (12.19)$$

$$m_j(\kappa R_0,\kappa d_R,d_{\max}/d_R,d_{\min}/d_R)=d_R\frac{\partial g_j(\kappa R_0,\kappa d_R,d_{\max}/d_R,d_{\min}/d_R)}{\partial d_R}, \quad (12.20)$$

and

$$m_{img}(n,\kappa R_0,\kappa d_R,d_{\max}/d_R,d_{\min}/d_R)=d_R\frac{\partial g_{img}(n,\kappa R_0,\kappa d_R,d_{\max}/d_R,d_{\min}/d_R)}{\partial d_R}. \quad (12.21)$$

We now will discuss Eq. (12.9). If we neglect the contribution from average elastic energy terms in Eq. (12.9) the solution is simple; we obtain

$$\theta_R^2=\frac{1}{2^{1/3}}\left(\frac{d_r}{l_p^b}\right)^{2/3}, \quad (12.22)$$

this is precisely the case given when $\beta_R=0$, as was obtained in [3]. In principle, as Eq. (12.9) is equivalent to a cubic equation in $\theta_R^2$, an analytical solution in terms of the cubic formula can be written. However, as opposed to a rather cumbersome exact expression, one can also find a very good approximation to solution of Eq. (12.9) of the form:

$$\theta_R\approx\left(\frac{d_R}{l_p}\right)^{1/3}\frac{1}{\left(4\Gamma^2+\frac{32}{9}\Gamma+2^{4/3}\right)^{1/8}}, \quad (12.23)$$

where

$$\Gamma=\left(\frac{d_R}{l_p}\right)^{4/3}\left(\frac{M_R l_p}{2R_0}\tilde{f}_2(R_0,d_R,d_{\max},d_{\min})\sin\left(\frac{\eta_0}{2}\right)^{-1}-\frac{2l_p^2}{R_0^2}\tilde{f}_1(R_0,d_R,d_{\max},d_{\min})\sin^2\left(\frac{\eta_0}{2}\right)\right),$$

(12.24)

which indeed recovers back the limiting case of Eq. (12.22) when $\Gamma=0$.

# 13. Modelling helix specific DNA-DNA interactions

### *13.1 Helix Non-Ideality*

Before discussing a particular model of helix specific DNA interactions, let us discuss a certain approximation that can be utilized to take account of irregular helix structure of DNA, depending only on helical structure of the molecule, not the underlying interaction model. This will give us the most general free energy to describe the braiding two DNA molecules with interactions included. We will then show explicit forms for various terms, using the mean-field electrostatic K-L theory for DNA-DNA interactions.

First we may write down for the helix densities $g_1^0(s_0)$ and $g_2^0(s_0)$, describing distorted irregular helices, for the molecules in their unstressed state, the following

$$g_1^0(s_0)=\bar{g}_0+\Delta g_1^0(s_0),\ \ g_2^0(s_0)=\bar{g}_0+\Delta g_2^0(s_0), \tag{13.1}$$

where $\bar{g}_0$ is the average twist density. The functions $\Delta g_1^0(s)$ and $\Delta g_2^0(s)$ describe deviations away from this average value. These functions depend on the base pair sequences of the molecules. In [14] it was shown that for a completely random base-pair text, over large enough length scales, consider $\Delta g_1^0(s)$ and $\Delta g_2^0(s)$ as random in such a way that

$$\left\langle \Delta g_1(s)\Delta g_1(s')\right\rangle_{g_1}=\left\langle \Delta g_2(s)\Delta g_2(s')\right\rangle_{g_1}=\frac{1}{\lambda_c^{(0)}}\delta(s-s'), \tag{13.2}$$

$$\left\langle \Delta g_1(s)\right\rangle_{g_1}=\left\langle \Delta g_2(s)\right\rangle_{g_2}=0, \tag{13.3}$$

where the subscripts $g_1$ and $g_2$ on the averaging bracket mean that they are ensemble averages over all possible realizations of $\Delta g_1(s)$ and $\Delta g_2(s)$. We can further assume that $\Delta g_1^0(s)$ and $\Delta g_1^0(s)$ are Gaussian distributed. We have a length scale $\lambda_c^{(0)}\approx 150\text{Å}$ [14] that describes the length over which, in the unstressed ground state of the molecules, the vectors $\hat{\mathbf{v}}_1(s_0)$ and $\hat{\mathbf{v}}_2(s_0)$ (see Section 2) fall out of alignment with each other for molecules with different base pair texts, i.e. the length over which phase $\Delta\Phi_0(s_0)=\xi_2(s_0)-\xi_1(s_0)$ can be assumed to be roughly constant (for a more detailed discussion of $\lambda_c^{(0)}$ see [15]).

Now, let us consider the two molecules having different base pairs at position $s_0$ along the braid, and suppose that the two distributions are uncorrelated. This allows us to write the average variational Free energy of the braid as a path integral with respect over $\Delta g^0(s_0)=\Delta g_1^0(s_0)-\Delta g_2^0(s_0)$, which is

$$\left\langle F_T[\Delta g^0(s_0)]\right\rangle_{\Delta g}=\frac{1}{Z_{\Delta g}}\int D\Delta g^0(s_0)\tilde{F}_T\left[\Delta\Phi_0\left[\frac{d\Delta g^0(s_0)}{ds}\right]\right]\exp\left(-\frac{\lambda_c^{(0)}}{4}\int_{-\infty}^{\infty}ds\Delta g^0(s_0)^2\right), \quad (13.4)$$

and

$$Z_{\Delta g}=\int D\Delta g^0(s_0)\exp\left(-\frac{\lambda_c^{(0)}}{4}\int_{-\infty}^{\infty}ds\Delta g^0(s_0)^2\right), \quad (13.5)$$

where $\tilde{F}_T$ is the minimum value of $F_T$ with respect to $\Delta\Phi_0(s)$, which is a functional of $\Delta g'^0(s_0)$ (the derivative of $\Delta g^0(s_0)$ with respect to its argument). Here, we have supposed that braid is sufficiently long that the limits of integration $-L/2$ and $L/2$ can be replaced with $-\infty$ and $\infty$. This functional dependence may be determined through Eq. (11.3). It is possible to reformulate Eq. (13.4) as a path integral over $\Delta\Phi_0(s)$ (see Appendix C), which may allow for an improvement on the approximation that will be used here, in later work. This approximation relies on a simpler approach, which has been seen to quite work well for two parallel molecules [16], and approximate a functional dependence of $\Delta\Phi_0(s)$ to $\Delta g^0(s)$ as

$$\Delta\Phi_0(s)\approx\Delta\bar{\Phi}+\frac{1}{2}\int_{-\infty}^{\infty}\frac{(s-s')}{|s-s'|}\Delta g(s')\exp\left(-\frac{|s-s'|}{\tilde{\lambda}_h}\right). \quad (13.6)$$

This choice becomes an exact solution to linearized version of Eq. (11.3); the linear response regime valid when $\lambda_c$ is sufficiently large. Now, $\tilde{\lambda}_h$ and $\Delta\bar{\Phi}$ are additional variational parameters chosen to minimize the average of a variational free energy when Eq. (13.6) is inserted into Eq. (10.26). This is allows for the approximation to come as close as possible to the exact average free energy. This approximation works better over smaller values of $\lambda_c^{(0)}$ than the linear response theory, more appropriate for DNA, since $\lambda_c^{(0)}\approx 150\text{Å}$ [14]. Using Eq. (13.6), the variational free energy (Eq. (10.26)) becomes

$$\frac{\tilde{F}_T}{k_B T} \approx \frac{L}{2}\left[\left(\frac{\alpha_x}{l_p}\right)^{1/2}+\left(\frac{\alpha_y}{l_p}\right)^{1/2}\right]+\frac{\alpha_\eta^{1/2}L}{2^{1/2}\tilde{l}_p^{1/2}}+\frac{d_R^2\alpha_H L}{2}+\frac{L}{4l_p\theta_R^2}+\frac{L}{4\lambda_h}+\frac{l_p\theta_R^4 L}{4d_R^2}-\frac{L}{16 l_p^{3/2}\cos\left(\frac{\eta_0}{2}\right)^4}\frac{1}{\left(\alpha_y^{1/2}+\alpha_x^{1/2}\right)}$$

$$-\frac{L\theta_R^2 l_p}{R_0^2}\tilde{f}_1(R_0,d_R,d_{\max},d_{\min})\sin^2\left(\frac{\eta_0}{2}\right)+\frac{4l_p L\tilde{f}_1(R_0,d_R,d_{\max},d_{\min})}{R_0^2}\sin^4\left(\frac{\eta_0}{2}\right)$$

$$+\frac{l_c}{16\tilde{\lambda}_h^2}\int_{-L/2}^{L/2}ds\int_{-L/2}^{L/2}ds'\int_{-L/2}^{L/2}ds''\frac{(s-s')}{|s-s'|}\frac{(s-s'')}{|s-s''|}\exp\left(-\frac{|s-s'|}{\tilde{\lambda}_h}\right)\exp\left(-\frac{|s-s''|}{\tilde{\lambda}_h}\right)\Delta g^0(s')\Delta g^0(s'')$$

$$+L\sum_{n=-\infty}^{\infty}\tilde{f}_{\mathrm{int}}^0(\eta_0,R_0,n,d_R,\theta_R)\exp\left(-\frac{n^2\lambda_h}{2l_c}\right)$$

$$-F_R L\cos\left(\frac{\eta_0}{2}\right)+\frac{2M_R L}{R_0}\sin\left(\frac{\eta_0}{2}\right)\tilde{f}_2(R_0,d_R,d_{\max},d_{\min})-\frac{M_R L}{4}\theta_R^2\frac{\tilde{f}_2(R_0,d_R,d_{\max},d_{\min})}{R_0}\sin\left(\frac{\eta_0}{2}\right)^{-1},$$

(13.7)

where Eq. (13.6) is substituted into the expressions for $\tilde{f}_{\mathrm{int}}^0(\eta_0,R_0,n,d_R,\theta_R)$, $\tilde{f}_{\mathrm{int}}^{\eta,\eta}(\eta_0,R_0,n,d_R,\theta_R)$ , $\tilde{f}_{\mathrm{int}}^{\eta',\eta'}(\eta_0,R_0,n,d_R,\theta_R)$, $\tilde{f}_{\mathrm{int}}^{x,x}(\eta_0,R_0,n,d_R,\theta_R)$ and $\tilde{f}_{\mathrm{int}}^{y,y}(\eta_0,R_0,n,d_R,\theta_R)$ (see Eqs. (10.11) and (10.12)) . We then can approximate the average value of the free energy, Eq. (13.4) as

$$\left\langle\frac{\tilde{F}_T}{k_B T}\right\rangle_{\Delta g} \approx \frac{L}{2}\left[\left(\frac{\langle\alpha_x\rangle_{\Delta g}}{l_p}\right)^{1/2}+\left(\frac{\langle\alpha_y\rangle_{\Delta g}}{l_p}\right)^{1/2}\right]+\frac{\langle\alpha_\eta\rangle_{\Delta g}^{1/2}L}{2^{1/2}\langle\tilde{l}_p\rangle_{\Delta g}^{1/2}}+\frac{d_R^2\alpha_H L}{2}+\frac{L}{4l_p\theta_R^2}+\frac{L}{4\lambda_h}+\frac{l_p\theta_R^4 L}{4d_R^2}$$

$$-\frac{M_R^2}{16 l_p^{3/2}}\frac{L}{\cos\left(\frac{\eta_0}{2}\right)^4}\frac{1}{\left(\langle\alpha_y\rangle_{\Delta g}^{1/2}+\langle\alpha_x\rangle_{\Delta g}^{1/2}\right)}-\frac{L\theta_R^2 l_p}{R_0^2}\tilde{f}_1(R_0,d_R,d_{\max},d_{\min})\sin^2\left(\frac{\eta_0}{2}\right)+\frac{l_c L}{8\tilde{\lambda}_h\lambda_c^{(0)}}$$

$$+\frac{4l_p L\tilde{f}_1(R_0,d_R,d_{\max},d_{\min})}{R_0^2}\sin^4\left(\frac{\eta_0}{2}\right)+L\sum_{n=-\infty}^{\infty}\tilde{p}_{\mathrm{int}}^0(\eta_0,R_0,n,d_R,\theta_R)\cos\left(n\Delta\bar{\Phi}\right)\exp\left(-n^2\left(\frac{\lambda_h}{2l_c}+\frac{\tilde{\lambda}_h}{4\lambda_c}\right)\right)$$

$$-F_R L\cos\left(\frac{\eta_0}{2}\right)+\frac{2M_R L}{R_0}\sin\left(\frac{\eta_0}{2}\right)\tilde{f}_2(R_0,d_R,d_{\max},d_{\min})-\frac{M_R L}{4}\theta_R^2\frac{\tilde{f}_2(R_0,d_R,d_{\max},d_{\min})}{R_0}\sin\left(\frac{\eta_0}{2}\right)^{-1},$$

(13.8)

where we have made use of the fact that

$$\langle\exp in\Delta\Phi_0(s_0)\rangle_{\Delta g}$$

$$=\frac{1}{Z_{\Delta g}}\int D\Delta g(s_0)\exp\left(-\frac{\lambda_c^{(0)}}{4}\int_{-\infty}^{\infty}ds_0\left(\Delta g(s_0)-\frac{i(s-s')}{\lambda_c^{(0)}|s-s'|}\exp\left(-\frac{|s-s'|}{\tilde{\lambda}_h}\right)\right)^2\right)$$

$$\exp\left(-\frac{1}{4\lambda_c^{(0)}}\int_{-\infty}^{\infty}ds_0\exp\left(-\frac{2|s-s'|}{\tilde{\lambda}_h}\right)\right)=\exp\left(-\frac{1}{4\lambda_c^{(0)}}\int_{-\infty}^{\infty}ds_0\exp\left(-\frac{2|s-s'|}{\tilde{\lambda}_h}\right)\right)=\exp\left(-\frac{\tilde{\lambda}_h}{4\lambda_c^{(0)}}\right),$$

(13.9)

as well as Eq. (13.2). We have also supposed that the interaction potential for cylindrical cross section of the DNA is symmetric about the minor groove, so that terms in the interaction are $n \to -n$ symmetric. An expression for $\tilde{p}^{0}_{\rm int}(\eta_0, R_0, n, d_R, \theta_R)$ is given by Eq. (11.4). The averages $\langle \alpha_x \rangle_{\Delta g}$, $\langle \alpha_y \rangle_{\Delta g}$, $\langle \alpha_\eta \rangle_{\Delta g}$ and $\langle \tilde{l}_p \rangle_{\Delta g}$ are given by the expressions

$$\langle \alpha_x \rangle_{\Delta g} = \frac{1}{2}\left( \sum_{n=-\infty}^{\infty} \tilde{p}^{x,x}_{\rm int}(\eta_0, R_0, n, d_R, \theta_R)\cos\left(n\Delta\bar{\Phi}\right)\exp\left(-n^2\left(\frac{\lambda_h}{2l_c}+\frac{\tilde{\lambda}_h}{4\lambda_c^{(0)}}\right)\right) + \frac{F_R}{\cos\left(\frac{\eta_0}{2}\right)}\right),$$

(13.10)

$$\langle \alpha_y \rangle_{\Delta g} = \frac{1}{2}\left( \sum_{n=-\infty}^{\infty} \tilde{p}^{y,y}_{\rm int}(\eta_0, R_0, n, d_R, \theta_R)\cos\left(n\Delta\bar{\Phi}\right)\exp\left(-n^2\left(\frac{\lambda_h}{2l_c}+\frac{\tilde{\lambda}_h}{4\lambda_c^{(0)}}\right)\right) + \frac{F_R}{\cos\left(\frac{\eta_0}{2}\right)}\right),$$

(13.11)

$$\langle \alpha_\eta \rangle_{\Delta g} = \left[ \frac{4l_p \tilde{f}_1(R_0, d_R, d_{\max}, d_{\min})}{R_0^2}\left(3\cos^2\left(\frac{\eta_0}{2}\right)\sin^2\left(\frac{\eta_0}{2}\right) - \sin^4\left(\frac{\eta_0}{2}\right)\right) + \frac{F_R}{4}\cos\left(\frac{\eta_0}{2}\right) \right.$$
$$\left. + \sum_{n=-\infty}^{\infty} \tilde{g}^{\eta,\eta}_{\rm int}(\eta_0, R_0, n, d_R, \theta_R)\cos\left(n\Delta\bar{\Phi}\right)\exp\left(-n^2\left(\frac{\lambda_h}{2l_c}+\frac{\tilde{\lambda}_h}{4\lambda_c^{(0)}}\right)\right) - \frac{M_R \tilde{f}_2(R_0, d_R, d_{\max}, d_{\min})}{2R_0}\sin\left(\frac{\eta_0}{2}\right)\right],$$

(13.12)

and

$$\langle \tilde{l}_p \rangle_{\Delta g} = l_p + 2\sum_{n=-\infty}^{\infty} \tilde{g}^{\eta',\eta'}_{\rm int}(\eta_0, R_0, n, d_R, \theta_R)\cos\left(n\Delta\bar{\Phi}\right)\exp\left(-n^2\left(\frac{\lambda_h}{2l_c}+\frac{\tilde{\lambda}_h}{4\lambda_c}\right)\right), \quad (13.13)$$

where $\tilde{p}^{x,x}_{\rm int}(\eta_0, R_0, n, d_R, \theta_R)$, $\tilde{p}^{y,y}_{\rm int}(\eta_0, R_0, n, d_R, \theta_R)$, and so on, are given by Eq. (11.5). Here, we have made two additional approximations. The first is that, on averaging, $\alpha_x$, $\alpha_y$, $\alpha_\eta$, $\tilde{l}_p$ can be approximated by their averages, which is valid provided that corrections to them from the interaction energy is sufficiently small. The second is the replacement of $g_1^0(s)$ and $g_2^0(s)$ with $\bar{g}_0$, which is valid provided that $\bar{g}_0\lambda_c^{(0)} \gg 1$, this approximation is indeed used in [3] and [8]. Following [2], it is convenient to change variables to

$$\frac{\lambda_h^*}{2\lambda_c} = \frac{\lambda_h}{2l_c} + \frac{\tilde{\lambda}_h}{4\lambda_c^{(0)}}, \qquad \frac{\tilde{\lambda}_h^*}{2\lambda_c} = \frac{\lambda_h}{2l_c} - \frac{\tilde{\lambda}_h}{4\lambda_c^{(0)}}, \quad (13.14)$$

where we define the combined helical coherence length

$$\frac{1}{\lambda_c}=\frac{1}{\lambda_c^{(0)}}+\frac{1}{l_c}. \tag{13.15}$$

Eqs. (13.14) and (13.15) allow us to rewrite the Free energy as

$$\begin{aligned}\left\langle\frac{\tilde{F}_T}{k_BT}\right\rangle_{\Delta g}\approx&\frac{L}{2}\left[\left(\frac{\langle\alpha_x\rangle_{\Delta g}}{l_p}\right)^{1/2}+\left(\frac{\langle\alpha_y\rangle_{\Delta g}}{l_p}\right)^{1/2}\right]+\frac{\langle\alpha_\eta\rangle_{\Delta g}^{1/2}L}{2^{1/2}\left\langle\tilde{l}_p\right\rangle_{\Delta g}^{1/2}}+\frac{d_R^2\alpha_H L}{2}+\frac{L}{4l_p\theta_R^2}+\frac{l_p\theta_R^4L}{4d_R^2}+\frac{L\lambda_c}{2l_c\left(\lambda_h^*+\tilde{\lambda}_h^*\right)}\\&-\frac{M_R^2}{16l_p^{3/2}}\frac{L}{\cos\left(\frac{\eta_0}{2}\right)^4}\frac{1}{\left(\langle\alpha_y\rangle_{\Delta g}^{1/2}+\langle\alpha_x\rangle_{\Delta g}^{1/2}\right)}-\frac{L\theta_R^2l_p}{R_0^2}\tilde{f}_1(R_0,d_R,d_{\max},d_{\min})\sin^2\left(\frac{\eta_0}{2}\right)+\frac{L\lambda_cl_c}{8\left(\lambda_c^{(0)}\right)^2\left(\lambda_h^*-\tilde{\lambda}_h^*\right)}\\&+\frac{4l_pL\tilde{f}_1(R_0,d_R,d_{\max},d_{\min})}{R_0^2}\sin^4\left(\frac{\eta_0}{2}\right)+L\sum_{n=-\infty}^{\infty}\tilde{p}_{\mathrm{int}}^0(\eta_0,R_0,n,d_R,\theta_R)\cos\left(n\Delta\bar{\Phi}\right)\exp\left(-\frac{n^2\lambda_h^*}{2\lambda_c}\right)\\&-F_RL\cos\left(\frac{\eta_0}{2}\right)+\frac{2M_RL}{R_0}\sin\left(\frac{\eta_0}{2}\right)\tilde{f}_2(R_0,d_R,d_{\max},d_{\min})-\frac{M_RL\theta_R^2}{4}\frac{\tilde{f}_2(R_0,d_R,d_{\max},d_{\min})}{R_0}\sin\left(\frac{\eta_0}{2}\right)^{-1}.\end{aligned}$$

(13.16)

Minimization of the average free energy with respect $\tilde{\lambda}_h^*$ then yields the equation

$$\frac{\lambda_c}{8\left(\lambda_c^{(0)}\right)^2\left(\lambda_h^*-\tilde{\lambda}_h^*\right)^2}=\frac{\lambda_c}{2(l_c)^2\left(\lambda_h^*+\tilde{\lambda}_h^*\right)^2}. \tag{13.17}$$

Using Eq. (13.14) to rewrite Eq. (13.17) back in terms of $\lambda_h$ and $\tilde{\lambda}_h$, we see that solution of Eq.(13.17) is simply $\tilde{\lambda}_h=\lambda_h$, which corresponds to

$$\tilde{\lambda}_h^*=\lambda_h^*\left(\frac{2\lambda_c^{(0)}-l_c}{2\lambda_c^{(0)}+l_c}\right). \tag{13.18}$$

On substitution of Eq. (13.18) back into Eq. (13.16) , and on rearrangement, we obtain

$$\left\langle \frac{\tilde{F}_T}{k_B T} \right\rangle_{\Delta g} \approx \frac{L}{2}\left[\left(\frac{\langle \alpha_x \rangle_{\Delta g}}{l_p}\right)^{1/2} + \left(\frac{\langle \alpha_y \rangle_{\Delta g}}{l_p}\right)^{1/2}\right] + \frac{\langle \alpha_\eta \rangle_{\Delta g}^{1/2} L}{2^{1/2} \langle \tilde{l}_p \rangle_{\Delta g}^{1/2}} + \frac{d_R^2 \alpha_H L}{2} + \frac{L}{4 l_p \theta_R^2} + \frac{l_p \theta_R^4 L}{4 d_R^2}$$
$$-\frac{L\theta_R^2 l_p}{R_0^2} \tilde{f}_1(R_0, d_R, d_{\max}, d_{\min}) \sin^2\left(\frac{\eta_0}{2}\right) + \frac{L(l_c+\lambda_c)^2}{16\lambda_h^* \lambda_c l_c} - \frac{M_R^2}{8 l_p^{3/2}} \frac{L}{\cos\left(\frac{\eta_0}{2}\right)^4} \frac{1}{\left(\langle \alpha_y \rangle_{\Delta g}^{1/2} + \langle \alpha_x \rangle_{\Delta g}^{1/2}\right)}$$
$$+\frac{4 l_p L \tilde{f}_1(R_0, d_R, d_{\max}, d_{\min})}{R_0^2} \sin^4\left(\frac{\eta_0}{2}\right) + L \sum_{n=-\infty}^{\infty} \tilde{p}_{\text{int}}^0(\eta_0, R_0, n, d_R, \theta_R) \cos\left(n\Delta\bar{\Phi}\right) \exp\left(-n^2 \frac{\lambda_h^*}{2\lambda_c}\right)$$
$$-F_R L \cos\left(\frac{\eta_0}{2}\right) + \frac{2M_R L}{R_0} \sin\left(\frac{\eta_0}{2}\right) \tilde{f}_2(R_0, d_R, d_{\max}, d_{\min}) - \frac{M_R L \theta_R^2}{4} \frac{\tilde{f}_2(R_0, d_R, d_{\max}, d_{\min})}{R_0} \sin\left(\frac{\eta_0}{2}\right)^{-1}.$$

(13.19)

The minimization of the average free energy (Eq. (13.19)) with respect to $\Delta\bar{\Phi}$ yields

$$0 = -\sum_{n=-\infty}^{\infty} n \sin\left(n\Delta\bar{\Phi}\right) \exp\left(-\frac{n^2 \lambda_h^*}{2\lambda_c}\right) \left[\left(\frac{1}{8} + \frac{M_R^2}{64 l_p \left(\langle \alpha_x \rangle_{\Delta g} + \langle \alpha_y \rangle_{\Delta g}\right)^2}\right)\right.$$
$$\left(\left(\frac{1}{l_p \langle \alpha_x \rangle_{\Delta g}}\right)^{1/2} \tilde{p}_{\text{int}}^{x,x}(\eta_0, R_0, n, d_R, \theta_R) + \left(\frac{1}{l_p \langle \alpha_y \rangle_{\Delta g}}\right)^{1/2} \tilde{p}_{\text{int}}^{y,y}(\eta_0, R_0, n, d_R, \theta_R)\right)$$
$$\left. + \frac{L}{2^{3/2} \langle \tilde{l}_p \rangle_{\Delta g}^{1/2} \langle \alpha_\eta \rangle_{\Delta g}^{1/2}} \tilde{p}_{\text{int}}^{\eta,\eta}(\eta_0, R_0, n, d_R, \theta_R) - \frac{\langle \alpha_\eta \rangle_{\Delta g}^{1/2} L}{2^{1/2} \langle \tilde{l}_p \rangle_{\Delta g}^{3/2}} \tilde{p}_{\text{int}}^{\eta',\eta'}(\eta_0, R_0, n, d_R, \theta_R) + \tilde{p}_{\text{int}}^0(\eta_0, R_0, n, d_R, \theta_R)\right].$$

(13.20)

On minimization with respect $R_0$, $\eta_0$, $d_R$, $\theta_R$, $\lambda_h^*$, Eqs. (11.10), (11.11), (11.20), (11.21) and (11.25) are modified to be

$$0 = \frac{1}{8} \sum_{n=-\infty}^{\infty} \cos\left(n\Delta\bar{\Phi}\right) \exp\left(-\frac{n^2 \lambda_h^*}{2\lambda_c}\right)$$
$$\left[\left(\frac{1}{\langle \alpha_x \rangle_{\Delta g} l_p}\right)^{1/2} \frac{\partial \tilde{p}_{\text{int}}^{x,x}(\eta_0, R_0, n, d_R, \theta_R)}{\partial R_0} + \left(\frac{1}{\langle \alpha_y \rangle_{\Delta g} l_p}\right)^{1/2} \frac{\partial \tilde{p}_{\text{int}}^{y,y}(\eta_0, R_0, n, d_R, \theta_R)}{\partial R_0}\right]$$
$$+\frac{1}{\langle \alpha_\eta \rangle_{\Delta g}^{1/2} 2^{3/2} \langle \tilde{l}_p \rangle_{\Delta g}^{1/2}} \frac{\partial \langle \alpha_\eta \rangle_{\Delta g}}{\partial R_0} + \frac{d_R^2}{2} \frac{\partial \alpha_H}{\partial R_0} - \frac{\langle \alpha_\eta \rangle_{\Delta g}^{1/2}}{2^{1/2} \langle \tilde{l}_p \rangle_{\Delta g}^{3/2}} \sum_{n=-\infty}^{\infty} \frac{\partial \tilde{p}_{\text{int}}^{\eta',\eta'}(\eta_0, R_0, n, d_R, \theta_R)}{\partial R_0} \cos\left(n\Delta\bar{\Phi}\right) \exp\left(-\frac{n^2 \lambda_h^*}{2\lambda_c}\right)$$
$$+\frac{l_p \theta_R^2}{R_0^3} \tilde{h}_1(R_0, d_R, d_{\max}, d_{\min}) \sin^2\left(\frac{\eta_0}{2}\right) - \frac{4 l_p \tilde{h}_1(R_0, d_R, d_{\max}, d_{\min})}{R_0^3} \sin^4\left(\frac{\eta_0}{2}\right)$$

$$+\sum_{n=-\infty}^{\infty}\frac{\partial\tilde{p}_{\text{int}}^{0}(\eta_0,R_0,n,d_R,\theta_R)}{\partial R_0}\cos\left(n\Delta\bar{\Phi}\right)\exp\left(-\frac{n^2\lambda_h^*}{2\lambda_c}\right)-\frac{2M_R}{R_0^2}\sin\left(\frac{\eta_0}{2}\right)\tilde{h}_2(R_0,d_R,d_{\max},d_{\min})$$

$$+\frac{M_R}{4}\theta_R^2\frac{\tilde{h}_2(R_0,d_R,d_{\max},d_{\min})}{R_0^2}\sin\left(\frac{\eta_0}{2}\right)^{-1}+\frac{M_R^2}{64l_p^{3/2}}\frac{1}{\cos\left(\frac{\eta_0}{2}\right)^4}\frac{1}{\left(\left\langle\alpha_y\right\rangle_{\Delta g}^{1/2}+\left\langle\alpha_x\right\rangle_{\Delta g}^{1/2}\right)^2}$$

$$\sum_{n=-\infty}^{\infty}\cos\left(n\Delta\bar{\Phi}\right)\exp\left(-\frac{n^2\lambda_h^*}{2\lambda_c}\right)\left(\frac{1}{\left\langle\alpha_y\right\rangle_{\Delta g}^{1/2}}\frac{\partial\tilde{p}_{\text{int}}^{y,y}(\eta_0,R_0,n,d_R,\theta_R)}{\partial R_0}+\frac{1}{\left\langle\alpha_x\right\rangle_{\Delta g}^{1/2}}\frac{\partial\tilde{p}_{\text{int}}^{x,x}(\eta_0,R_0,n,d_R,\theta_R)}{\partial R_0}\right)$$

(13.21)

$$0=\frac{1}{16}\left[\left(\frac{1}{\left\langle\alpha_x\right\rangle_{\Delta g}l_p}\right)^{1/2}+\left(\frac{1}{\left\langle\alpha_y\right\rangle_{\Delta g}l_p}\right)^{1/2}\right]\frac{F_R\sin\left(\frac{\eta_0}{2}\right)}{\cos\left(\frac{\eta_0}{2}\right)^2}+\frac{d_R^2}{2}\frac{\partial\alpha_H}{\partial\eta_0}+\frac{1}{\left\langle\alpha_\eta\right\rangle_{\Delta g}^{1/2}2^{3/2}\left\langle\tilde{l}_p\right\rangle_{\Delta g}^{1/2}}\frac{\partial\left\langle\alpha_\eta\right\rangle_{\Delta g}}{\partial\eta_0}+\frac{F_R}{2}\sin\left(\frac{\eta_0}{2}\right)$$

$$+\frac{1}{8}\sum_{n=-\infty}^{\infty}\cos\left(n\Delta\bar{\Phi}\right)\exp\left(-\frac{n^2\lambda_h^*}{2\lambda_c}\right)\left[\left(\frac{1}{\left\langle\alpha_x\right\rangle_{\Delta g}l_p}\right)^{1/2}\frac{\partial\tilde{p}_{\text{int}}^{x,x}(\eta_0,R_0,n,d_R,\theta_R)}{\partial\eta_0}+\left(\frac{1}{\left\langle\alpha_y\right\rangle_{\Delta g}l_p}\right)^{1/2}\frac{\partial\tilde{p}_{\text{int}}^{x,x}(\eta_0,R_0,n,d_R,\theta_R)}{\partial\eta_0}\right]$$

$$-\frac{\left\langle\alpha_\eta\right\rangle_{\Delta g}^{1/2}}{2^{1/2}\left\langle\tilde{l}_p\right\rangle_{\Delta g}^{3/2}}\sum_{n=-\infty}^{\infty}\frac{\partial\tilde{p}_{\text{int}}^{\eta',\eta'}(\eta_0,R_0,n,d_R,\theta_R)}{\partial\eta_0}\cos\left(n\Delta\bar{\Phi}\right)\exp\left(-\frac{n^2\lambda_h^*}{2\lambda_c}\right)$$

$$-\frac{l_p\theta_R^2}{R_0^2}\tilde{f}_1(R_0,d_R,d_{\max},d_{\min})\sin\left(\frac{\eta_0}{2}\right)\cos\left(\frac{\eta_0}{2}\right)+\frac{8l_p\tilde{f}_1(R_0,d_R,d_{\max},d_{\min})}{R_0^2}\cos\left(\frac{\eta_0}{2}\right)\sin^3\left(\frac{\eta_0}{2}\right)$$

$$+\sum_{n=-\infty}^{\infty}\frac{\partial\tilde{p}_{\text{int}}^{0}(\eta_0,R_0,n,d_R,\theta_R)}{\partial\eta_0}\cos\left(n\Delta\bar{\Phi}\right)\exp\left(-\frac{n^2\lambda_h^*}{2\lambda_c}\right)+\frac{M_R}{R_0}\cos\left(\frac{\eta_0}{2}\right)\tilde{f}_2(R_0,d_R,d_{\max},d_{\min})$$

$$+\frac{M_R\theta_R^2}{8}\frac{\tilde{f}_2(R_0,d_R,d_{\max},d_{\min})}{R_0}\cos\left(\frac{\eta_0}{2}\right)\sin\left(\frac{\eta_0}{2}\right)^{-2}+\frac{M_R^2}{128l_p^{3/2}}\frac{F_R\sin\left(\frac{\eta_0}{2}\right)}{\cos\left(\frac{\eta_0}{2}\right)^6}\frac{1}{\left(\left\langle\alpha_y\right\rangle_{\Delta g}^{1/2}+\left\langle\alpha_x\right\rangle_{\Delta g}^{1/2}\right)\left\langle\alpha_y\right\rangle_{\Delta g}^{1/2}\left\langle\alpha_x\right\rangle_{\Delta g}^{1/2}}$$

$$+\frac{M_R^2}{64l_p^{3/2}}\frac{1}{\cos\left(\frac{\eta_0}{2}\right)^4\left(\left\langle\alpha_y\right\rangle_{\Delta g}^{1/2}+\left\langle\alpha_x\right\rangle_{\Delta g}^{1/2}\right)^2}\sum_{n=-\infty}^{\infty}\cos\left(\Delta\bar{\Phi}\right)\exp\left(-\frac{n^2\lambda_h^*}{2\lambda_c}\right)\left(\frac{1}{\left\langle\alpha_y\right\rangle_{\Delta g}^{1/2}}\frac{\partial\tilde{p}_{\text{int}}^{y,y}(\eta_0,R_0,n,d_R,\theta_R)}{\partial\eta_0}\right.$$

$$\left.+\frac{1}{\left\langle\alpha_x\right\rangle_{\Delta g}^{1/2}}\frac{\partial\tilde{p}_{\text{int}}^{x,x}(\eta_0,R_0,n,d_R,\theta_R)}{\partial\eta_0}\right)-\frac{M_R^2}{8l_p^{3/2}}\frac{1}{\cos\left(\frac{\eta_0}{2}\right)^5\left(\left\langle\alpha_y\right\rangle_{\Delta g}^{1/2}+\left\langle\alpha_x\right\rangle_{\Delta g}^{1/2}\right)},$$

(13.22)

$$0=-\frac{1}{2l_p\theta_R^3}+\frac{l_p\theta_R^3}{4d_R^2}-\frac{2l_p\theta_R}{R_0^2}\tilde{f}_1(R_0,d_R,d_{\max},d_{\min})\sin^2\left(\frac{\eta_0}{2}\right)-\frac{M_R\theta_R}{2}\frac{\tilde{f}_2(R_0,d_R,d_{\max},d_{\min})}{R_0}\sin\left(\frac{\eta_0}{2}\right)^{-1}$$
$$+\sum_{n=-\infty}^{\infty}\cos\left(n\Delta\bar{\Phi}\right)\exp\left(-\frac{n^2\lambda_h^*}{2\lambda_c}\right)\left(\frac{1}{2^{3/2}\left\langle\alpha_\eta\right\rangle_{\Delta g}^{1/2}\left\langle\tilde{l}_p\right\rangle_{\Delta g}^{1/2}}\frac{\partial\tilde{p}_{\text{int}}^{\eta,\eta}(\eta_0,R_0,n,d_R,\theta_R)}{\partial\theta_R}-\frac{\left\langle\alpha_\eta\right\rangle_{\Delta g}^{1/2}}{2^{1/2}\left\langle\tilde{l}_p\right\rangle_{\Delta g}^{3/2}}\frac{\partial\tilde{p}_{\text{int}}^{\eta',\eta'}(\eta_0,R_0,n,d_R,\theta_R)}{\partial\theta_R}\right)$$
$$+\frac{1}{8}\sum_{n=-\infty}^{\infty}\cos\left(n\Delta\bar{\Phi}\right)\exp\left(-\frac{n^2\lambda_h^*}{2\lambda_c}\right)\left[\left(\frac{1}{\left\langle\alpha_x\right\rangle_{\Delta g}l_p}\right)^{1/2}\frac{\partial\tilde{p}_{\text{int}}^{x,x}(\eta_0,R_0,n,d_R,\theta_R)}{\partial\theta_R}+\left(\frac{1}{\left\langle\alpha_y\right\rangle_{\Delta g}l_p}\right)^{1/2}\frac{\partial\tilde{p}_{\text{int}}^{y,y}(\eta_0,R_0,n,d_R,\theta_R)}{\partial\theta_R}\right]$$
$$+\sum_{n=-\infty}^{\infty}\frac{\partial\tilde{p}_{\text{int}}^{0}(\eta_0,R_0,n,d_R,\theta_R)}{\partial\theta_R}\cos\left(n\Delta\bar{\Phi}\right)\exp\left(-\frac{n^2\lambda_h^*}{2\lambda_c}\right)$$
$$+\frac{M_R^2}{64l_p^{3/2}}\frac{1}{\cos\left(\frac{\eta_0}{2}\right)^4}\frac{1}{\left(\left\langle\alpha_y\right\rangle_{\Delta g}^{1/2}+\left\langle\alpha_x\right\rangle_{\Delta g}^{1/2}\right)^2}\sum_{n=-\infty}^{\infty}\cos\left(n\Delta\bar{\Phi}\right)\exp\left(-\frac{n^2\lambda_h^*}{2\lambda_c}\right)\left(\frac{1}{\left\langle\alpha_y\right\rangle_{\Delta g}^{1/2}}\frac{\partial\tilde{p}_{\text{int}}^{y,y}(\eta_0,R_0,n,d_R,\theta_R)}{\partial\theta_R}\right.$$
$$\left.+\frac{1}{\left\langle\alpha_x\right\rangle_{\Delta g}^{1/2}}\frac{\partial\tilde{p}_{\text{int}}^{x,x}(\eta_0,R_0,n,d_R,\theta_R)}{\partial\theta_R}\right),$$

(13.23)

$$0=\frac{1}{2^{3/2}\left\langle\alpha_\eta\right\rangle_{\Delta g}^{1/2}\left\langle\tilde{l}_p\right\rangle_{\Delta g}^{1/2}}\frac{\partial\left\langle\alpha_\eta\right\rangle_{\Delta g}}{\partial d_R}-\frac{\left\langle\alpha_\eta\right\rangle_{\Delta g}^{1/2}}{2^{1/2}\left\langle\tilde{l}_p\right\rangle_{\Delta g}^{3/2}}\sum_{n=-\infty}^{\infty}\frac{\partial\tilde{p}_{\text{int}}^{\eta',\eta'}(\eta_0,R_0,n,d_R,\theta_R)}{\partial d_R}\cos\left(n\Delta\bar{\Phi}\right)\exp\left(-\frac{n^2\lambda_h^*}{2\lambda_c}\right)$$
$$+\frac{1}{8}\sum_{n=-\infty}^{\infty}\cos\left(n\Delta\bar{\Phi}\right)\exp\left(-\frac{n^2\lambda_h^*}{2\lambda_c}\right)\left[\left(\frac{1}{\left\langle\alpha_x\right\rangle_{\Delta g}l_p}\right)^{1/2}\frac{\partial\tilde{p}_{\text{int}}^{x,x}(\eta_0,R_0,n,d_R,\theta_R)}{\partial d_R}+\left(\frac{1}{\left\langle\alpha_y\right\rangle_{\Delta g}l_p}\right)^{1/2}\frac{\partial\tilde{p}_{\text{int}}^{y,y}(\eta_0,R_0,n,d_R,\theta_R)}{\partial d_R}\right]$$
$$-\frac{l_p\theta_R^2}{d_RR_0^2}\tilde{l}_1(R_0,d_R,d_{\max},d_{\min})\sin^2\left(\frac{\eta_0}{2}\right)+d_R\alpha_H-\frac{l_p\theta_R^4}{2d_R^3}$$
$$+\frac{4l_p}{d_RR_0^2}\tilde{l}_1(R_0,d_R,d_{\max},d_{\min})\sin^4\left(\frac{\eta_0}{2}\right)+\sum_{n=-\infty}^{\infty}\frac{\partial\tilde{p}_{\text{int}}^{0}(\eta_0,R_0,n,d_R,\theta_R)}{\partial d_R}\cos\left(n\Delta\bar{\Phi}\right)\exp\left(-\frac{n^2\lambda_h^*}{2\lambda_c}\right)$$
$$+\frac{2M_R}{d_RR_0}\tilde{l}_2(R_0,d_R,d_{\max},d_{\min})\sin\left(\frac{\eta_0}{2}\right)-\frac{M_R\theta_R^2}{4d_RR_0}\tilde{l}_2(R_0,d_R,d_{\max},d_{\min})\sin\left(\frac{\eta_0}{2}\right)^{-1}$$
$$+\frac{M_R^2}{64l_p^{3/2}}\frac{1}{\cos\left(\frac{\eta_0}{2}\right)^4}\frac{1}{\left(\left\langle\alpha_y\right\rangle_{\Delta g}^{1/2}+\left\langle\alpha_x\right\rangle_{\Delta g}^{1/2}\right)^2}\sum_{n=-\infty}^{\infty}\cos\left(n\Delta\bar{\Phi}\right)\exp\left(-\frac{n^2\lambda_h^*}{2\lambda_c}\right)\left(\frac{1}{\left\langle\alpha_y\right\rangle_{\Delta g}^{1/2}}\frac{\partial\tilde{p}_{\text{int}}^{y,y}(\eta_0,R_0,n,d_R,\theta_R)}{\partial d_R}\right.$$
$$\left.+\frac{1}{\left\langle\alpha_x\right\rangle_{\Delta g}^{1/2}}\frac{\partial\tilde{p}_{\text{int}}^{x,x}(\eta_0,R_0,n,d_R,\theta_R)}{\partial d_R}\right),$$

(13.24)

and

$$
\begin{aligned}
0=&-\sum_{n=-\infty}^{\infty}\frac{n^2}{2\lambda_c}\cos\left(n\Delta\bar{\Phi}\right)\exp\left(-\frac{n^2\lambda_h^*}{2\lambda_c}\right)\Bigg(\tilde{p}^{0}_{\rm int}(\eta_0,R_0,n,d_R,\theta_R)-\frac{1}{\langle\alpha_\eta\rangle_{\Delta g}^{1/2}2^{3/2}\langle\tilde{l}_p\rangle_{\Delta g}^{1/2}}\tilde{p}^{\eta,\eta}_{\rm int}(\eta_0,R_0,n,d_R,\theta_R)\\
&-\frac{\langle\alpha_\eta\rangle_{\Delta g}^{1/2}}{2^{1/2}\langle\tilde{l}_p\rangle_{\Delta g}^{3/2}}\tilde{p}^{\eta',\eta'}_{\rm int}(\eta_0,R_0,n,d_R,\theta_R)+\frac{1}{8}\left(\frac{1}{\langle\alpha_x\rangle_{\Delta g}^{1/2}l_p}\right)^{1/2}\tilde{p}^{x,x}_{\rm int}(\eta_0,R_0,n,d_R,\theta_R)+\frac{1}{8}\left(\frac{1}{\langle\alpha_y\rangle_{\Delta g}^{1/2}l_p}\right)^{1/2}\tilde{p}^{x,x}_{\rm int}(\eta_0,R_0,n,d_R,\theta_R)\Bigg)\\
&-\frac{M^2}{64l_p^{3/2}}\frac{1}{\cos\left(\frac{\eta_0}{2}\right)^4}\frac{1}{\left(\langle\alpha_x\rangle_{\Delta g}^{1/2}+\langle\alpha_y\rangle_{\Delta g}^{1/2}\right)^2}\sum_{n=-\infty}^{\infty}\frac{n^2}{2\lambda_c}\cos\left(n\Delta\bar{\Phi}\right)\exp\left(-\frac{n^2\lambda_h^*}{2\lambda_c}\right)\Bigg(\frac{1}{\langle\alpha_y\rangle_{\Delta g}^{1/2}}\tilde{p}^{y,y}_{\rm int}(\eta_0,R_0,n,d_R,\theta_R)\\
&+\frac{1}{\langle\alpha_x\rangle_{\Delta g}^{1/2}}\tilde{p}^{x,x}_{\rm int}(\eta_0,R_0,n,d_R,\theta_R)\Bigg)-\frac{\left(l_c+\lambda_c\right)^2}{16\lambda_c l_c(\lambda_h^*)^2}.
\end{aligned}
$$

(13.25)

### *13.2 Including interaction terms for the KL theory*

We will now examine what $\tilde{g}^{0}_{\rm int}(\eta_0,R_0,n,d_R,\theta_R)$ $\tilde{g}^{\eta,\eta}_{\rm int}(\eta_0,R_0,n,d_R,\theta_R)$ should be in terms of the KL theory [12]. We neglect $\tilde{g}^{\eta',\eta'}_{\rm int}(\eta_0,R_0,n,d_R,\theta_R), \tilde{g}^{x,x}_{\rm int}(\eta_0,R_0,n,d_R,\theta_R)$ and $\tilde{g}^{y,y}_{\rm int}(\eta_0,R_0,n,d_R,\theta_R)$, which should be valid sufficiently large enough values $l_p$ and pulling force $F$. Following the approach given in [5] these corrections may be calculated for the braid. For the KL theory we have

$$
\begin{aligned}
\bar{\varepsilon}^{0}_{\rm int}(\eta_0,R(s_0),0,R'(s_0),\bar{g}_0,\bar{g}_0)\approx&\frac{2l_B(1-\theta_c)^2}{l_e^2}\frac{K_0(\kappa_D R(s_0))}{(a\kappa_D)^2K_1(a\kappa_D)^2}\\
&-\frac{2l_B}{l_e^2}\sum_{n=-\infty}^{\infty}\sum_{j=-\infty}^{\infty}\xi_n(\theta_c,f_1,f_2)\frac{K_{n-j}(\kappa_n R(s_0))K_{n-j}(\kappa_n R(s_0))}{(a\kappa_n)^2K_n'(a\kappa_n)^2}\frac{K_j'(a\kappa_n)}{I_j'(a\kappa_n)},
\end{aligned}
\quad (13.26)
$$

$$
\bar{\varepsilon}^{0}_{\rm int}(\eta_0,R(s_0),n,R'(s_0),\bar{g}_0,\bar{g}_0)\approx\frac{2l_B}{l_e^2}\frac{(-1)^n\xi_n(\theta_c,f_1,f_2)^2}{(a\kappa_n)^2K_n'(a\kappa_n)^2}\left(K_0(\kappa_n R(s_0))+\frac{n^2\bar{g}_0}{\kappa_n}K_1(\kappa_n R(s_0))\sin\eta_0\right),
$$

(13.27)

$$
\bar{\varepsilon}^{\eta,\eta}_{\rm int}(\eta_0,R(s_0),n,R'(s_0),\bar{g}_0,\bar{g}_0)\approx-\frac{2n^2\bar{g}_0 l_B}{\kappa_n l_e^2}\frac{(-1)^n\xi_n(\theta_c,f_1,f_2)^2}{(a\kappa_n)^2K_n'(a\kappa_n)^2}K_1(\kappa_n R(s_0))\sin\eta_0.
$$

(13.28)

for $n=-2,-1,1,2$, and

$$
\kappa_n=\sqrt{\kappa_D^2+n^2\bar{g}_0^2},\qquad \xi_n(\theta_c,f_1,f_2)=\theta_c\left(1-f_1-f_2\right)\delta_{n,0}+\theta_c f_1+\left(-1\right)^n\theta_c f_2-\cos\left(n\tilde{\phi}_s\right).\quad (13.29)
$$

We neglect $|n|>2$ in the sums in the free energy and the equations that minimize it, as these are found to be small. Here, the parameters $f_1$ and $f_2$ are the effective proportions of ions localized in the minor and major grooves of the DNA. The angle $\tilde{\phi}_s \approx 0.4\pi$ is the half width of the minor groove of the DNA molecule.

Eqs. (13.26), (13.27) and (13.28) allow us to recast Eq (13.19) as

$$\left\langle \frac{\tilde{F}_T}{k_B T} \right\rangle_{\Delta g} \approx L\left(\frac{F_R}{2l_p\cos(\eta_0/2)}\right)^{1/2} + \frac{\langle\alpha_\eta\rangle_{\Delta g}^{1/2} L}{2^{1/2} l_p} + \frac{d_R^2\alpha_H L}{2} + \frac{L}{4l_p\theta_R^2} + \frac{l_p\theta_R^4 L}{4d_R^2} - \frac{M_R^2}{8l_p^{3/2}} \frac{L}{\cos\left(\frac{\eta_0}{2}\right)^4} \frac{1}{\left(\langle\alpha_y\rangle_{\Delta g}^{1/2} + \langle\alpha_x\rangle_{\Delta g}^{1/2}\right)}$$
$$-\frac{Ll_p\theta_R^2}{R_0^2}\tilde{f}_1(R_0,d_R,d_{\max},d_{\min})\sin^2\left(\frac{\eta_0}{2}\right) + \frac{L(l_c+\lambda_c)^2}{16\lambda_h^*\lambda_c l_c} + \frac{4Ll_p\tilde{f}_1(R_0,d_R,d_{\max},d_{\min})}{R_0^2}\sin^4\left(\frac{\eta_0}{2}\right)$$
$$+\frac{2l_B L(1-\theta_c)^2}{l_e^2(a\kappa_D)^2 K_1(a\kappa_D)^2} g_0\left(\kappa_D R_0, \kappa_D d_R, \frac{d_{\max}}{d_R}, \frac{d_{\min}}{d_R}\right) - F_R L\cos\left(\frac{\eta_0}{2}\right) + \frac{2M_R L}{R_0}\sin\left(\frac{\eta_0}{2}\right)\tilde{f}_2(R_0,d_R,d_{\max},d_{\min})$$
$$-\frac{4l_B\xi_1(\theta_c,f_1,f_2)^2}{l_e^2}\frac{L\cos\left(\Delta\bar{\Phi}\right)\exp\left(-\frac{\lambda_h^*}{2\lambda_c}\right)}{(a\kappa_1)^2 K_1'(a\kappa_1)^2}\left(g_0\left(\kappa_1 R_0,\kappa_1 d_R,\frac{d_{\max}}{d_R},\frac{d_{\min}}{d_R}\right) + \frac{\bar{g}_0}{\kappa_1} g_1\left(\kappa_1 R_0,\kappa_1 d_R,\frac{d_{\max}}{d_R},\frac{d_{\min}}{d_R}\right)\sin\eta_0\right)$$
$$+\frac{4l_B\xi_2(\theta_c,f_1,f_2)^2}{l_e^2}\frac{L\cos\left(2\Delta\bar{\Phi}\right)\exp\left(-\frac{2\lambda_h^*}{\lambda_c}\right)}{(a\kappa_2)^2 K_2'(a\kappa_2)^2}\left(g_0\left(\kappa_2 R_0,\kappa_2 d_R,\frac{d_{\max}}{d_R},\frac{d_{\min}}{d_R}\right) + \frac{4\bar{g}_0}{\kappa_2} g_1\left(\kappa_2 R_0,\kappa_2 d_R,\frac{d_{\max}}{d_R},\frac{d_{\min}}{d_R}\right)\sin\eta_0\right)$$
$$-\frac{M_R L\theta_R^2}{4}\frac{\tilde{f}_2(R_0,d_R,d_{\max},d_{\min})}{R_0}\sin\left(\frac{\eta_0}{2}\right)^{-1} + \frac{2l_B L}{l_e^2}\sum_{n=-\infty}^{\infty}\frac{g_{img}(n,\kappa_n R_0,\kappa_n d_R, d_{\max}/d_R, d_{\min}/d_R;a)}{(\kappa_n a K_n'(\kappa_n a))^2}\xi_n(f_1,f_{2,}\theta)^2, \tag{13.30}$$

where

$$\langle\alpha_\eta\rangle_{\Delta g} = \left[\frac{4l_p\tilde{f}_1(R_0,d_R,d_{\max},d_{\min})}{R_0^2}\left(3\cos^2\left(\frac{\eta_0}{2}\right)\sin^2\left(\frac{\eta_0}{2}\right) - \sin^4\left(\frac{\eta_0}{2}\right)\right) + \frac{F_R}{4}\cos\left(\frac{\eta_0}{2}\right)\right.$$
$$+\frac{4l_B\xi_1(\theta_c,f_1,f_2)^2}{l_e^2}\frac{\cos\left(\Delta\bar{\Phi}\right)}{(a\kappa_1)^2 K_1'(a\kappa_1)^2}\frac{\bar{g}_0}{\kappa_1} g_1(\kappa_1 R_0,\kappa_1 d_R, d_{\max}/d_R, d_{\min}/d_R)\sin\eta_0\exp\left(-\frac{\lambda_h^*}{2\lambda_c}\right)$$
$$-\frac{16l_B\xi_2(\theta_c,f_1,f_2)^2}{l_e^2}\frac{\cos\left(2\Delta\bar{\Phi}\right)}{(a\kappa_2)^2 K_2'(a\kappa_2)^2}\frac{\bar{g}_0}{\kappa_2} g_1(\kappa_2 R_0,\kappa_2 d_R, d_{\max}/d_R, d_{\min}/d_R)\sin\eta_0\exp\left(-\frac{2\lambda_h^*}{\lambda_c}\right)$$
$$\left.-\frac{M_R\tilde{f}_2(R_0,d_R,d_{\max},d_{\min})}{2R_0}\sin\left(\frac{\eta_0}{2}\right)\right]. \tag{13.31}$$

The equations that minimize Eq. (13.30) with respect to $\Delta\bar{\Phi}$, $R_0$ and $\eta_0$ are

$$0=\frac{4l_B\xi_1(\theta_c,f_1,f_2)^2}{l_e^2}\frac{\sin\left(\Delta\bar{\Phi}\right)\exp\left(-\frac{\lambda_h^*}{2\lambda_c}\right)}{(a\kappa_1)^2K_1'(a\kappa_1)^2}$$

$$\left(g_0\left(\kappa_1R_0,\kappa_1d_R,\frac{d_{\max}}{d_R},\frac{d_{\min}}{d_R}\right)+\frac{\bar{g}_0}{\kappa_1}\left(1-\frac{1}{2^{3/2}l_p^{1/2}\left\langle\alpha_\eta\right\rangle_{\Delta g}^{1/2}}\right)g_1\left(\kappa_1R_0,\kappa_1d_R,\frac{d_{\max}}{d_R},\frac{d_{\min}}{d_R}\right)\sin\eta_0\right)$$

$$-\frac{8l_B\xi_2(\theta_c,f_1,f_2)^2}{l_e^2}\frac{\sin\left(2\Delta\bar{\Phi}\right)\exp\left(-\frac{2\lambda_h^*}{\lambda_c}\right)}{(a\kappa_2)^2K_2'(a\kappa_2)^2}$$

$$\left(g_0\left(\kappa_2R_0,\kappa_2d_R,\frac{d_{\max}}{d_R},\frac{d_{\min}}{d_R}\right)+\frac{4\bar{g}_0}{\kappa_2}\left(1-\frac{1}{2^{3/2}l_p^{1/2}\left\langle\alpha_\eta\right\rangle_{\Delta g}^{1/2}}\right)g_1\left(\kappa_2R_0,\kappa_2d_R,\frac{d_{\max}}{d_R},\frac{d_{\min}}{d_R}\right)\sin\eta_0\right),$$

(13.32)

$$0=\frac{2l_B(1-\theta_c)^2}{l_e^2(a\kappa_D)^2R_0K_1(a\kappa_D)^2}q_0\left(\kappa_DR_0,\kappa_Dd_R,\frac{d_{\max}}{d_R},\frac{d_{\min}}{d_R}\right)+\frac{1}{2^{3/2}\left\langle\alpha_\eta\right\rangle_{\Delta g}^{1/2}l_p^{1/2}}\frac{\partial\left\langle\alpha_\eta\right\rangle_{\Delta g}}{\partial R_0}+\frac{d_R^2}{2}\frac{\partial\alpha_H}{\partial R_0}$$

$$-\frac{4l_B\xi_1(\theta_c,f_1,f_2)^2}{R_0l_e^2}\frac{\cos\left(\Delta\bar{\Phi}\right)\exp\left(-\frac{\lambda_h^*}{2\lambda_c}\right)}{(a\kappa_1)^2K_1'(a\kappa_1)^2}\left(q_0\left(\kappa_1R_0,\kappa_1d_R,\frac{d_{\max}}{d_R},\frac{d_{\min}}{d_R}\right)+\frac{\bar{g}_0}{\kappa_1}q_1\left(\kappa_1R_0,\kappa_1d_R,\frac{d_{\max}}{d_R},\frac{d_{\min}}{d_R}\right)\sin\eta_0\right)$$

$$+\frac{4l_B\xi_2(\theta_c,f_1,f_2)^2}{R_0l_e^2}\frac{\cos\left(2\Delta\bar{\Phi}\right)\exp\left(-\frac{2\lambda_h^*}{\lambda_c}\right)}{(a\kappa_2)^2K_2'(a\kappa_2)^2}\left(q_0\left(\kappa_2R_0,\kappa_2d_R,\frac{d_{\max}}{d_R},\frac{d_{\min}}{d_R}\right)+\frac{4\bar{g}_0}{\kappa_2}q_1\left(\kappa_2R_0,\kappa_2d_R,\frac{d_{\max}}{d_R},\frac{d_{\min}}{d_R}\right)\sin\eta_0\right)$$

$$+\frac{l_p\theta_R^2}{R_0^3}\tilde{h}_1(R_0,d_R,d_{\max},d_{\min})\sin^2\left(\frac{\eta_0}{2}\right)-\frac{4l_p\tilde{h}_1(R_0,d_R,d_{\max},d_{\min})}{R_0^3}\sin^4\left(\frac{\eta_0}{2}\right)-\frac{2M_R}{R_0^2}\sin\left(\frac{\eta_0}{2}\right)\tilde{h}_2(R_0,d_R,d_{\max},d_{\min})$$

$$+\frac{M_R\theta_R^2}{4}\frac{\tilde{h}_2(R_0,d_R,d_{\max},d_{\min})}{R_0^2}\sin\left(\frac{\eta_0}{2}\right)^{-1}+\frac{2l_B}{l_e^2R_0}\sum_{n=-\infty}^{\infty}\frac{q_{img}(n,\kappa_nR_0,\kappa_nd_r,d_{\max}/d_R,d_{\min}/d_R;a)}{(\kappa_na)^2K_n'(\kappa_na)^2}\xi_n(f_1,f_2,\theta)^2,$$

(13.33)

and

$$0=\frac{1}{4}\left(\frac{1}{2l_pF_R}\right)^{1/2}\frac{\sin\left(\frac{\eta_0}{2}\right)}{\cos\left(\frac{\eta_0}{2}\right)^{3/2}}-\frac{7M_R^2}{64\left(2F_R\right)^{1/2}l_p^{3/2}}\frac{\sin\left(\frac{\eta_0}{2}\right)}{\cos\left(\frac{\eta_0}{2}\right)^{9/2}}$$

$$-\frac{4l_B\xi_1(\theta_c,f_1,f_2)^2}{l_e^2}\frac{L\cos\left(\Delta\bar{\Phi}\right)\exp\left(-\frac{\lambda_h^*}{2\lambda_c}\right)}{(a\kappa_1)^2K_1'(a\kappa_1)^2}\frac{\bar{g}_0}{\kappa_1}g_1(\kappa_1R_0,\kappa_1d_R,d_{\max}/d_R,d_{\min}/d_R)\cos\eta_0$$

$$+\frac{16l_B\xi_2(\theta_c,f_1,f_2)^2}{l_e^2}\frac{L\cos\left(2\Delta\bar{\Phi}\right)\exp\left(-\frac{2\lambda_h^*}{\lambda_c}\right)}{(a\kappa_2)^2K_2'(a\kappa_2)^2}\frac{\bar{g}_0}{\kappa_2}g_1(\kappa_2R_0,\kappa_2d_R,d_{\max}/d_R,d_{\min}/d_R)\cos\eta_0$$

$$+\frac{d_R^2}{2}\frac{\partial\alpha_H}{\partial\eta_0}+\frac{1}{2^{3/2}\left\langle\alpha_\eta\right\rangle_{\Delta g}^{1/2}l_p^{1/2}}\frac{\partial\left\langle\alpha_\eta\right\rangle_{\Delta g}}{\partial\eta_0}+\frac{M_R}{R_0}\cos\left(\frac{\eta_0}{2}\right)\tilde{f}_2(R_0,d_R,d_{\max},d_{\min})+\frac{F_R}{2}\sin\left(\frac{\eta_0}{2}\right) \quad (13.34)$$

$$-\frac{l_p\theta_R^2}{R_0^2}\tilde{f}_1(R_0,d_R,d_{\max},d_{\min})\sin\left(\frac{\eta_0}{2}\right)\cos\left(\frac{\eta_0}{2}\right)+\frac{8l_p\tilde{f}_1(R_0,d_R,d_{\max},d_{\min})}{R_0^2}\cos\left(\frac{\eta_0}{2}\right)\sin^3\left(\frac{\eta_0}{2}\right)$$

$$+\frac{M_R\theta_R^2}{8}\frac{\tilde{f}_2(R_0,d_R,d_{\max},d_{\min})}{R_0}\cos\left(\frac{\eta_0}{2}\right)\sin\left(\frac{\eta_0}{2}\right)^{-2}.$$

We have for the partial derivatives of $\left\langle\alpha_\eta\right\rangle_{\Delta g}$ with respect to $R_0$ and $\eta_0$

$$\frac{\partial\left\langle\alpha_\eta\right\rangle_{\Delta g}}{\partial R_0}=\left[-\frac{4l_p\tilde{h}_1(R_0,d_R,d_{\max},d_{\min})}{R_0^3}\left(3\cos^2\left(\frac{\eta_0}{2}\right)\sin^2\left(\frac{\eta_0}{2}\right)-\sin^4\left(\frac{\eta_0}{2}\right)\right)\right.$$

$$+\frac{M_R\tilde{h}_2(R_0,d_R,d_{\max},d_{\min})}{2R_0^2}\sin\left(\frac{\eta_0}{2}\right)$$

$$+\frac{4l_B\xi_1(\theta_c,f_1,f_2)^2}{R_0l_e^2}\frac{\bar{g}_0}{\kappa_1}\frac{\cos\left(\Delta\bar{\Phi}\right)\exp\left(-\frac{\lambda_h^*}{2\lambda_c}\right)}{(a\kappa_1)^2K_1'(a\kappa_1)^2}q_1\left(\kappa_1R_0,\kappa_1d_R,\frac{d_{\max}}{d_R},\frac{d_{\min}}{d_R}\right)\sin\eta_0 \quad (13.35)$$

$$\left.-\frac{16l_B\xi_2(\theta_c,f_1,f_2)^2}{R_0l_e^2}\frac{\bar{g}_0}{\kappa_2}\frac{\cos\left(2\Delta\bar{\Phi}\right)\exp\left(-\frac{2\lambda_h^*}{\lambda_c}\right)}{(a\kappa_2)^2K_2'(a\kappa_2)^2}q_1\left(\kappa_2R_0,\kappa_2d_R,\frac{d_{\max}}{d_R},\frac{d_{\min}}{d_R}\right)\sin\eta_0\right],$$

$$\frac{\partial\left\langle\alpha_\eta\right\rangle_{\Delta g}}{\partial\eta_0}=\left[\frac{4l_p\tilde{f}_1(R_0,d_R,d_{\max},d_{\min})}{R_0^2}\left(3\cos^3\left(\frac{\eta_0}{2}\right)\sin\left(\frac{\eta_0}{2}\right)-5\sin^3\left(\frac{\eta_0}{2}\right)\cos\left(\frac{\eta_0}{2}\right)\right)-\frac{F_R}{8}\sin\left(\frac{\eta_0}{2}\right)\right.$$

$$+\frac{4l_B\xi_1(\theta_c,f_1,f_2)^2}{l_e^2}\frac{\cos\left(\Delta\bar{\Phi}\right)}{(a\kappa_1)^2K_1'(a\kappa_1)^2}\frac{\bar{g}_0}{\kappa_1}g_1(\kappa_1R_0,\kappa_1d_R,d_{\max}/d_R,d_{\min}/d_R)\cos\eta_0\exp\left(-\frac{\lambda_h^*}{2\lambda_c}\right)$$

$$-\frac{16l_B\xi_2(\theta_c,f_1,f_2)^2}{l_e^2}\frac{\cos\left(2\Delta\bar{\Phi}\right)}{(a\kappa_2)^2K_2'(a\kappa_2)^2}\frac{\bar{g}_0}{\kappa_2}g_1(\kappa_2R_0,\kappa_2d_R,d_{\max}/d_R,d_{\min}/d_R)\cos\eta_0\exp\left(-\frac{2\lambda_h^*}{\lambda_c}\right)$$

$$\left.-\frac{M_R\tilde{f}_2(R_0,d_R,d_{\max},d_{\min})}{4R_0}\cos\left(\frac{\eta_0}{2}\right)\right].$$

(13.36)

The equation for the minimization of the free energy with respect to $\theta_R$ remains as Eq. (12.9), for which the approximate solution Eq. (12.23) can be used. On minimization of Eq. (13.30), with respect to $d_R$ and $\lambda_h^*$ we may write

$$0=\frac{1}{2^{3/2}\left\langle\alpha_\eta\right\rangle_{\Delta g}^{1/2} l_p^{1/2}}\frac{\partial\left\langle\alpha_\eta\right\rangle_{\Delta g}}{\partial d_R}+\frac{2l_B L}{l_e^2 d_R}\sum_{n=-\infty}^{\infty}\frac{m_{img}(n,\kappa_n R_0,\kappa_n d_R,d_{\max}/d_R,d_{\min}/d_R;a)}{(\kappa_n a)^2 K_n'(\kappa_n a))^2}\xi_n(f_1,f_2,\theta)^2$$

$$+\frac{2l_B(1-\theta_c)^2}{d_R l_e^2}\frac{1}{(a\kappa_D)^2 K_1(a\kappa_D)^2}m_0\left(\kappa_D R_0,\kappa_D d_R,\frac{d_{\max}}{d_R},\frac{d_{\min}}{d_R}\right)$$

$$-\frac{4l_B\xi_1(\theta_c,f_1,f_2)^2}{l_e^2 d_R}\frac{\cos\left(\Delta\bar{\Phi}\right)\exp\left(-\frac{\lambda_h^*}{2\lambda_c}\right)}{(a\kappa_1)^2 K_1'(a\kappa_1)^2}\left(m_0\left(\kappa_1 R_0,\kappa_1 d_R,\frac{d_{\max}}{d_R},\frac{d_{\min}}{d_R}\right)+\frac{\bar{g}_0}{\kappa_1}m_1\left(\kappa_1 R_0,\kappa_1 d_R,\frac{d_{\max}}{d_R},\frac{d_{\min}}{d_R}\right)\sin\eta_0\right)$$

$$+\frac{4l_B\xi_2(\theta_c,f_1,f_2)^2}{l_e^2 d_R}\frac{\cos\left(2\Delta\bar{\Phi}\right)\exp\left(-\frac{2\lambda_h^*}{\lambda_c}\right)}{(a\kappa_2)^2 K_2'(a\kappa_2)^2}\left(m_0\left(\kappa_2 R_0,\kappa_2 d_R,\frac{d_{\max}}{d_R},\frac{d_{\min}}{d_R}\right)+\frac{4\bar{g}_0}{\kappa_2}m_1\left(\kappa_2 R_0,\kappa_2 d_R,\frac{d_{\max}}{d_R},\frac{d_{\min}}{d_R}\right)\sin\eta_0\right)$$

$$-\frac{l_p\theta_R^2}{d_R R_0^2}\tilde{l}_1(R_0,d_R,d_{\max},d_{\min})\sin^2\left(\frac{\eta_0}{2}\right)+d_R\alpha_H-\frac{l_p\theta_R^4}{2d_R^3}+\frac{4l_p}{d_R R_0^2}\tilde{l}_1(R_0,d_R,d_{\max},d_{\min})\sin^4\left(\frac{\eta_0}{2}\right)$$

$$+\frac{2M_R}{d_R R_0}\tilde{l}_2(R_0,d_R,d_{\max},d_{\min})\sin\left(\frac{\eta_0}{2}\right)-\frac{M_R\theta_R^2}{4d_R R_0}\tilde{l}_2(R_0,d_R,d_{\max},d_{\min})\sin\left(\frac{\eta_0}{2}\right)^{-1},$$

(13.37)

and

$$0=-\frac{1}{16}\frac{(l_c+\lambda_c)^2}{(\lambda_h^*)^2 l_c}+\frac{2l_B\xi_1(\theta_c,f_1,f_2)^2}{l_e^2}\frac{\cos\left(\Delta\bar{\Phi}\right)\exp\left(-\frac{\lambda_h^*}{2\lambda_c}\right)}{(a\kappa_1)^2 K_1'(a\kappa_1)^2}$$

$$\left(g_0\left(\kappa_1 R_0,\kappa_1 d_R,\frac{d_{\max}}{d_R},\frac{d_{\min}}{d_R}\right)+\frac{\bar{g}_0}{\kappa_1}\left(1-\frac{1}{2^{3/2}l_p^{1/2}\left\langle\alpha_\eta\right\rangle_{\Delta g}^{1/2}}\right)g_1\left(\kappa_1 R_0,\kappa_1 d_R,\frac{d_{\max}}{d_R},\frac{d_{\min}}{d_R}\right)\sin\eta_0\right)$$

(13.38)

$$-\frac{8l_B\xi_2(\theta_c,f_1,f_2)^2}{l_e^2}\frac{\cos\left(2\Delta\bar{\Phi}\right)\exp\left(-\frac{2\lambda_h^*}{\lambda_c}\right)}{(a\kappa_2)^2 K_2'(a\kappa_2)^2}$$

$$\left(g_0\left(\kappa_2 R_0,\kappa_2 d_R,\frac{d_{\max}}{d_R},\frac{d_{\min}}{d_R}\right)+\frac{4\bar{g}_0}{\kappa_n}\left(1-\frac{1}{2^{3/2}l_p^{1/2}\left\langle\alpha_\eta\right\rangle_{\Delta g}^{1/2}}\right)g_1\left(\kappa_2 R_0,\kappa_2 d_R,\frac{d_{\max}}{d_R},\frac{d_{\min}}{d_R}\right)\sin\eta_0\right)$$

The partial derivatives of $\left\langle\alpha_\eta\right\rangle_{\Delta g}$ with respect to $d_R$ is given by

$$
\frac{\partial \langle \alpha_\eta \rangle_{\Delta g}}{\partial d_R} = \left[ \frac{4 l_p \tilde{l}_1(R_0, d_R, d_{\max}, d_{\min})}{d_R R_0^2} \left( 3\cos^2\left(\frac{\eta_0}{2}\right)\sin^2\left(\frac{\eta_0}{2}\right) - \sin^4\left(\frac{\eta_0}{2}\right) \right) \right.
$$

$$
+ \frac{4 l_B \xi_1(\theta_c, f_1, f_2)^2}{d_R l_e^2} \frac{\cos(\Delta\bar{\Phi})}{(a\kappa_1)^2 K_1'(a\kappa_1)^2} \frac{\bar{g}_0}{\kappa_1} m_1(\kappa_1 R_0, \kappa_1 d_R, d_{\max}/d_R, d_{\min}/d_R) \sin\eta_0 \exp\left(-\frac{\lambda_h^*}{2\lambda_c}\right)
$$

$$
- \frac{16 l_B \xi_2(\theta_c, f_1, f_2)^2}{d_R l_e^2} \frac{\cos(2\Delta\bar{\Phi})}{(a\kappa_2)^2 K_2'(a\kappa_2)^2} \frac{\bar{g}_0}{\kappa_2} m_1(\kappa_2 R_0, \kappa_2 d_R, d_{\max}/d_R, d_{\min}/d_R) \sin\eta_0 \exp\left(-\frac{2\lambda_h^*}{\lambda_c}\right)
$$

$$
\left. - \frac{M_R \tilde{l}_2(R_0, d_R, d_{\max}, d_{\min})}{2R_0} \sin\left(\frac{\eta_0}{2}\right) \right].
$$

(13.39)

This system of Equations exhibits three different roots corresponding to different braiding states. One root is where $\lambda = \infty$, where only Eqs. (13.32), (13.33) and (13.37) matter, where helix effects only show up image charge contribution. For the parameter range appropriate, for mono-valent salt solution, we fit the solution to available experimental data [17]. A second root is a finite value of $\lambda_h^*$ with the trivial solution of Eq. (13.32), $\Delta\bar{\Phi} = 0$. The last solution is where $\Delta\bar{\Phi} \neq 0$ and finite $\lambda_h^*$, where the molecules are tightly braided. The numerical results of solving Eqs (13.32), (13.33), (13.34) , (13.37) and (13.38) are presented in [18] for a range of parameter values, as well as the solution to the system of equations presented in Section 12. For the first two states we use our first choice of $d_{\max} = -d_{\min} = R_0 - 2a$ , whereas for the third root, corresponding to a tight braid we use, perhaps, the more appropriate choice of $-d_{\min} = R_0 - 2a$ and $d_{\max}$ given by Eq. (5.11) in $\alpha_H$ and set to $\infty$ elsewhere.

## Acknowledgements

D.J. Lee would like to acknowledge useful discussions with R. Cortini , A. Korte, A. A. Korynshev, E. L. Starostin and G.H.M. van der Heijden, and. This work was initially inspired by joint work that has been supported by the United Kingdom Engineering and Physical Sciences Research Council (grant EP/H004319/1). He would also like to acknowledge the support of the Human Frontiers Science Program (grant RGP0049/2010-C102).

## Appendix A: Generalized Gaussian Averaging

We want to calculate averages of the form

$$
\left\langle F\left( X(s), \frac{dX(s)}{ds} \right) \right\rangle_X = \frac{\int DX(s) F\left( X(s), \frac{dX(s)}{ds} \right) \exp\left[-E[X(s)]\right]}{\int DX(s) \exp\left[-E[X(s)]\right]}, \tag{A.1}
$$

where

$$E[X(s)] = \int_{-\infty}^{\infty} ds \int_{-\infty}^{\infty} ds' X(s) G^{-1}(s-s') X(s'). \qquad \text{(A.2)}$$

To perform the average in Eq. (A.2) we can first start by writing

$$\left\langle F\left( X(s), \frac{dX(s)}{ds} \right) \right\rangle_X = \int_{-\infty}^{\infty} dx \int_{-\infty}^{\infty} dx' F\left(x, x'\right) \left\langle \delta(x - X(s)) \delta\left( x' - \frac{dX(s)}{ds} \right) \right\rangle_X . \qquad \text{(A.3)}$$

We can recast Eq. (A.3) as

$$\left\langle F\left( X(s), \frac{dX(s)}{ds} \right) \right\rangle_X = \frac{1}{(2\pi)^2} \int_{-\infty}^{\infty} dx \int_{-\infty}^{\infty} dx' \int_{-\infty}^{\infty} dp \int_{-\infty}^{\infty} dp' F\left(x, x'\right) \exp(-ipx) \exp(-ip'x')$$
$$\left\langle \exp\left(ipX(s)\right) \exp\left( ip' \frac{dX(s)}{ds} \right) \right\rangle_X . \qquad \text{(A.4)}$$

We are then left with computing the average within Eq. (A.4), which is

$$\left\langle \exp\left(ipX(s)\right) \exp\left( ip' \frac{dX(s)}{ds} \right) \right\rangle_X$$
$$= \frac{\int D\tilde{X}(k) \exp\left( \frac{i}{2\pi} \int_{-\infty}^{\infty} dk \left( p + ikp' \right) \tilde{X}(k) \exp\left(ikx\right) \right) \exp\left( -\frac{1}{4\pi} \int_{-\infty}^{\infty} dk \tilde{X}(k) G^{-1}(k) \tilde{X}(-k) \right)}{\int D\tilde{X}(k) \exp\left( -\frac{1}{4\pi} \int_{-\infty}^{\infty} dk \tilde{X}(k) G^{-1}(k) \tilde{X}(-k) \right)} . \qquad \text{(A.5)}$$

This can be rewritten as

$$\left\langle \exp\left(ipX(s)\right) \exp\left( ip' \frac{dX(s)}{ds} \right) \right\rangle_X$$
$$= \frac{\int D\tilde{X}(k) \exp\left( -\frac{1}{4\pi} \int_{-\infty}^{\infty} dk \left( \tilde{X}(k) - G(k)\left(p + ikp'\right) \exp\left(ikx\right) \right) G^{-1}(k) \left( \tilde{X}(-k) - G(k)\left(p - ikp'\right) \exp\left(-ikx\right) \right) \right)}{\int D\tilde{X}(k) \exp\left( -\frac{1}{4\pi} \int_{-\infty}^{\infty} dk \tilde{X}(k) G^{-1}(k) \tilde{X}(-k) \right)}$$
$$\times \exp\left( \frac{1}{4\pi} \int_{-\infty}^{\infty} dk G(k) \left( p^2 + k^2 p'^2 \right) \right) = \exp\left( \frac{1}{4\pi} \int_{-\infty}^{\infty} dk G(k) \left( p^2 + k^2 p'^2 \right) \right). \qquad \text{(A.6)}$$

Finally, substitution of Eq. (A.6) into Eq. (A.4) yields the general formula

$$\left\langle F\left(X(s),\frac{dX(s)}{ds}\right)\right\rangle_X = \frac{1}{(2\pi)^2}\int_{-\infty}^{\infty}dx\int_{-\infty}^{\infty}dx'\int_{-\infty}^{\infty}dp\int_{-\infty}^{\infty}dp'F(x,x')\exp(ipx)\exp(ip'x')\exp\left(-\frac{p^2d_X^2}{2}\right)\exp\left(-\frac{p^2\theta_X^2}{2}\right)$$

$$= \frac{1}{2\pi d_X\theta_X}\int_{-\infty}^{\infty}dx\int_{-\infty}^{\infty}dx'F(x,x')\exp\left(-\frac{1}{2}\frac{x^2}{d_X^2}\right)\exp\left(-\frac{1}{2}\frac{x'^2}{\theta_X^2}\right), \tag{A.7}$$

where

$$d_X^2 = \frac{1}{2\pi}\int_{-\infty}^{\infty}dkG(k), \qquad \theta_X^2 = \frac{1}{2\pi}\int_{-\infty}^{\infty}dkk^2G(k). \tag{A.8}$$

## Appendix B Evaluation of the integrals appearing in Eqs. (10.3) and (10.4)

Here we concern ourselves with the evaluation of the integrals

$$I_1 = \frac{1}{2\pi}\int_{-\infty}^{\infty}dk\frac{1}{k^4+1+\gamma k^2} \quad \text{and} \quad I_2 = \frac{1}{2\pi}\int_{-\infty}^{\infty}dk\frac{k^2}{k^4+1+\gamma k^2}. \tag{B.1}$$

We first may factorize the denominators of these integrals so that

$$I_1 = \frac{1}{2\pi}\int_{-\infty}^{\infty}dk\frac{1}{(k^2-K^+)(k^2-K^-)} \quad \text{and} \quad I_2 = \frac{1}{2\pi}\int_{-\infty}^{\infty}dk\frac{k^2}{(k^2-K^+)(k^2-K^-)}, \tag{B.2}$$

where

$$K^{\pm} = \frac{-\gamma \pm \sqrt{\gamma^2-4}}{2}. \tag{B.3}$$

In our analysis we need to examine two separate cases. These are the case where $-2<\gamma<2$ , where $K^+$ and $K^+$ are complex, and the case $\gamma \geq 2$ where the roots are negative and real. To complete the factorization of the denominator appearing in the integrals in Eq. (B.2), we need to find the four roots to the equations $k^2 = K^+$ and $k^2 = K^-$ , $k_1$, $k_2$, $k_3$ and $k_4$. This allows us to write

$$I_1 = \frac{1}{2\pi}\int_{-\infty}^{\infty}dk\frac{1}{(k-k_1)(k-k_2)(k-k_3)(k-k_4)} \quad \text{and} \quad I_2 = \frac{1}{2\pi}\int_{-\infty}^{\infty}dk\frac{k^2}{(k-k_1)(k-k_2)(k-k_3)(k-k_4)}, \tag{B.4}$$

where for $-2<\gamma<2$

$$k_1 = \frac{1}{\sqrt{2}}\left(\sqrt{1-\frac{\gamma}{2}} + i\sqrt{1+\frac{\gamma}{2}}\right) = -k_2 \quad \text{and} \quad k_3 = \frac{1}{\sqrt{2}}\left(-\sqrt{1-\frac{\gamma}{2}} + i\sqrt{1+\frac{\gamma}{2}}\right) = -k_4, \tag{B.5}$$

and for $\gamma \geq 2$

$$k_1 = \frac{i}{\sqrt{2}}\left(\gamma + \sqrt{\gamma^2 - 4}\right)^{1/2} = -k_2 \quad \text{and} \quad k_3 = \frac{i}{\sqrt{2}}\left(\gamma + \sqrt{\gamma^2 - 4}\right)^{1/2} = -k_4. \tag{B.6}$$

We can then use the standard method of complex contour integration to evaluate these integrals, where we close the contours of integration in the top half of the complex plane, by adding the negligible contribution from the integration contour along the perimeter of a semi-circle connecting the points $k = -\infty$ and $k = \infty$. This technique can be found in any university level mathematics textbook. We pick up contributions from the residues of the poles at $k_1$ and $k_3$. This allows us to write, through the residue theorem

$$I_1 = \frac{i}{(k_1 - k_2)(k_1 - k_3)(k_1 - k_4)} + \frac{i}{(k_3 - k_2)(k_3 - k_1)(k_3 - k_4)}, \tag{B.7}$$

$$I_2 = \frac{i(k_1)^2}{(k_1 - k_2)(k_1 - k_3)(k_1 - k_4)} + \frac{i(k_3)^2}{(k_3 - k_2)(k_3 - k_1)(k_3 - k_4)}. \tag{B.8}$$

Substituting in Eq. (B.5) for $-2 < \gamma < 2$ and Eq. (B.6) for $\gamma \geq 2$ into Eqs. (B.7) and (B.8) yields on algebraic manipulation

$$I_1 = I_2 = \frac{1}{2\sqrt{\gamma + 2}} \qquad \text{for } \gamma > -2. \tag{B.9}$$

This result is then substituted into Eqs. (10.3) and (10.4) to produce Eq. (10.6).

## Appendix C: Path integral reformulation of ensemble averaging over random field $\Delta g^0(s)$

We will suppose that functional dependence of $\Delta\Phi_0 = \Delta\Phi_0\left[\frac{d\Delta g^0(s)}{ds}\right]$ is determined through the equation of the form (for instance Eq. (11.3))

$$\frac{l_c}{2}\frac{d\Delta g^0(s)}{ds} = \frac{l_c}{2} f[\Delta\Phi(s)] \equiv \frac{l_c}{2}\left(\frac{d^2\Delta\Phi_0(s)}{ds^2}\right) - \frac{\partial h(\Delta\Phi_0(s))}{\partial\Delta\Phi_0(s)}, \tag{C.1}$$

where $\Delta g^0(s)$ is a Gaussian distributed random field of the form discussed in Section. 13. Eq. (C.1) minimizes a free energy of the form

$$F_T\left[\Delta\Phi_0, \frac{d\Delta g^0(s)}{ds}\right] = \frac{l_c}{4}\int_{-L/2}^{L/2} ds\left(\frac{d\Delta\Phi_0(s)}{ds} - \Delta g^0(s)\right)^2 + \int_{-L/2}^{L/2} dsh(\Delta\Phi_0(s)), \tag{C.2}$$

where provided that $L$ is sufficiently large we can take the limit $L \to \infty$ on most limits of integration in $s$. Therefore, the functional of $\Delta\Phi_0\left[\frac{d\Delta g^0(s)}{ds}\right]$ that is the minimum value of Eq. (C.2) can be written as

$$\tilde{F}_T\left[\Delta\Phi_0\left[\frac{d\Delta g^0(s)}{ds}\right]\right] = \frac{1}{4l_c}\int_{-L/2}^{L/2} ds\int_{-\infty}^{\infty} ds'\int_{-\infty}^{\infty} ds'' \frac{\partial h(\Delta\Phi_0(s'))}{\partial(\Delta\Phi_0(s'))}\frac{\partial h(\Delta\Phi_0(s''))}{\partial(\Delta\Phi_0(s''))}\mathrm{sgn}(z-z')\,\mathrm{sgn}(z-z'')$$
$$+\int_{-L/2}^{L/2} dsh(\Delta\Phi_0(s)). \tag{C.3}$$

Now let us consider the ensemble averaged free energy. The reads as

$$\left\langle F_T[\Delta g^0(s)]\right\rangle_{\Delta g} = \frac{1}{Z_{\Delta g}}\int D\Delta g^0(s)\tilde{F}_T\left[\Delta\Phi_0\left[\frac{d\Delta g^0(s)}{ds}\right]\right]\exp\left(-\frac{\lambda_c^{(0)}}{4}\int_{-\infty}^{\infty} ds\Delta g^0(s)^2\right), \tag{C.4}$$

where

$$Z_{\Delta g} = \int D\Delta g^0(s)\exp\left(-\frac{\lambda_c^{(0)}}{4}\int_{-\infty}^{\infty} ds\Delta g^0(s)^2\right). \tag{C.5}$$

Our goal here is to reformulate Eqs. (C.4) and (C.5) as path integrals over $\Delta\Phi_0(s)$. We are first able to rewrite Eq. (C.4)

$$\left\langle F_T[\Delta g^0(s)]\right\rangle_{\Delta g} = \frac{1}{\tilde{Z}_{\Delta g}}\int D\Delta g^0(s)\int Df(s)\tilde{F}_T[\Delta\Phi_0[f(s)]]\delta^\infty\left(f(s) - \frac{d\Delta g^0(s)}{ds}\right)\exp\left(-\frac{\lambda_c^{(0)}}{4}\int_{-\infty}^{\infty} ds\Delta g^0(s)^2\right), \tag{C.6}$$

with

$$\tilde{Z}_{\Delta g} = \int D\Delta g^0(s)\int Df(s)\delta^\infty\left(f(s) - \frac{d\Delta g^0(s)}{ds}\right)\exp\left(-\frac{\lambda_c^{(0)}}{4}\int_{-\infty}^{\infty} ds\Delta g^0(s)^2\right), \tag{C.7}$$

where $\delta^\infty\left(f(s) - \frac{d\Delta g^0(s)}{ds}\right)$ is the functional delta function, which is a product of delta functions each taken for the variables $f(s)$ and $\frac{d\Delta g^0(s)}{ds}$ associated with a particular point $s$. We can then write Eq. (C.6) and (C.7) as

$$\left\langle F_T[\Delta g^0(s)]\right\rangle_{\Delta g} \propto \int D\Delta g^0(s)\int Df(s)\int Dp(s)\tilde{F}_T[\Delta\Phi_0[f(s)]]\exp\left(i\int_{-\infty}^{\infty} dsp(s)\left(f(s)-\frac{d\Delta g^0(s)}{ds}\right)\right)$$

$$\exp\left(-\frac{\lambda_c^{(0)}}{4}\int_{-\infty}^{\infty} ds\Delta g^0(s)^2\right) \tag{C.8}$$

$$\propto \int Dp(s)\int Df(s)\tilde{F}_T[\Delta\Phi_0[f(s)]]\exp\left(i\int_{-\infty}^{\infty} dsp(s)f(s)\right)\exp\left(-\frac{2}{\lambda_c^{(0)}}\int_{-\infty}^{\infty} ds\left(\frac{dp(s)}{ds}\right)^2\right),$$

and

$$\tilde{Z}_{\Delta g} \propto \int Dp(s)\int Df(s)\exp\left(i\int_{-\infty}^{\infty} dsp(s)f(s)\right)\exp\left(-\frac{2}{\lambda_c^{(0)}}\int_{-\infty}^{\infty} ds\left(\frac{dp(s)}{ds}\right)^2\right). \tag{C.9}$$

We now make a functional change of variables from $f(s)$ to $\Delta\Phi_0(s)$ through Eq.(C.1) . This variable change can be rewritten as

$$\left\langle F_T[\Delta g^0(s)]\right\rangle_{\Delta g} = \frac{1}{\bar{Z}_{\Delta\Phi}}\int Dp(s)\int D\Delta\Phi_0(s)J[\Delta\Phi_0(s)]\tilde{F}_T[\Delta\Phi_0(s)]$$

$$\exp\left(i\int_{-\infty}^{\infty} dsp(s)\left[\left(\frac{d^2\Delta\Phi_0(s)}{ds^2}\right)-\frac{2}{l_c}\frac{\partial h(\Delta\Phi_0(s))}{\partial\Delta\Phi_0(s)}\right]\right)\exp\left(-\frac{2}{\lambda_c^{(0)}}\int_{-\infty}^{\infty} ds\left(\frac{dp(s)}{ds}\right)^2\right), \tag{C.10}$$

$$\bar{Z}_{\Delta\Phi} = \int Dp(s)\int D\Delta\Phi_0(s)J[\Delta\Phi_0(s)]$$

$$\exp\left(i\int_{-\infty}^{\infty} dsp(s)\left[\left(\frac{d^2\Delta\Phi_0(s)}{ds^2}\right)-\frac{2}{l_c}\frac{\partial h(\Delta\Phi_0(s))}{\partial\Delta\Phi_0(s)}\right]\right)\exp\left(-\frac{2}{\lambda_c^{(0)}}\int_{-\infty}^{\infty} ds\left(\frac{dp(s)}{ds}\right)^2\right). \tag{C.11}$$

Here $J[\Delta\Phi_0(s)]$ is the Jacobian of the functional variable change, which we will now evaluate. We start by discretising $s$ into $N$ equally spaced points. We define the distance between two adjacent points as $\varepsilon$, so that we may write

$$f_i \equiv \frac{2\Delta\Phi_{0,i}-\Delta\Phi_{0,i-1}-\Delta\Phi_{0,i+1}}{\varepsilon^2}-\frac{2}{l_c}\frac{\partial h(\Delta\Phi_{0,i})}{\partial\Delta\Phi_{0,i}}, \tag{C.12}$$

which becomes Eq. (C.1) in the limit $\varepsilon \to 0$. We also note that by definition of path integration we have that

$$\int Df(s) \propto \lim_{\substack{\varepsilon\to 0\\ N\to\infty}}\prod_{i=-1}^{N}\int_{-\infty}^{\infty} df_i \quad \text{and} \quad \int D\Delta\Phi_0(s) \propto \lim_{\substack{\varepsilon\to 0\\ N\to\infty}}\prod_{i=1}^{N}\int_{-\infty}^{\infty} d\Delta\Phi_{0,i}. \tag{C.13}$$

Therefore, we may write the functional Jacobian as

$$J[\Delta\Phi_0(s)] \propto \lim_{\substack{\varepsilon\to 0\\ N\to\infty}}\left|D_N\right|, \tag{C.14}$$

where $D_N$ is the determinate of a $N \times N$ tri-diagonal matrix, namely

$$D_N = \begin{vmatrix} \frac{2}{\varepsilon^2} - \frac{2}{l_c}\frac{\partial^2 h(\Delta\Phi_{0,1})}{\partial\Delta\Phi_{0,1}^2} & -\frac{1}{\varepsilon^2} & 0 & & & \\ -\frac{1}{\varepsilon^2} & \frac{2}{\varepsilon^2} - \frac{2}{l_c}\frac{\partial^2 h(\Delta\Phi_{0,2})}{\partial\Delta\Phi_{0,2}^2} & \ddots & \ddots & & \\ 0 & \ddots & \ddots & \ddots & 0 & \\ & \ddots & \ddots & \frac{2}{\varepsilon^2} - \frac{2}{l_c}\frac{\partial^2 h(\Delta\Phi_{0,N-1})}{\partial\Delta\Phi_{0,N-1}^2} & -\frac{1}{\varepsilon^2} \\ & & 0 & -\frac{1}{\varepsilon^2} & \frac{2}{\varepsilon} - \frac{2}{l_c}\frac{\partial^2 h(\Delta\Phi_{0,N})}{\partial\Delta\Phi_{0,N}^2} \end{vmatrix}. \quad \text{(C.15)}$$

Therefore, $D_N$ satisfies the recursion relation

$$D_N = \left(\frac{2}{\varepsilon^2} - \frac{2}{l_c}\frac{\partial^2 h[\Delta\Phi_{0,N}]}{\partial\Delta\Phi_{0,N}^2}\right)D_{N-1} - \frac{1}{\varepsilon^4}D_{N-2}. \quad \text{(C.16)}$$

The values of $D_2$ and $D_1$, the determinates of $2\times 2$ matrix and $1\times 1$ matrix are boundary conditions on Eq. (C.16). However, they can be freely chosen to specify particular boundary condition choices on $\frac{\partial^2 h(\Delta\Phi_{0,1})}{\partial\Delta\Phi_{0,1}^2}$ and $\frac{\partial^2 h(\Delta\Phi_{0,2})}{\partial\Delta\Phi_{0,2}^2}$. We will make a choice for large $N$ and small $\varepsilon$ that makes functional Jacobian finite in the limit $L \to \infty$.

Let us suppose that we can write for large $N$ and small $\varepsilon$

$$D_N \approx \left(\frac{1}{\varepsilon}\right)^{2N} \prod_{i=1}^{N}\left(A + \varepsilon^2 B_i\right). \quad \text{(C.17)}$$

Substitution of Eq. (C.17) into (C.16) then yields

$$\left(A + \varepsilon^2 B_N\right)\left(A + \varepsilon^2 B_{N-1}\right) \approx \left(2 - \frac{2\varepsilon^2}{l_c}\frac{\partial^2 h[\Delta\Phi_{0,N}]}{\partial\Delta\Phi_{0,N}^2}\right)\left(A + \varepsilon^2 B_{N-1}\right) - 1. \quad \text{(C.18)}$$

Expanding Eq. (C.18) up to $O(\varepsilon^2)$ yields

$$A^2 + \varepsilon^2 A\left(B_N + B_{N-1}\right) \approx 2A - 1 - \frac{2A\varepsilon^2}{l_c}\frac{\partial^2 h[\Delta\Phi_{0,N}]}{\partial\Delta\Phi_{0,N}^2} + 2\varepsilon^2 B_{N-1}. \quad \text{(C.19)}$$

By comparing the leading order terms, we find that have the solution $A = 1$ (the other solution is $A = -1$, which we do not choose) and so we have from Eq. (C.19)

$$B_N \approx -\frac{2}{l_c}\frac{\partial^2 h[\Delta\Phi_{0,N}]}{\partial\Delta\Phi_{0,N}^2} + B_{N-1}. \tag{C.20}$$

A solution to Eq. (C.20) is

$$B_i \approx -\frac{2}{l_c}\sum_{j=1}^{i}\frac{\partial^2 h[\Delta\Phi_{0,i}]}{\partial\Delta\Phi_{0,i}^2} - \frac{2}{l_c}\tilde{B}. \tag{C.21}$$

Using Eqs. (C.17) and (C.21), we can write down a solution to Eq. (C.16)

$$D_N \approx \left(\frac{1}{\varepsilon}\right)^{2N}\prod_{i=1}^{N}\left(1-\varepsilon^2\left(\frac{2}{l_c}\sum_{j=1}^{i}\frac{\partial^2 h[\Delta\Phi_{0,i}]}{\partial\Delta\Phi_{0,i}^2} + \frac{2}{l_c}\tilde{B}\right)\right). \tag{C.22}$$

Since $\varepsilon$ is small, we can rewrite Eq. (C.22) as

$$D_N \approx \left(\frac{1}{\varepsilon}\right)^{2N}\exp\left(-\frac{2\varepsilon^2}{l_c}\sum_{i=1}^{N}\left(\sum_{j=1}^{N}\theta(\varepsilon i-\varepsilon j)\frac{\partial^2 h[\Delta\Phi_{0,i}]}{\partial\Delta\Phi_{0,i}^2} + \tilde{B}\right)\right). \tag{C.23}$$

Now, we choose

$$\tilde{B} = -\sum_{j=-N}^{N}\frac{\partial^2 h[\Delta\Phi_{0,i}]}{\partial\Delta\Phi_{0,i}^2}, \tag{C.24}$$

to insure that $J[\phi]$ is finite in the limit $L\to\infty$. Using, Eqs. (C.14), (C.23) and (C.24) we therefore find for the Jacobian

$$J[\Delta\Phi_0(s)] = \exp\left(-\frac{1}{l_c}\int_{-\infty}^{\infty}ds\int_{-\infty}^{\infty}ds'\,\mathrm{sgn}(s-s')\frac{\partial^2 h[\Delta\Phi_0(s')]}{\partial\Delta\Phi_0(s')^2}\right). \tag{C.25}$$

Using Eqs. (C.10) and (C.25), we can then express the average free energy as

$$\begin{aligned}
&\left\langle F_T[\Delta g^0(s)]\right\rangle_{\Delta g} = \frac{1}{\bar{Z}_{\Delta\Phi}}\int Dp(s)\int D\Delta\Phi_0(s)\tilde{F}_T[\Delta\Phi_0(s)]\\
&\exp\left(-\frac{1}{l_c}\int_{-\infty}^{\infty}ds\int_{-\infty}^{\infty}ds'\,\mathrm{sgn}(s-s')\frac{\partial h(\Delta\Phi_0(s'))}{\partial\Delta\Phi_0(s')}\right)\exp\left(-\frac{2}{\lambda_c^{(0)}}\int_{-\infty}^{\infty}ds\left(\frac{dp(s)}{ds}\right)^2\right)\\
&\exp\left(i\int_{-\infty}^{\infty}dsp(s)\left[\left(\frac{d^2\Delta\Phi_0(s)}{ds^2}\right)-\frac{2}{l_c}\frac{\partial h(\Delta\Phi_0(s))}{\partial\Delta\Phi_0(s)}\right]\right).
\end{aligned} \tag{C.26}$$

We can integrate out $p(s)$ yielding the final expressions

$$\left\langle F_T[\Delta g^0(s)]\right\rangle_{\Delta g} = \frac{1}{\bar{Z}_{\Delta\Phi}}\int D\Delta\Phi_0(s)\tilde{F}_T[\Delta\Phi_0(s)]\exp\left(-\frac{1}{l_c}\int_{-\infty}^{\infty} ds\int_{-\infty}^{\infty} ds'\,\mathrm{sgn}(s-s')\frac{\partial^2 h(\Delta\Phi_0(s'))}{\partial\Delta\Phi_0(s')^2}\right)$$
$$\exp\left(-\frac{\lambda_c^{(0)}}{4}\int_{-\infty}^{\infty} ds\int_{-\infty}^{\infty} ds'\left|s-s'\right|\left[\left(\frac{d^2\Delta\Phi_0(s)}{ds^2}\right)-\frac{2}{l_c}\frac{\partial h(\Delta\Phi_0(s))}{\partial\Delta\Phi_0(s)}\right]\left[\left(\frac{d^2\Delta\Phi_0(s')}{ds'^2}\right)-\frac{2}{l_c}\frac{\partial h(\Delta\Phi_0(s'))}{\partial\Delta\Phi_0(s')}\right]\right),$$

(C.27)

$$\bar{Z}_{\Delta\Phi} = \int D\Delta\Phi_0(s)\exp\left(-\frac{1}{l_c}\int_{-\infty}^{\infty} ds\int_{-\infty}^{\infty} ds'\,\mathrm{sgn}(s-s')\frac{\partial^2 h(\Delta\Phi_0(s'))}{\partial\Delta\Phi_0(s')^2}\right)$$
$$\exp\left(-\frac{\lambda_c^{(0)}}{4}\int_{-\infty}^{\infty} ds\int_{-\infty}^{\infty} ds'\left|s-s'\right|\left[\left(\frac{d^2\Delta\Phi_0(s)}{ds^2}\right)-\frac{2}{l_c}\frac{\partial h(\Delta\Phi_0(s))}{\partial\Delta\Phi_0(s)}\right]\left[\left(\frac{d^2\Delta\Phi_0(s')}{ds'^2}\right)-\frac{2}{l_c}\frac{\partial h(\Delta\Phi_0(s'))}{\partial\Delta\Phi_0(s')}\right]\right).$$
(C.28)

The linear response theory expression Eq. (13.6), that is used as the basis of the variational approximation used in the main text, corresponds to setting

$$\frac{2}{l_c}\frac{\partial h(\Delta\Phi_0(s))}{\partial\Delta\Phi_0(s)} = \frac{\Delta\Phi_0(s)}{\tilde{\lambda}_h} \qquad \text{and} \qquad \frac{2}{l_c}\frac{\partial^2 h(\Delta\Phi_0(s'))}{\partial\Delta\Phi_0(s')^2} = \frac{1}{\tilde{\lambda}_h} \tag{C.29}$$

in Eqs. (C.3), (C.27) and (C.28). A better variational approximation might be obtained by constructing an approximation for $\ln\bar{Z}_\phi$ in the same manner, exploiting the Gibbs-Bogoliubov inequality, as was considered for thermal fluctuations (for example see Eq. (6.8) ), using a trial functional of the form

$$\Psi[\Delta\Phi_0(s)] = \frac{\lambda_c^{(0)}}{4}\int_{-\infty}^{\infty} ds\int_{-\infty}^{\infty} ds'\left|s-s'\right|\left[\left(\frac{d^2\Delta\Phi_0(s)}{ds^2}\right)-\frac{\Delta\Phi_0(s)}{\tilde{\lambda}_h}\right]\left[\left(\frac{d^2\Delta\Phi_0(s')}{ds'^2}\right)-\frac{\Delta\Phi_0(s')}{\tilde{\lambda}_h}\right],$$

(C.30)

and expanding Eq. (C.28) around the partition function

$$Z_0 = \int D\Delta\Phi_0(s)\exp\left(-\Psi[\Delta\Phi_0(s)]\right). \tag{C.31}$$

---

[1] W. Helfrich, W. Harbcih, Chem. Scr. **25,** 32 (1985)
[2] A. A. Kornyshev, D. J. Lee, S. Leikin, A. Wynveen, J. Phys. Chem. B **114,** 11668 (2010)
[3] D. J. Lee, Phys. Rev. E. **88,** 022719 (2013)
[4] E. Starostin and G. Van der Heijden; http://www.ucl.ac.uk/~ucesgvd/braids.html
[5] D. J. Lee; http://arxiv.org/abs/1302.1420
[6] A. A. Kornyshev, D. J. Lee, S. Leikin, A. Wynveen, Rev. Mod. Phys. **79,** 943 (2007); and references contained therein.
[7] J. D. Moroz, P. Nelson, Proc. Natl. Acad. Sci. USA, **94**, 14418 (1997)
[8] D. J. Lee, R. Cortini, A. Korte, E. Starostin, G. Van der Heijden, A. A. Kornyshev, Soft Matter**, 9**, 9833 (2013)
[9] J. F. Marko, Phys. Rev. E, **55**, 1758 (1997)
[10] D. J. Lee, http://arxiv.org/abs/1312.5581
[11] F. Oosawa, Biopolymers **6**, 1633 (1968).

[12] A. A. Kornyshev, S. Leikin, J. Chem Phys **107,** 3656 (1997)
[13] D. J. Lee, J. Phys. Cond. Matt. **23,** 105102 (2011).
[14] A. Wynveen, D. J. Lee, A. Kornyshev, S. Leikin, Nucl. Acids Res. , **36**, 5540 (2008)
[15] A. Wynveen, D. J. Lee, A. Kornyshev, S. Leikin, Nucl. Acids Res., **39**, 7289 (2011)
[16] D. J. Lee, A. Wynveen, A. Kornyshev, Phys. Rev. E, **70**, 051913 (2004)
[17] D. J. Lee, in preparation.
[18] D. J. Lee, in preparation.